\documentclass[%
 reprint,
superscriptaddress,
preprintnumbers,
nofootinbib,
 amsmath,amssymb,
 aps, prl,
]{revtex4-2}

\usepackage{graphicx}% Include figure files
\usepackage{subcaption} % For subfigures (PRL allows this; avoid the older subfigure pkg)
\usepackage{dcolumn}% Align table columns on decimal point
\usepackage{bm}% bold math
\usepackage{comment}
\usepackage{xcolor}
\usepackage{rotating}
\usepackage{hyperref}% add hypertext capabilities

\begin{document}

\preprint{FERMILAB-PUB-26-0220-PPD}

\title{\textbf{Measurement of Inclusive Charged-Current $\bar{\nu}_{\mu}$ Scattering on C, CH, Fe, and Pb at $\langle E_{\bar{\nu}}\rangle \sim$~6~GeV with MINERvA} 
}

%% List of institution addresses, in command form.
\newcommand{\Rutgers}{Rutgers, The State University of New Jersey, Piscataway, New Jersey 08854, USA}
\newcommand{\Hampton}{Hampton University, Dept. of Physics, Hampton, VA 23668, USA}
\newcommand{\Dortmund}{Institute of Physics, Dortmund University, 44221, Germany }
\newcommand{\Otterbein}{Department of Physics, Otterbein University, 1 South Grove Street, Westerville, OH, 43081 USA}
\newcommand{\JMU}{James Madison University, Harrisonburg, Virginia 22807, USA}
\newcommand{\Florida}{University of Florida, Department of Physics, Gainesville, FL 32611}
\newcommand{\UCIrvine}{Department of Physics and Astronomy, University of California, Irvine, Irvine, California 92697-4575, USA}
\newcommand{\CBPF}{Centro Brasileiro de Pesquisas F\'{i}sicas, Rua Dr. Xavier Sigaud 150, Urca, Rio de Janeiro, Rio de Janeiro, 22290-180, Brazil}
\newcommand{\PUCP}{Secci\'{o}n F\'{i}sica, Departamento de Ciencias, Pontificia Universidad Cat\'{o}lica del Per\'{u}, Apartado 1761, Lima, Per\'{u}}
\newcommand{\INRM}{Institute for Nuclear Research of the Russian Academy of Sciences, 117312 Moscow, Russia}
\newcommand{\Jlab}{Jefferson Lab, 12000 Jefferson Avenue, Newport News, VA 23606, USA}
\newcommand{\Pittsburgh}{Department of Physics and Astronomy, University of Pittsburgh, Pittsburgh, Pennsylvania 15260, USA}
\newcommand{\Guanajuato}{Campus Le\'{o}n y Campus Guanajuato, Universidad de Guanajuato, Lascurain de Retana No. 5, Colonia Centro, Guanajuato 36000, Guanajuato M\'{e}xico.}
\newcommand{\Athens}{Department of Physics, University of Athens, GR-15771 Athens, Greece}
\newcommand{\Tufts}{Physics Department, Tufts University, Medford, Massachusetts 02155, USA}
\newcommand{\WM}{Department of Physics, William \& Mary, Williamsburg, Virginia 23187, USA}
\newcommand{\FNAL}{Fermi National Accelerator Laboratory, Batavia, Illinois 60510, USA}
\newcommand{\Purdue}{Department of Chemistry and Physics, Purdue University Calumet, Hammond, Indiana 46323, USA}
\newcommand{\MCLA}{Massachusetts College of Liberal Arts, 375 Church Street, North Adams, MA 01247}
\newcommand{\UMD}{Department of Physics, University of Minnesota -- Duluth, Duluth, Minnesota 55812, USA}
\newcommand{\Northwestern}{Northwestern University, Evanston, Illinois 60208}
\newcommand{\UNI}{Facultad de Ciencias F\'{i}sicas, Universidad Nacional Mayor de San Marcos, CP 15081, Lima, Per\'{u}}
\newcommand{\Rochester}{Department of Physics and Astronomy, University of Rochester, Rochester, New York 14627 USA}
\newcommand{\Austin}{Department of Physics, University of Texas, 1 University Station, Austin, Texas 78712, USA}
\newcommand{\USM}{Departamento de F\'{i}sica, Universidad T\'{e}cnica Federico Santa Mar\'{i}a, Avenida Espa\~{n}a 1680 Casilla 110-V, Valpara\'{i}so, Chile}
\newcommand{\Geneva}{University of Geneva, 1211 Geneva 4, Switzerland}
\newcommand{\Chicago}{Enrico Fermi Institute, University of Chicago, Chicago, IL 60637 USA}
\newcommand{\hired}{}
\newcommand{\OregonState}{Department of Physics, Oregon State University, Corvallis, Oregon 97331, USA}
\newcommand{\oxford}{Oxford University, Department of Physics, Oxford, OX1 3PJ United Kingdom}
\newcommand{\umiss}{University of Mississippi, Oxford, Mississippi 38677, USA}
\newcommand{\upenn}{Department of Physics and Astronomy, University of Pennsylvania, Philadelphia, PA 19104}
\newcommand{\AMU}{Department of Physics, Aligarh Muslim University, Aligarh, Uttar Pradesh 202002, India}
\newcommand{\wroclaw}{University of Wroclaw, plac Uniwersytecki 1, 50-137 Wroa\l{}aw, Poland}
\newcommand{\Mohali}{Department of Physical Sciences, IISER Mohali, Knowledge City, SAS Nagar, Mohali - 140306, Punjab, India}
\newcommand{\CINVESTAV}{Departamento de Fisica Col. San Pedro Zacatenco, 07360 Mexico, DF, Av. Instituto PolitÃ©cnico Nacional, Mexico}
\newcommand{\york}{York University, Department of Physics and Astronomy, Toronto, Ontario, M3J 1P3 Canada}
\newcommand{\ND}{Department of Physics and Astronomy, University of Notre Dame, Notre Dame, Indiana 46556, USA}
\newcommand{\ICL}{The Blackett Laboratory,  Imperial College London,  London SW7 2BW, United Kingdom}
\newcommand{\warwick}{Department of Physics, University of Warwick, Coventry, CV4 7AL, UK}
\newcommand{\qmul}{G O Jones Building, Queen Mary University of London, 327 Mile End Road, London E1 4NS, UK}
\newcommand{\LLNL}{Nuclear and Chemical Sciences Division, Lawrence Livermore National Laboratory, Livermore, CA 94550, USA}
\newcommand{\ricfregianThanks}{now at Department of Physics and Astronomy, University of California at Davis, Davis, CA 95616, USA}
\newcommand{\kleykampThanks}{now at Department of Physics and Astronomy, University of Mississippi, Oxford, MS 38677}
\newcommand{\adrianThanks}{Now at Department of Physics, Drexel University, Philadelphia, Pennsylvania 19104, USA}
\newcommand{\byaeggyThanks}{Now at Department of Physics, University of Cincinnati,  Cincinnati, Ohio 45221, USA}
\newcommand{\lazazuetareyesThanks}{now at Syracuse University, Syracuse, NY 13244, USA}

% -----------------------------------------------------------------------------
% AUTHORS
\author{A.~Klustov\'{a}}\email{Contact author: a.klustova20@imperial.ac.uk}                  \affiliation{\ICL}
\author{S.~Akhter}                        \affiliation{\AMU}
\author{Z.~Ahmad~Dar}                     \affiliation{\WM}  \affiliation{\AMU}
\author{M.~Sajjad~Athar}                  \affiliation{\AMU}
\author{G.~Caceres}\thanks{\ricfregianThanks}  \affiliation{\CBPF}
\author{H.~da~Motta}                      \affiliation{\CBPF}
\author{J.~Felix}                         \affiliation{\Guanajuato}
\author{P.K.Gaur}                         \affiliation{\AMU}
\author{R.~Gran}                          \affiliation{\UMD}
\author{E.Granados}                       \affiliation{\Guanajuato}  \affiliation{\Guanajuato}
\author{D.A.~Harris}                      \affiliation{\york}  \affiliation{\FNAL}
\author{A.L.~Hart}                        \affiliation{\qmul}
\author{J.~Kleykamp}\thanks{\kleykampThanks}  \affiliation{\Rochester}
\author{M.~Kordosky}                      \affiliation{\WM}
\author{D.~Last}                          \affiliation{\Rochester}  \affiliation{\upenn}
\author{A.~Lozano}\thanks{\adrianThanks}  \affiliation{\CBPF}
\author{S.~Manly}                         \affiliation{\Rochester}
\author{W.A.~Mann}                        \affiliation{\Tufts}
\author{K.S.~McFarland}                   \affiliation{\Rochester}
\author{M.~Mehmood}                       \affiliation{\york}
\author{O.~Moreno}                        \affiliation{\WM}  \affiliation{\Guanajuato}
\author{J.G.~Morf\'{i}n}                  \affiliation{\FNAL}
\author{V.~Paolone}                       \affiliation{\Pittsburgh}
\author{G.N.~Perdue}                      \affiliation{\FNAL}  \affiliation{\Rochester}
\author{C.~Pernas}                        \affiliation{\WM}
\author{M.A.~Ram\'{i}rez}                 \affiliation{\upenn}  \affiliation{\Guanajuato}
\author{N.~Roy}                           \affiliation{\york}
\author{D.~Ruterbories}                   \affiliation{\Rochester}
\author{H.~Schellman}                     \affiliation{\OregonState}
\author{C.J.~Solano~Salinas}              \affiliation{\UNI}
\author{D.S.~Correia}                     \affiliation{\CBPF}
\author{A.~Srivastava}                    \affiliation{\UMD}
\author{V.S.~Syrotenko}                   \affiliation{\Tufts}
\author{N.H.~Vaughan}                     \affiliation{\OregonState}
\author{A.V.~Waldron}                     \affiliation{\qmul}  \affiliation{\ICL}
\author{M.O.~Wascko}                      \affiliation{\oxford}  \affiliation{\ICL}
\author{B.~Yaeggy}\thanks{\byaeggyThanks}  \affiliation{\USM}
\author{L.~Zazueta}\thanks{\lazazuetareyesThanks}  \affiliation{\WM}

\collaboration{The MINER$\nu$A Collaboration}

\date{\today}

\begin{abstract}
We report MINERvA's first measurement of inclusive charged-current $\bar{\nu}_\mu$ cross sections on carbon, hydrocarbon, iron, and lead, and their ratios to the cross section on hydrocarbon, as functions of the antimuon transverse momentum, $p_{\mathrm{T}}$. Using a wide-band $\bar{\nu}_\mu$ beam with mean energy $\sim 6~\text{GeV}$, these measurements probe all interaction modes, including the transition from resonance production to deep-inelastic scattering. The total uncertainties are typically $5$--$10\%$ for the absolute cross sections and $2$--$5\%$ for the ratios. Comparisons with multiple neutrino interaction models reveal significant discrepancies in the $p_{\mathrm{T}}$ dependence, particularly for heavier nuclei. The disagreements are most pronounced at low $p_{\mathrm{T}}$ but extend across the full $p_{\mathrm{T}}$ range, indicating missing or mis-modelled nuclear effects.
\end{abstract}

\maketitle

% -----------------------------------------------------------------------------

Precision studies in current and future neutrino–nucleus experiments require an accurate description of neutrino–nucleus interactions. Uncertainties in these interactions are expected to be a leading source of systematic uncertainty, reflecting our incomplete understanding of how the nuclear environment affects neutrino scattering~\cite{Alvarez_Ruso_2018}. Nuclear effects modify the kinematics and identity of the outgoing particles and smear the reconstructed neutrino energy in a target-dependent way. Neutrino oscillation experiments~\cite{PhysRevD.108.072011, x53y-2b86, DUNE_LOI, HyperK_LOI}, which aim to determine the neutrino mass ordering and measure charge-parity ($CP$) violation, rely on precise neutrino energy reconstruction and are therefore particularly sensitive to these effects.

Comparisons between neutrino and antineutrino channels are central to $CP$-violation measurements. Antineutrino interactions are particularly important because their cross sections on matter are smaller and sensitive to the magnitude of the vector–axial interference term, which changes sign between neutrinos and antineutrinos, as well as to different resonance channels and quark contributions~\cite{Formaggio2012}. High-statistics antineutrino measurements on multiple nuclei are therefore required to improve interaction models and control systematic uncertainties. The Main Injector Experiment for $\nu$–A (MINERvA)~\cite{MINERvAWebsite, ALIAGA2014130} addresses this by measuring neutrino and antineutrino interactions on multiple different nuclear targets within a single detector exposed to the same beam, enabling direct studies of nuclear effects and stringent tests of neutrino event generators.

This Letter presents MINERvA's first measurements of inclusive charged-current $\bar{\nu}_\mu$ cross sections on carbon, hydrocarbon, iron, and lead, as well as their ratios to the high-statistics hydrocarbon cross section, as a function of the antimuon momentum transverse to the neutrino beam, $p_{\mathrm{T}}$. With a wide-band antineutrino beam of average energy 6~GeV, these measurements span all interaction modes and are dominated by resonance production and the transition to deep-inelastic scattering, i.e., shallow inelastic scattering (SIS), a kinematic region of direct interest to DUNE~\cite{DUNE_LOI} and the higher-energy atmospheric samples in Hyper-Kamiokande~\cite{HyperK_LOI}. Comparing results across targets tests how neutrino--nucleus event generators connect interaction regimes and treat nuclear effects over the kinematic range relevant to future oscillation analyses. While earlier MINERvA work studied neutrino--inclusive interactions at $\sim$~3~GeV across various targets~\cite{PhysRevLett.112.231801}, this analysis focuses on antineutrino interactions in a wider energy regime, using a factor of ten more events and exploring previously uncharted kinematic phase space.

% -----------------------------------------------------------------------------

MINERvA operated in the Neutrinos at the Main Injector (NuMI) beam at Fermi National Accelerator Laboratory~\cite{Adamson_2016}, produced by 120~GeV$/c$ protons impinging on a graphite target. Charged pions and kaons were focused by two magnetic horns into a 675~m-long decay pipe, where they decayed predominantly to muon neutrinos or antineutrinos, depending on horn polarity. This measurement uses data collected between 2016 and 2019 using the NuMI medium-energy beam with horns set to focus negatively charged mesons.
An exposure of $1.12 \times 10^{21}$ protons on target~(POT) was analysed, producing a predominantly antineutrino flux with a spectral peak near 6~GeV.

The MINERvA detector consists of hexagonal planes arranged perpendicular to the beam direction, and begins with a 1.25~m upstream nuclear-target region in which passive targets (C, Fe, Pb, H$_2$O) are interspersed with polystyrene scintillator planes. The C, Fe, and Pb targets are central to this analysis; additional details of the target geometry are provided in Ref.~\cite{ALIAGA2014130}. Downstream of the target region is the active tracker, comprising 124 consecutive planes of polystyrene scintillator, which serve as the hydrocarbon (CH) target material for this measurement. The tracker is followed by electromagnetic and hadronic calorimeters, consisting of interleaved lead–scintillator and steel–scintillator layers, respectively. Each tracking plane contains 127 triangular scintillator strips oriented perpendicular to the beam and arranged in three alternating orientations across successive planes, enabling precise three-dimensional track reconstruction. The full detector spans approximately 5~m in length. The magnetised MINOS Near Detector~\cite{Michael_2008}, located about 2~m downstream, serves as a muon range stack and spectrometer.

% -----------------------------------------------------------------------------
% EVENT SELECTION (CH, PB) AND ALL CROSS SECTIONS FIGURE
\begin{figure*}[t]
  \centering
  % ---- TOP ROW ----
  \begin{subfigure}[b]{0.75\textwidth}
    \centering
    \includegraphics[width=\linewidth]{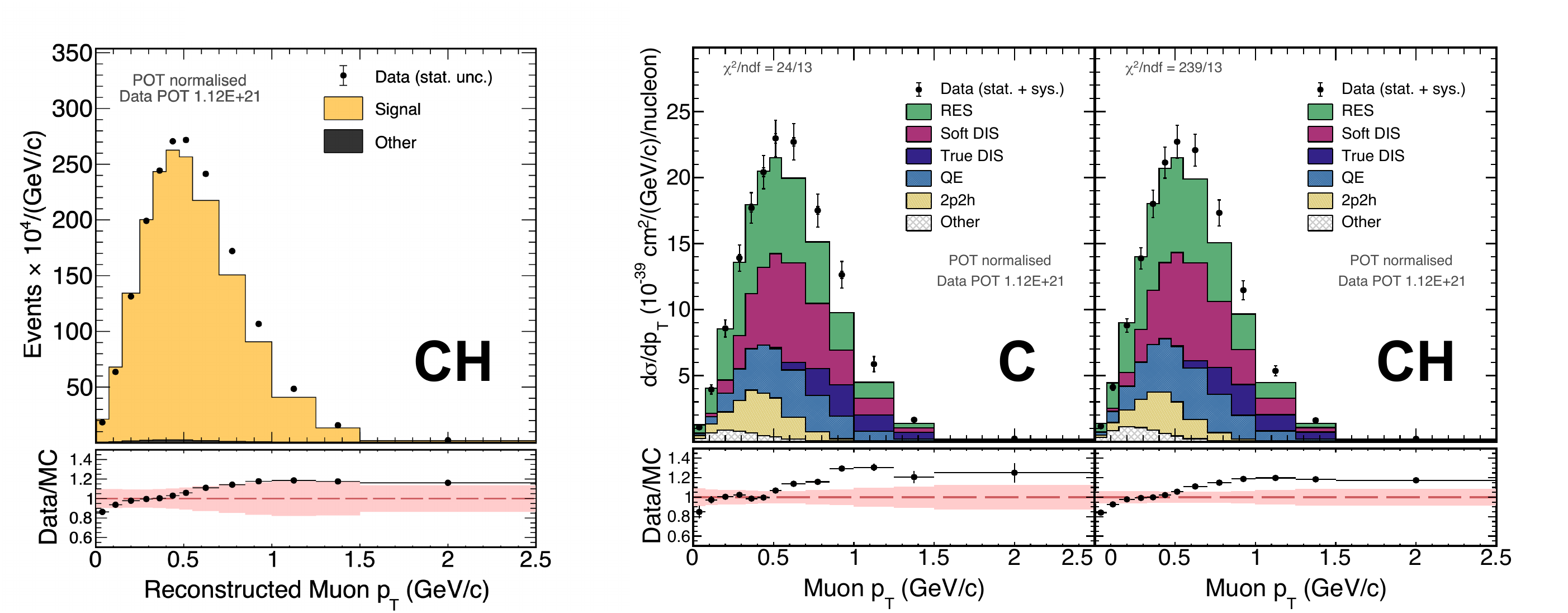}
  \end{subfigure}%
  % ---- BOTTOM ROW ----
  \vspace{-0.1\baselineskip}
  \begin{subfigure}[b]{0.75\textwidth}
    \centering
    \includegraphics[width=\linewidth]{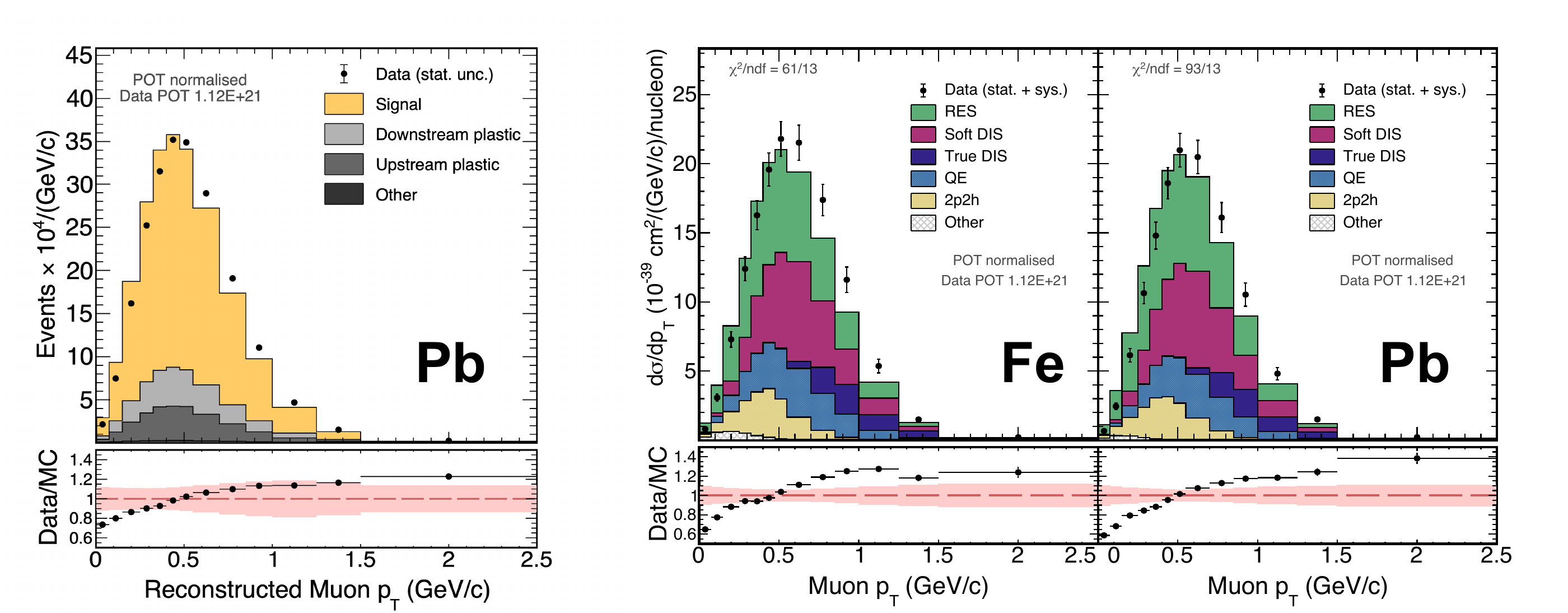}
  \end{subfigure}
  \vspace{-0.4\baselineskip}
  \caption{
    \textit{Left:} Selected events in CH and Pb as a function of $p_{\mathrm{T}}$. For each target, the upper panel shows data with statistical uncertainties and stacked simulated signal and background; the lower panel shows the data-to-simulation ratio, with a band for simulation systematics, error bars for statistical uncertainties, and horizontal bars indicating bin widths.
    \textit{Right:} Cross sections for C, CH, Fe, and Pb versus
    $p_{\mathrm{T}}$, with simulation predictions decomposed by true interaction channel. For each target, the upper panel shows data with statistical (inner) and total (outer) uncertainties (statistical uncertainty is smaller than the marker size); the lower panel shows data-to-simulation ratio, with a band indicating the systematic uncertainty on the data and points indicating statistical uncertainties.}
  \label{fig:pt_evsel_and_xsec}
\end{figure*}
% -----------------------------------------------------------------------------

A detailed simulation of the MINERvA experiment is used to estimate backgrounds and correct for detector efficiency, acceptance, and resolution. The antineutrino flux is predicted with a GEANT4-based simulation~\cite{AGOSTINELLI2003250} of the NuMI beamline~\cite{PhysRevD.94.092005}, with hadron production reweighted to external data from NA49~\cite{na49data} and other experiments~\cite{PhysRevD.90.032001}. Additional \textit{in situ} flux constraints come from a combined analysis of (anti)neutrino–electron ($\nu$--e) scattering and inverse muon decay (IMD)~\cite{PhysRevD.107.012001,PhysRevD.100.092001,PhysRevD.104.092010}, and from MINERvA’s low-energy-transfer (low-$\nu$) flux measurement in the medium-energy configuration~\cite{Fine:2020knh,Srivastava:2023zyx}, applied to the antineutrino flux. Above $7.5~\text{GeV}$, the low-$\nu$ parametrisation provides stronger leverage than the $\nu$--e+IMD constraint, increasing the predicted high-energy flux by up to 17\% in the $\sim13$--$26~\text{GeV}$ range.

Antineutrino interactions are simulated with GENIE v2.12.6~\cite{ANDREOPOULOS201087} using tuned configuration (``MINERvA Tune v4.3.0''), developed to describe previous MINERvA cross-section measurements~\cite{Bercellie2023_PhysRevLett.131.011801}. The underlying model employs a relativistic Fermi gas with a Bodek--Ritchie tail~\cite{RFG_SMITH1972605, BodekRitchie1981_PhysRevD.24.1400}, Llewellyn-Smith quasi-elastic scattering~\cite{LLEWELLYNSMITH1972261}, Val\`{e}ncia 2-particle 2-hole (2p2h) contribution from correlated nucleons~\cite{PhysRevC.83.045501, PhysRevD.88.113007}, Rein--Sehgal resonance and coherent pion production~\cite{REIN198179, coherent_REIN198329, coherentMass_REIN2007207}, and a modified Bodek--Yang prescription at low $Q^2$ for nonresonant and inelastic scattering~\cite{BodekYang_2003, BODEK2005113}. Hadronisation is modelled by AGKY~\cite{AGKY_Yang_2009, KNO_KOBA1972317} and PYTHIA/JETSET~\cite{Pythia_Sjostrand_2001} at higher hadronic invariant masses, and intranuclear final-state interactions by INTRANUKE-hA~\cite{Dytman:2009zz}. 

In the MINERvA tune, quasi-elastic scattering is modified by a random phase approximation correction from the Val\`{e}ncia model~\cite{RPANieves2004_PhysRevC.70.055503, gran2017model}. The 2p2h contribution is enhanced, resulting in an overall rate roughly 50\% higher than the nominal GENIE prediction, based on fits to MINERvA low-energy neutrino data~\cite{Rodrigues2016_PhysRevLett.116.071802, Gran2018_PhysRevLett.120.221805}. Nonresonant pion production is reduced to 43\% of its nominal value. The resonant axial mass and normalisation are tuned to $M_A^{\text{RES}} = 0.94~\text{GeV}/c^2$ and 1.15, respectively, based on a reanalysis of pion production data from deuterium bubble chambers~\cite{pion_Rodrigues_2016}. Coherent pion production is reweighted to match MINERvA's medium-energy measurement~\cite{Ramirez2023_PhysRevLett.131.051801}, with an additional 43.7\% normalisation increase to account for diffractive pion production on hydrogen~\cite{PhysRevD.85.073003}. For this antineutrino analysis, single $\pi^-$ production~(resonant and nonresonant) with $W < 1.4~\text{GeV}/c^2$ is subject to an additional low-$Q^2$ suppression derived from the neutrino charged-current $1\pi^+$ measurement~\cite{Bercellie2023_PhysRevLett.131.011801}, applied to all nuclear targets except hydrogen under the assumption that the effect is nuclear in origin and approximately isospin symmetric.

The detector response is simulated with GEANT4 version 4.9.3.p6 using the QGSP\_BERT physics list~\cite{AGOSTINELLI2003250, KAIDALOV1982459}, including optical and electronics effects. The absolute energy scale is calibrated with through-going muons~\cite{ALIAGA2014130}, and the charged-hadron response is tuned to test-beam measurements with a scaled-down MINERvA detector~\cite{ALIAGA201528}. The neutron inelastic cross section is reweighted to more recent neutron-interaction data across all passive materials in the detector, as described in Ref.~\cite{PhysRevD.100.052002}.

% -----------------------------------------------------------------------------
% ANALYSIS, RESULTS, DISCUSSION

In both data and simulation, we apply a common selection for charged-current $\bar{\nu}_\mu$ interactions on C, Fe, and Pb in the MINERvA passive targets, excluding the most upstream target layer, which has higher contamination from events originating in the target hall walls. The same selection is applied to interactions on CH in the active scintillator tracker. Signal events must include a positively charged muon, $\mu^+$, with an interaction in the target material and $2 < E_{\mu^+} < 20~\mathrm{GeV}$ and $\theta_{\mu^+} < 17^\circ$ relative to the beam. These kinematic constraints ensure reliable matching of the antimuon to the downstream MINOS detector, where its momentum and charge are measured.

The interaction location is reconstructed using a deep convolutional neural network~\cite{Akbar_2022}. A minimum probability threshold of 0.2 is applied to maximise statistics while assigning events to the correct target material and maintaining consistent performance between data and simulation. For nuclear targets, the interaction point must lie within the target plane or an adjacent scintillator plane. The reconstructed vertical and horizontal positions identify the material. CH events are required to originate within the tracker region. Additional cuts ensure good material separation, accurate antimuon reconstruction in MINOS, and that events lie within the MINERvA fiducial volume, as detailed in Ref.~\cite{ALIAGA2014130}.

% -----------------------------------------------------------------------------
% CROSS SECTIONS WITH MODEL COMPARISONS FIGURE
\begin{figure}[t]
  \centering
  \includegraphics[width=1.0\columnwidth]{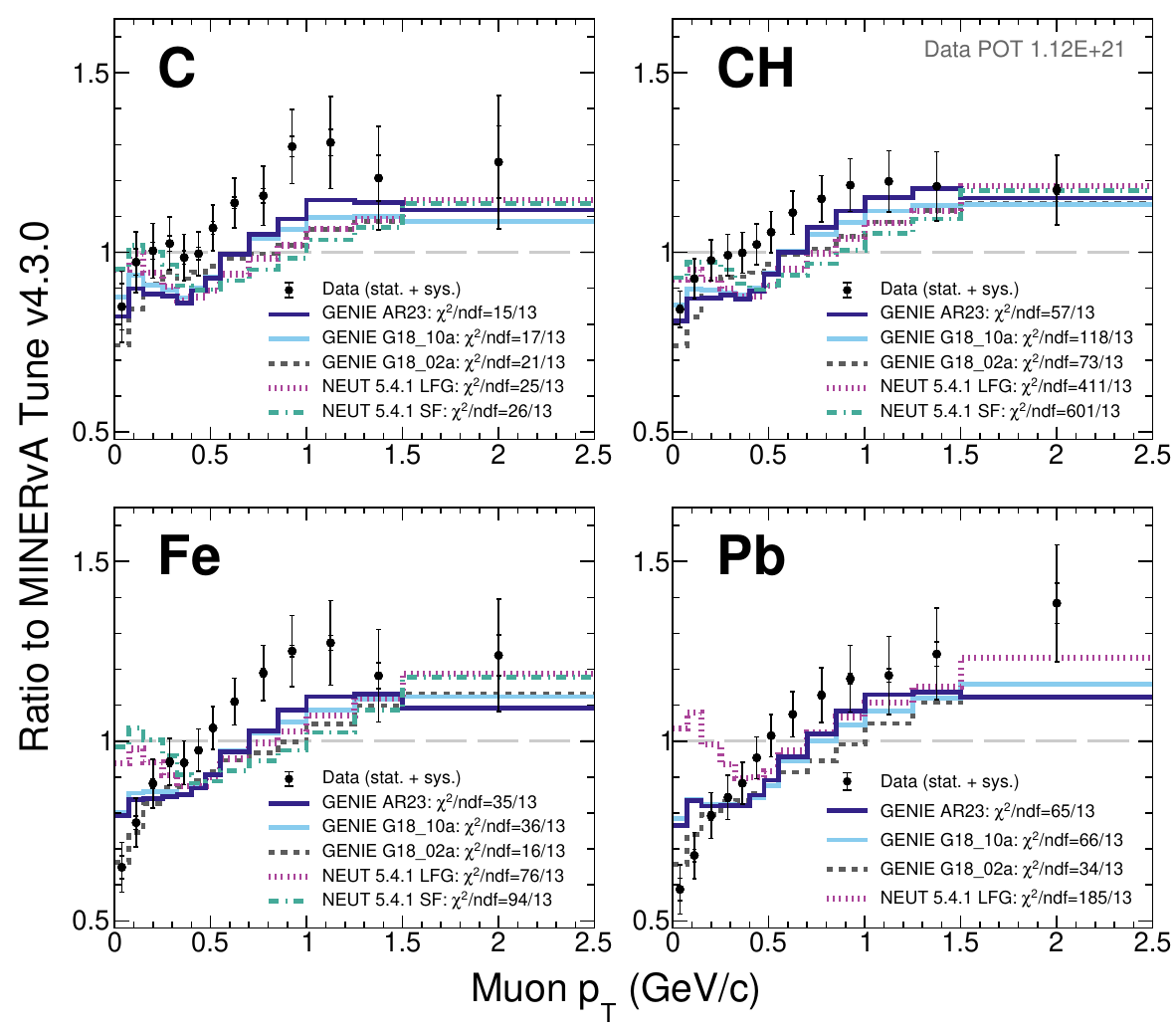}
  \caption{Ratio of data and alternative generator predictions to the baseline simulation for C, CH, Fe, and Pb cross sections versus $p_{\mathrm{T}}$. Data points show statistical (inner) and total (outer) uncertainties.}
  \label{fig:pt_xsec_models}
\end{figure}
% -----------------------------------------------------------------------------

The analysis selects $63{,}956$ events on C, $192{,}694$ on~Fe, $225{,}313$ on Pb, and $1{,}897{,}567$ on CH. Figure~\ref{fig:pt_evsel_and_xsec}~(left) shows the $p_{\mathrm{T}}$ distributions of selected events for CH (left-upper) and Pb (left-lower). Background in the CH sample consists of out-of-fiducial events, out-of-energy-range events, and neutrino contamination, each contributing at roughly equal levels and totalling $\sim 1\%$. This background is subtracted using the simulation. For the nuclear targets, the background is at the $\sim 25\%$ level and is dominated by events originating in the upstream and downstream plastic scintillator planes surrounding the target but reconstructed in the target, as shown for Pb in the lower panel of Figure~\ref{fig:pt_evsel_and_xsec}~(left). To constrain this background, we define upstream and downstream background-rich sidebands of up to six planes surrounding each target, excluding the planes immediately adjacent, which are included in the signal fiducial volume. Simulation predictions in these sidebands are simultaneously fitted to data using a $\chi^2$ minimisation as a function of plane number for each material. The resulting per-material normalisation factors for upstream and downstream contributions, which increase the background by $3$--$7\%$, are applied in the signal region, and the scaled background is subtracted. As in the CH sample, ``Other'' backgrounds (out-of-fiducial events, out-of-energy-range events, and neutrino contamination) contribute at the $\sim 1\%$ level and are subtracted using the simulation prediction.

The background-subtracted reconstructed $p_{\mathrm{T}}$ distributions are unfolded using the D’Agostini iterative method~\cite{DAGOSTINI1995487} with three iterations, chosen using pseudodataset studies to balance residual model-dependent bias against amplified statistical noise. To account for detector inefficiencies, the unfolded distributions are corrected by an efficiency factor, defined as the number of true signal events satisfying the selection criteria divided by the number of reconstructed and selected simulated events. Events originating outside the selected muon angular phase space can migrate into the reconstructed sample; no truth-level angular requirement is applied to the numerator, so such migrations are naturally absorbed into the correction rather than treated as a separate background subtraction. The efficiency decreases with $p_{\mathrm{T}}$, ranging from 40--50\% in C, Fe, and Pb to approximately 65\% in CH. Differential cross sections are obtained by normalising to the antineutrino flux integrated over 0--100~GeV for each target, the number of target nucleons, and the bin width.

% -----------------------------------------------------------------------------
% CROSS-SECTION RATIOS WITH MODEL COMPARISONS FIGURE
\begin{figure*}[t]
  \centering
    \includegraphics[width=0.8\textwidth]{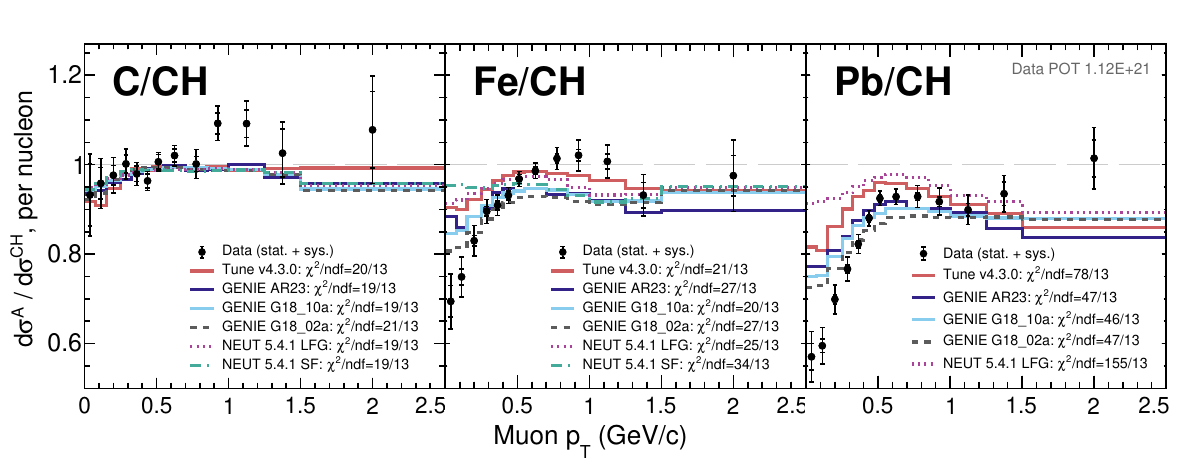}
  \caption{C/CH, Fe/CH, and Pb/CH cross-section ratios as a function of $p_{\mathrm{T}}$ compared to the baseline simulation and various generator predictions. Data points show statistical (inner) and total (outer) uncertainties.}
  \label{fig:pt_xsec_ratio_models}
\end{figure*}
% -----------------------------------------------------------------------------

The differential cross sections per nucleon for each target as a function of $p_{\mathrm{T}}$ are shown in the four panels in Figure~\ref{fig:pt_evsel_and_xsec} (right). For the inclusive signal, MINERvA Tune v4.3.0 predicts resonant pion production (RES) as the dominant interaction (40.1\% in Pb), followed by ``Soft deep inelastic scattering (Soft DIS)'' (27.7\%, $Q^2 < 1~(\mathrm{GeV}/c)^2$ or $W < 2~\mathrm{GeV}/c^2$) and ``True DIS'' (4.6\%, both $Q^2 > 1~(\mathrm{GeV}/c)^2$ and $W > 2~\mathrm{GeV}/c^2$)\footnote{The terms ``Soft DIS'' and ``True DIS'' use the GENIE definition of ``DIS,'' which is not based on the kinematics of the interaction. Contemporary definitions~\cite{lozano2025measurementchargedcurrentnumubarnumu} relabel these as ``Soft SIS'' and ``Multiquark SIS,'' reserving kinematic DIS for $Q^2 \gtrsim 4~(\mathrm{GeV}/c)^2$.}. ``Other'' includes coherent pion production and diffractive scattering on hydrogen in CH. Note that $p_{\mathrm{T}}$ serves as a proxy for $Q^2$, with low-$p_{\mathrm{T}}$ events corresponding to low-$Q^2$ interactions, such as quasi-elastic (QE) and 2p2h processes. The reference simulation broadly describes the peak region for all targets. For C and CH, it agrees reasonably well at low $p_{\mathrm{T}}$ but underpredicts the data at higher $p_{\mathrm{T}}$. For Fe and Pb, a similar high-$p_{\mathrm{T}}$ underprediction is accompanied by an overprediction at low $p_{\mathrm{T}}$, with the size of the discrepancy increasing for heavier nuclei.

Systematic uncertainties are evaluated using a multi-universe approach~\cite{fine2022datapreservationminerva}, in which each source is varied and the cross section re-extracted to account for bin-to-bin and target correlations. The dominant contributions arise from flux ($\sim5\%$ fractional uncertainty) and muon reconstruction (ranging from $\sim1\%$ in the peak of the distribution to $\sim7\%$ in the tail). The interaction uncertainties are dominated by the GENIE parameter $M_{\rm A}^{\rm RES}$ and conservative MINERvA tune reweights, particularly for nuclear targets. Individual contributions to the systematic uncertainty on the cross section are provided in the Supplemental Material~(SM)~\cite{supp}. Total uncertainties on the measured cross sections are 5--6\% in the peak region and typically below 10\% across all targets.

The measured cross sections can be used to probe the robustness of interaction models. We compare to commonly used generator predictions using \textsc{NUISANCE}~\cite{Stowell:2016jfr}, as shown in Figure~\ref{fig:pt_xsec_models}. These include five GENIE3~\cite{ANDREOPOULOS201087} and two NEUT 5.4.1~\cite{NEUT2021} predictions, with configurations \cite{BETANCOURT20181,PhysRevD.105.092004,ANDREOPOULOS201087,RFG_SMITH1972605,BodekRitchie1981_PhysRevD.24.1400,LLEWELLYNSMITH1972261,10.1063/1.4919465,PhysRevD.76.113004,PhysRevD.79.053003,BodekYang_2003,BODEK2005113,AGKY_Yang_2009,KNO_KOBA1972317,Pythia_Sjostrand_2001,Dytman:2009zz,PhysRevD.104.072009,RPANieves2004_PhysRevC.70.055503,gran2017model,PhysRevC.83.045501,PhysRevD.88.113007,NEUT2021,REIN198179,BENHAR1994493,2006NuPhS.159..211J} detailed in the~SM~\cite{supp}.
For clarity, Figure~\ref{fig:pt_xsec_models} shows GENIE3 predictions using the h$A$ intranuclear cascade model only; the corresponding h$N$-based predictions (G18\_02b and G18\_10b) and their $\chi^2$ values, which would overlap closely with the h$A$ curves, are provided in the~SM~\cite{supp}. Among the predictions, GENIE3 AR23~(currently used by DUNE~\cite{DUNE_LOI} and the SBN programme \cite{icaruscollaboration2026measurementmuonantineutrinochargedcurrent}) and G18\_02 yield the lowest $\chi^2$ most consistently across targets. GENIE3 reproduces the low-$p_{\mathrm{T}}$ shape best, although additional suppression is needed for heavier targets. Differences between GENIE3 and NEUT are observed at low $p_{\mathrm{T}}$ in Fe and Pb, reflecting differences in $A$-scaling for pion-producing processes. For example, the coherent cross section in GENIE scales as $A^{2/3}$ whereas in NEUT it scales as $A$.

In the $p_{\mathrm{T}}$ range of roughly 0.4--1.0~GeV/$c$, all generators underpredict the cross section. Similar underprediction was seen in an exclusive QE measurement~\cite{PhysRevD.108.032018}, attributed to the dipole approximation underestimating the QE axial form factor at higher $Q^2$~\cite{NatureMINERvA}. However, GENIE AR23, which uses a $z$-expansion axial form factor fit to neutrino-deuterium data, performs only marginally better and still shows similar underprediction. Recent studies show QE lattice quantum chromodynamics (LQCD) form factors~\cite{meyer2026nucleonaxialformfactor} increase the QE cross section by $\sim10\%$ over the deuterium fit~\cite{Liu:2026wlw}, possibly accounting for some of the discrepancy; similar effects are expected for resonance form factors, where no LQCD predictions yet exist. At higher $p_{\mathrm{T}}$, other generators describe the data reasonably well, while the baseline continues to underpredict it. This likely reflects the updated $Q^2$ dependence of resonance form factors in GENIE3 and NEUT~\cite{PhysRevD.77.053001}, along with the baseline's nonresonant pion suppression at all $Q^2$, even though the deuterium data used to constrain it are limited to $W < 1.2$~GeV/$c^2$ and $Q^2 < 1$~(GeV/$c$)$^2$~\cite{pion_Rodrigues_2016}. These high-$p_{\mathrm{T}}$ bins are dominated by ``Soft DIS'' and ``True DIS'' events, a region not yet constrained by ongoing MINERvA measurements~\cite{lozano2025measurementchargedcurrentnumubarnumu}.

Ratios of C, Fe, and Pb cross sections to CH are formed to cancel shared systematic uncertainties, such as flux and muon reconstruction uncertainties,  and isolate nuclear effects. Their construction requires careful treatment of the antineutrino flux, which varies by a few percent along the longitudinal and radial axes due to a slight offset between the tracker symmetry and beam axes. Additionally, the passive nuclear targets are asymmetrically positioned, causing target-dependent flux differences, particularly near 8~GeV at the focusing-peak edge. To ensure a common flux in numerator and denominator, the CH target is divided into twelve equally spaced angular regions, and a weighted linear combination is formed to reproduce the flux incident on C, Fe, and Pb~\cite{PhysRevLett.130.161801}.

The resulting cross-section ratios, shown in Figure~\ref{fig:pt_xsec_ratio_models}, have total uncertainties below 5\% and $\sim$2\% in the peak region, with individual contributions provided in the~SM~\cite{supp}. The C/CH ratio serves as a cross-check: the small low-$p_{\mathrm{T}}$ depletion in both data and simulation is consistent with diffractive scattering on hydrogen in CH. For Fe/CH and Pb/CH, all models predict an $A$-dependent suppression at low $p_{\mathrm{T}}$ but underestimate its magnitude, with GENIE3 G18\_10 yielding the lowest $\chi^2$ most consistently across all three ratios. Since the low-$p_{\mathrm{T}}$ region is predominantly resonance-dominated, the systematic underprediction across all generators points to missing or underestimated nuclear effects in the resonance regime, though mismodelling in other contributing processes may also play a role.

% -----------------------------------------------------------------------------
% SUMMARY

In summary, the inclusive charged-current $\bar{\nu}_\mu$ cross sections on C, CH, Fe, and Pb, and their ratios C/CH, Fe/CH, and Pb/CH exhibit strong suppression at low $p_{\mathrm{T}}$ and high $A$. Most current interaction models capture the qualitative behaviour but systematically underestimate the strength of this suppression observed in MINERvA data. The cross-section ratios clearly demonstrate that the nuclear dependence at low $p_{\mathrm{T}}$ in iron and lead exceeds the predictions of the models. Additional discrepancies are present across the full $p_{\mathrm{T}}$ range. These measurements span the antineutrino energy range most relevant for DUNE and for higher-energy atmospheric neutrinos in Hyper-Kamiokande, covering a nuclear mass range pertinent to oscillation experiments including argon. They provide strong benchmarks for neutrino--nucleus interaction models, testing generator predictions across the quasi-elastic, resonance, and shallow-inelastic (transition to deep-inelastic) regimes for different nuclei, and constraining interaction modelling for future oscillation analyses.

% -----------------------------------------------------------------------------
% ACKNOWLEDGEMENTS
\section*{Acknowledgements}

This document was prepared by members of the MINERvA Collaboration using the resources of the Fermi National Accelerator Laboratory (Fermilab), a U.S. Department of Energy, Office of Science, Office of High Energy Physics HEP User Facility. Fermilab is managed by Fermi Forward Discovery Group, LLC, acting under Contract No. 89243024CSC000002. These resources included support for the MINERvA construction project, and support for construction also was granted by the United States National Science Foundation under Award No. PHY-0619727 and by the University of Rochester. Support for participating scientists was provided by NSF and DOE (USA); by CAPES and CNPq (Brazil); by CoNaCyT (Mexico); by ANID PIA / APOYO AFB180002, CONICYT PIA ACT1413, and Fondecyt 3170845 and 11130133 (Chile); by CONCYTEC (Consejo Nacional de Ciencia, Tecnolog\'ia e Innovaci\'on Tecnol\'ogica), DGI-PUCP (Direcci\'on de Gesti\'on de la Investigaci\'on  - Pontificia Universidad Cat\'olica del Peru), and VRI-UNI (Vice-Rectorate for Research of National University of Engineering) (Peru); NCN Opus Grant No. 2016/21/B/ST2/01092 (Poland); by Science and Technology Facilities Council (UK); by EU Horizon 2020 Marie Skłodowska-Curie Action; by a Cottrell Postdoctoral Fellowship from the Research Corporation for Scientific Advancement; by an Imperial College London President's PhD Scholarship.  We thank the MINOS Collaboration for use of its near detector data. Finally, we thank the staff of
Fermilab for support of the beam line, the detector, and computing infrastructure.

% -----------------------------------------------------------------------------
% DATA AVAILABILITY
\section*{Data availability}
The data that support the findings of this Letter are available as follows. The results are based on neutrino--nucleus interaction data collected by the MINERvA experiment, which are publicly available through the MINERvA Open Data release~\cite{MINERvAOpenData}. The analysis presented here is based on an earlier version of the MINERvA Open Data product. The numerical results of this measurement, including cross sections, cross-section ratios, covariances, correlations, and fluxes, are openly available in Zenodo~\cite{ZenodoRelease}. The analysis uses the public MINERvA software framework~\cite{MINERvASoftware}. The exact analysis code is available from the corresponding author upon request.

% -----------------------------------------------------------------------------
% BIBLIOGRAPHY
\bibliographystyle{apsrev4-1}
\bibliography{biblio}

@PREAMBLE{
 "\providecommand{\noopsort}[1]{}" 
 # "\providecommand{\singleletter}[1]{#1}%" 
}

@article{PhysRevD.108.072011,
  title = {{Updated T2K measurements of muon neutrino and antineutrino disappearance using $3.6\ifmmode\times\else\texttimes\fi{}{10}^{21}$ protons on target}},
  author = {Abe, K. and others},
  collaboration = {T2K Collaboration},
  journal = {Phys. Rev. D},
  volume = {108},
  issue = {7},
  pages = {072011},
  numpages = {10},
  year = {2023},
  month = {Oct},
  publisher = {American Physical Society},
  doi = {10.1103/PhysRevD.108.072011},
  url = {https://link.aps.org/doi/10.1103/PhysRevD.108.072011}
}

@article{x53y-2b86,
  title = {{Precision Measurement of Neutrino Oscillation Parameters with 10 Years of Data from the NOvA Experiment}},
  author = {Abubakar, S. and others},
  collaboration = {NOvA Collaboration},
  journal = {Phys. Rev. Lett.},
  volume = {136},
  issue = {1},
  pages = {011802},
  numpages = {10},
  year = {2026},
  month = {Jan},
  publisher = {American Physical Society},
  doi = {10.1103/x53y-2b86},
  url = {https://link.aps.org/doi/10.1103/x53y-2b86}
}

@article{DUNE_LOI,
  author        = {R. Acciarri and others},
  collaboration = {DUNE Collaboration},
  title         = {Long-Baseline Neutrino Facility (LBNF) and Deep Underground Neutrino Experiment (DUNE)},
  journal       = {arXiv preprint},
  year          = {2015},
  eprint        = {1512.06148},
  archivePrefix = {arXiv},
  primaryClass  = {physics.ins-det}
}

@article{HyperK_LOI,
  author        = {K. Abe and others},
  collaboration = {Hyper-K Collaboration},
  title         = {Letter of Intent: The Hyper-Kamiokande Experiment --- Detector Design and Physics Potential},
  journal       = {arXiv preprint},
  year          = {2011},
  eprint        = {1109.3262},
  archivePrefix = {arXiv},
  primaryClass  = {hep-ex}
}

@article{PhysRevLett.112.231801,
  title = {Measurement of Ratios of ${\ensuremath{\nu}}_{\ensuremath{\mu}}$ Charged-Current Cross Sections on C, Fe, and Pb to CH at Neutrino Energies 2--20 GeV},
  author = {Tice, B. G. and others},
  collaboration = {MINER$\nu$A Collaboration},
  journal = {Phys. Rev. Lett.},
  volume = {112},
  issue = {23},
  pages = {231801},
  numpages = {6},
  year = {2014},
  publisher = {American Physical Society},
  doi = {10.1103/PhysRevLett.112.231801},
  url = {https://link.aps.org/doi/10.1103/PhysRevLett.112.231801}
}

@article{Alvarez_Ruso_2018,
   title={NuSTEC White Paper: Status and challenges of neutrino–nucleus scattering},
   volume={100},
   ISSN={0146-6410},
   url={http://dx.doi.org/10.1016/j.ppnp.2018.01.006},
   DOI={10.1016/j.ppnp.2018.01.006},
   journal={Prog. Part. Nucl. Phys.},
   publisher={Elsevier BV},
   author={Alvarez-Ruso, L. and others},
   year={2018},
   month=may, pages={1} }

@article{Formaggio2012,
  author  = {Formaggio, J. A. and Zeller, G. P.},
  title   = {From eV to EeV: Neutrino Cross Sections across Energy Scales},
  journal = {Rev. Mod. Phys.},
  volume  = {84},
  pages   = {1307--1341},
  year    = {2012},
  doi     = {10.1103/RevModPhys.84.1307},
  url     = {https://doi.org/10.1103/RevModPhys.84.1307}
}

@article{AGOSTINELLI2003250,
title = {{GEANT4—a simulation toolkit}},
journal = {Nucl. Instrum. Methods Phys. Res., Sect. A},
volume = {506},
number = {3},
pages = {250-303},
year = {2003},
issn = {0168-9002},
doi = {https://doi.org/10.1016/S0168-9002(03)01368-8},
url = {https://www.sciencedirect.com/science/article/pii/S0168900203013688},
author = {S. Agostinelli and others},
}

@article{PhysRevD.94.092005,
  title = {{Neutrino flux predictions for the NuMI beam}},
  author = {Aliaga, L. and others},
  collaboration = {MINER\ensuremath{\nu}A Collaboration},
  journal = {Phys. Rev. D},
  volume = {94},
  issue = {9},
  pages = {092005},
  numpages = {10},
  year = {2016},
  month = {Nov},
  publisher = {American Physical Society},
  doi = {10.1103/PhysRevD.94.092005},
  url = {https://link.aps.org/doi/10.1103/PhysRevD.94.092005}
}

@article{na49data,
	author = {Alt, C. and others},
    collaboration = {NA Collaboration},
	journal = {Eur. Phys. J. C. },
	number = {4},
	pages = {897--917},
	title = {{Inclusive production of charged pions in p + C collisions at 158 GeV/c beam momentum}},
    doi = {10.1140/epjc/s10052-006-0165-7},
    url = {https://link.springer.com/article/10.1140/epjc/s10052-006-0165-7},
	volume = {49},
	year = {2007}}

@article{PhysRevD.90.032001,
  title = {Measurement of charged pion production yields off the NuMI target},
  author = {Paley, J. M. and others},
  collaboration = {MIPP Collaboration},
  journal = {Phys. Rev. D},
  volume = {90},
  issue = {3},
  pages = {032001},
  numpages = {14},
  year = {2014},
  publisher = {American Physical Society},
  doi = {10.1103/PhysRevD.90.032001},
  url = {https://link.aps.org/doi/10.1103/PhysRevD.90.032001}
}

@article{PhysRevD.100.092001,
  title = {{Constraint of the $\mathrm{MINER}\ensuremath{\nu}$A medium energy neutrino flux using neutrino-electron elastic scattering}},
  author = {Valencia, E. and others},
  collaboration = {$\mathrm{MINER}\ensuremath{\nu}$A Collaboration},
  journal = {Phys. Rev. D},
  volume = {100},
  issue = {9},
  pages = {092001},
  numpages = {14},
  year = {2019},
  month = {Nov},
  publisher = {American Physical Society},
  doi = {10.1103/PhysRevD.100.092001},
  url = {https://link.aps.org/doi/10.1103/PhysRevD.100.092001}
}

@article{PhysRevD.104.092010,
  title = {{Constraining the NuMI neutrino flux using inverse muon decay reactions in MINERvA}},
  author = {Ruterbories, D. and others},
  collaboration = {$\mathrm{MINER}\ensuremath{\nu}\mathrm{A}$ Collaboration},
  journal = {Phys. Rev. D},
  volume = {104},
  issue = {9},
  pages = {092010},
  numpages = {9},
  year = {2021},
  month = {Nov},
  publisher = {American Physical Society},
  doi = {10.1103/PhysRevD.104.092010},
  url = {https://link.aps.org/doi/10.1103/PhysRevD.104.092010}
}

@article{PhysRevD.107.012001,
  title = {{Improved constraint on the $\mathrm{M}\mathrm{I}\mathrm{N}\mathrm{E}\mathrm{R}\ensuremath{\nu}\mathrm{A}$ medium energy neutrino flux using $\overline{\ensuremath{\nu}}{e}^{\ensuremath{-}}\ensuremath{\rightarrow}\overline{\ensuremath{\nu}}{e}^{\ensuremath{-}}$ data}},
  author = {Zazueta, L. and others},
  collaboration = {$\mathrm{M}\mathrm{I}\mathrm{N}\mathrm{E}\mathrm{R}\ensuremath{\nu}\mathrm{A}$ Collaboration},
  journal = {Phys. Rev. D},
  volume = {107},
  issue = {1},
  pages = {012001},
  numpages = {12},
  year = {2023},
  month = {Jan},
  publisher = {American Physical Society},
  doi = {10.1103/PhysRevD.107.012001},
  url = {https://link.aps.org/doi/10.1103/PhysRevD.107.012001}
}

@phdthesis{Fine:2020knh,
    author = "Fine, Robert",
    title = "{Measurement of the Medium Energy NuMI Flux Using the Low-$\nu$ and High-$\nu$ Methods at MINERvA}",
    reportNumber = "FERMILAB-THESIS-2020-14",
    school = "Rochester U.",
    year = "2020"
}

@mastersthesis{Srivastava:2023zyx,
    author = "Srivastava, Asit",
    title = "{Energy Dependence Analysis of Neutrino-Nucleus Interactions in Scintillator for MINERvA}",
    reportNumber = "FERMILAB-MASTERS-2023-01",
    school = "Minnesota U.",
    year = "2023"
}

@article{PhysRevLett.130.161801,
  title = {{Simultaneous Measurement of ${\ensuremath{\nu}}_{\ensuremath{\mu}}$ Quasielasticlike Cross Sections on CH, C, ${\mathrm{H}}_{2}\mathrm{O}$, Fe, and Pb as a Function of Muon Kinematics at MINERvA}},
  author = {Kleykamp, J. and others},
  collaboration = {$\mathrm{M}\mathrm{I}\mathrm{N}\mathrm{E}\mathrm{R}\ensuremath{\nu}\mathrm{A}$ Collaboration},
  journal = {Phys. Rev. Lett.},
  volume = {130},
  issue = {16},
  pages = {161801},
  numpages = {7},
  year = {2023},
  publisher = {American Physical Society},
  doi = {10.1103/PhysRevLett.130.161801},
  url = {https://link.aps.org/doi/10.1103/PhysRevLett.130.161801}
}

@article{ANDREOPOULOS201087,
    title = {{The GENIE neutrino Monte Carlo generator}},
    journal = {Nucl. Instrum. Methods Phys.~Res., Sect.~A},
    volume = {614},
    number = {1},
    pages = {87-104},
    year = {2010},
    issn = {0168-9002},
    doi = {https://doi.org/10.1016/j.nima.2009.12.009},
    url = {https://www.sciencedirect.com/science/article/pii/S0168900209023043},
    author = {C. Andreopoulos and others},

}

@article{RFG_SMITH1972605,
    title = {Neutrino reactions on nuclear targets},
    journal = {Nucl. Phys. B.},
    volume = {43},
    pages = {605-622},
    year = {1972},
    issn = {0550-3213},
    doi = {https://doi.org/10.1016/0550-3213(72)90040-5},
    url = {https://www.sciencedirect.com/science/article/pii/0550321372900405},
    author = {R.A. Smith and E.J. Moniz}, 
}

@article{BodekRitchie1981_PhysRevD.24.1400,
  title = {{Further studies of Fermi-motion effects in lepton scattering from nuclear targets}},
  author = {Bodek, A. and Ritchie, J. L.},
  journal = {Phys. Rev. D},
  volume = {24},
  issue = {5},
  pages = {1400--1402},
  numpages = {0},
  year = {1981},
  month = {Sep},
  publisher = {American Physical Society},
  doi = {10.1103/PhysRevD.24.1400},
  url = {https://link.aps.org/doi/10.1103/PhysRevD.24.1400}
}

@article{LLEWELLYNSMITH1972261,
    title = {Neutrino reactions at accelerator energies},
    journal = {Phys. Rep.},
    volume = {3},
    number = {5},
    pages = {261-379},
    year = {1972},
    issn = {0370-1573},
    doi = {https://doi.org/10.1016/0370-1573(72)90010-5},
    url = {https://www.sciencedirect.com/science/article/pii/0370157372900105},
    author = {C.H. {Llewellyn Smith}}
}

@article{PhysRevC.83.045501,
  title = {Inclusive charged-current neutrino-nucleus reactions},
  author = {Nieves, J. and Simo, I. Ruiz and Vacas, M. J. Vicente},
  journal = {Phys. Rev. C},
  volume = {83},
  issue = {4},
  pages = {045501},
  numpages = {19},
  year = {2011},
  month = {Apr},
  publisher = {American Physical Society},
  doi = {10.1103/PhysRevC.83.045501},
  url = {https://link.aps.org/doi/10.1103/PhysRevC.83.045501}
}

@article{PhysRevD.88.113007,
  title = {Neutrino-nucleus quasi-elastic and 2p2h interactions up to 10 GeV},
  author = {Gran, R. and Nieves, J. and Sanchez, F. and Vacas, M. J. Vicente},
  journal = {Phys. Rev. D},
  volume = {88},
  issue = {11},
  pages = {113007},
  numpages = {10},
  year = {2013},
  month = {Dec},
  publisher = {American Physical Society},
  doi = {10.1103/PhysRevD.88.113007},
  url = {https://link.aps.org/doi/10.1103/PhysRevD.88.113007}
}

@article{REIN198179,
    title = {Neutrino-excitation of baryon resonances and single pion production},
    journal = {Ann. Phys. (N. Y.)},
    volume = {133},
    number = {1},
    pages = {79-153},
    year = {1981},
    issn = {0003-4916},
    doi = {https://doi.org/10.1016/0003-4916(81)90242-6},
    url = {https://www.sciencedirect.com/science/article/pii/0003491681902426},
    author = {Dieter Rein and Lalit M Sehgal},
}

@article{coherent_REIN198329,
    title = {Coherent $\pi^0$ production in neutrino reactions},
    journal = {Nucl. Phys. B},
    volume = {223},
    number = {1},
    pages = {29-44},
    year = {1983},
    issn = {0550-3213},
    doi = {https://doi.org/10.1016/0550-3213(83)90090-1},
    url = {https://www.sciencedirect.com/science/article/pii/0550321383900901},
    author = {Dieter Rein and Lalit M. Sehgal},
}

@article{coherentMass_REIN2007207,
    title = {{PCAC and the deficit of forward muons in $\pi^+$ production by neutrinos}},
    journal = {Phys. Lett. B},
    volume = {657},
    number = {4},
    pages = {207-209},
    year = {2007},
    issn = {0370-2693},
    doi = {https://doi.org/10.1016/j.physletb.2007.10.025},
    url = {https://www.sciencedirect.com/science/article/pii/S0370269307012580},
    author = {D. Rein and L.M. Sehgal},
}

@article{BodekYang_2003,
doi = {10.1088/0954-3899/29/8/369},
url = {https://dx.doi.org/10.1088/0954-3899/29/8/369},
year = {2003},
volume = {29},
number = {8},
pages = {1899},
author = {A. Bodek and U. K. Yang},
title = {{Higher twist, $\xi_w$ scaling, and effective LO PDFs for lepton scattering in the few GeV region}},
journal = {J. Phys. G: Nucl. Part. Phys.},
}

@article{BODEK2005113,
title = {{Improved low Q2 model for neutrino and electron nucleon cross sections in few GeV region}},
journal = {Nucl. Phys. Proc. Suppl.},
volume = {139},
pages = {113-118},
year = {2005},
doi = {https://doi.org/10.1016/j.nuclphysbps.2004.11.208},
url = {https://www.sciencedirect.com/science/article/pii/S0920563204007492},
author = {Arie Bodek and Inkyu Park and Un ki Yang},
}

@article{AGKY_Yang_2009,
  title={{A hadronization model for few-GeV neutrino interactions}},
  volume={63},
  ISSN={1434-6052},
  url={http://dx.doi.org/10.1140/epjc/s10052-009-1094-z},
  DOI={10.1140/epjc/s10052-009-1094-z},
  number={1},
  journal={Eur. Phys. J. C},
  publisher={Springer Science and Business Media LLC},
  author={Yang, T. and Andreopoulos, C. and Gallagher, H. and Hofmann, K. and Kehayias, P.},
  year={2009},
  pages={1–10} 
}

@article{KNO_KOBA1972317,
  title = {Scaling of multiplicity distributions in high energy hadron collisions},
  journal = {Nucl. Phys. B},
  volume = {40},
  pages = {317-334},
  year = {1972},
  issn = {0550-3213},
  doi = {https://doi.org/10.1016/0550-3213(72)90551-2},
  url = {https://www.sciencedirect.com/science/article/pii/0550321372905512},
  author = {Z. Koba and H.B. Nielsen and P. Olesen},
}

@article{Pythia_Sjostrand_2001,
  title={{High-energy-physics event generation with PYTHIA 6.1}},
  volume={135},
  ISSN={0010-4655},
  url={http://dx.doi.org/10.1016/S0010-4655(00)00236-8},
  DOI={10.1016/s0010-4655(00)00236-8},
  number={2},
  journal={Comput. Phys. Commun.},
  publisher={Elsevier BV},
  author={Sjöstrand, Torbjörn and others},
  year={2001},
  pages={238} 
}

@article{Dytman:2009zz,
  title = {Final State Interactions in Neutrino-Nucleus Experiments},
  author = {S. Dytman},
  journal = {Acta Phys. Pol. B},
  volume = {40},
  issue = {2245},
  year = {2009},
  url = {https://www.actaphys.uj.edu.pl/R/40/9/2445/pdf}
}

@article{RPANieves2004_PhysRevC.70.055503,
  title = {Inclusive quasielastic charged-current neutrino-nucleus reactions},
  author = {Nieves, J. and Amaro, J. E. and Valverde, M.},
  journal = {Phys. Rev. C},
  volume = {70},
  issue = {5},
  pages = {055503},
  numpages = {23},
  year = {2004},
  publisher = {American Physical Society},
  doi = {10.1103/PhysRevC.70.055503},
  url = {https://link.aps.org/doi/10.1103/PhysRevC.70.055503}
}

@misc{gran2017model,
      title={{Model uncertainties for Valencia RPA effect for MINERvA}}, 
      author={Richard Gran},
      year={2017},
      eprint={1705.02932},
      archivePrefix={arXiv},
      primaryClass={hep-ex},
      url={https://arxiv.org/abs/1705.02932}, 
}

@article{Rodrigues2016_PhysRevLett.116.071802,
  title = {{Identification of Nuclear Effects in Neutrino-Carbon Interactions at Low Three-Momentum Transfer}},
  author = {Rodrigues, P. A. and others},
  collaboration = {MINER\ensuremath{\nu}A Collaboration},
  journal = {Phys. Rev. Lett.},
  volume = {116},
  issue = {7},
  pages = {071802},
  numpages = {6},
  year = {2016},
  month = {Feb},
  publisher = {American Physical Society},
  doi = {10.1103/PhysRevLett.116.071802},
  url = {https://link.aps.org/doi/10.1103/PhysRevLett.116.071802}
}

@article{Gran2018_PhysRevLett.120.221805,
  title = {{Antineutrino Charged-Current Reactions on Hydrocarbon with Low Momentum Transfer}},
  author = {Gran, R. and others},
  collaboration = {MINER\ensuremath{\nu}A Collaboration},
  journal = {Phys. Rev. Lett.},
  volume = {120},
  issue = {22},
  pages = {221805},
  numpages = {7},
  year = {2018},
  month = {Jun},
  publisher = {American Physical Society},
  doi = {10.1103/PhysRevLett.120.221805},
  url = {https://link.aps.org/doi/10.1103/PhysRevLett.120.221805}
}

@article{pion_Rodrigues_2016,
	author = {Rodrigues, Philip and Wilkinson, Callum and McFarland, Kevin},
	date = {2016/08/24},
	doi = {10.1140/epjc/s10052-016-4314-3},
	id = {Rodrigues2016},
	isbn = {1434-6052},
	journal = {Eur. Phys. J. C. },
	number = {8},
	pages = {474},
	title = {{Constraining the GENIE model of neutrino-induced single pion production using reanalyzed bubble chamber data}},
	url = {https://doi.org/10.1140/epjc/s10052-016-4314-3},
	volume = {76},
	year = {2016}}

@article{Ramirez2023_PhysRevLett.131.051801,
  title = {{Neutrino-Induced Coherent ${\ensuremath{\pi}}^{+}$ Production in C, CH, Fe, and Pb at $⟨{E}_{\ensuremath{\nu}}⟩\ensuremath{\sim}6\text{ }\text{ }\mathrm{GeV}$}},
  author = {Ram\'{\i}rez, M. A. and others},
  collaboration = {MINER\ensuremath{\nu}A Collaboration},
  journal = {Phys. Rev. Lett.},
  volume = {131},
  issue = {5},
  pages = {051801},
  numpages = {8},
  year = {2023},
  publisher = {American Physical Society},
  doi = {10.1103/PhysRevLett.131.051801},
  url = {https://link.aps.org/doi/10.1103/PhysRevLett.131.051801}
}

@article{PhysRevD.85.073003,
  title = {{Diffractive neutrino production of pions on nuclei: Adler relation within the color-dipole description}},
  author = {Kopeliovich, B. Z. and Schmidt, Iv\'an and Siddikov, M.},
  journal = {Phys. Rev. D},
  volume = {85},
  issue = {7},
  pages = {073003},
  numpages = {8},
  year = {2012},
  publisher = {American Physical Society},
  doi = {10.1103/PhysRevD.85.073003},
  url = {https://link.aps.org/doi/10.1103/PhysRevD.85.073003}
}

@article{Bercellie2023_PhysRevLett.131.011801,
  title = {{Simultaneous Measurement of Muon Neutrino ${\ensuremath{\nu}}_{\ensuremath{\mu}}$ Charged-Current Single ${\ensuremath{\pi}}^{+}$ Production in CH, C, ${\mathrm{H}}_{2}\mathrm{O}$, Fe, and Pb Targets in MINERvA}},
  author = {Bercellie, A. and others},
  collaboration = {MINER\ensuremath{\nu}A Collaboration},
  journal = {Phys. Rev. Lett.},
  volume = {131},
  issue = {1},
  pages = {011801},
  numpages = {7},
  year = {2023},
  publisher = {American Physical Society},
  doi = {10.1103/PhysRevLett.131.011801},
  url = {https://link.aps.org/doi/10.1103/PhysRevLett.131.011801}
}

@article{KAIDALOV1982459,
    title = {The quark-gluon structure of the pomeron and the rise of inclusive spectra at high energies},
    journal = {Phys. Lett. B},
    volume = {116},
    number = {6},
    pages = {459-463},
    year = {1982},
    issn = {0370-2693},
    doi = {https://doi.org/10.1016/0370-2693(82)90168-X},
    url = {https://www.sciencedirect.com/science/article/pii/037026938290168X},
    author = {A.B. Kaidalov},

}

@article{ALIAGA2014130,
    title = {{Design, calibration, and performance of the MINERvA detector}},
    journal = {Nucl. Instrum. Methods Phys.~Res., Sect.~A},
    collaboration = {MINER\ensuremath{\nu}A Collaboration},
    volume = {743},
    pages = {130-159},
    year = {2014},
    issn = {0168-9002},
    doi = {https://doi.org/10.1016/j.nima.2013.12.053},
    url = {https://www.sciencedirect.com/science/article/pii/S0168900214000035},
    author = {L. Aliaga and others},

}

@article{ALIAGA201528,
    title = {{MINERvA neutrino detector response measured with test beam data}},
    journal = {Nucl. Instrum. Methods Phys.~Res., Sect.~A},
    collaboration = {MINER\ensuremath{\nu}A Collaboration},
    volume = {789},
    pages = {28-42},
    year = {2015},
    issn = {0168-9002},
    doi = {https://doi.org/10.1016/j.nima.2015.04.003},
    url = {https://www.sciencedirect.com/science/article/pii/S0168900215004568},
    author = {L. Aliaga and others},

}

@article{PhysRevD.100.052002,
  title = {Neutron measurements from antineutrino hydrocarbon reactions},
  author = {Elkins, M. and others},
  collaboration = {MINER\ensuremath{\nu}A Collaboration},
  journal = {Phys. Rev. D},
  volume = {100},
  issue = {5},
  pages = {052002},
  numpages = {20},
  year = {2019},
  publisher = {American Physical Society},
  doi = {10.1103/PhysRevD.100.052002},
  url = {https://link.aps.org/doi/10.1103/PhysRevD.100.052002}
}

@article{Michael_2008,
   title={The magnetized steel and scintillator calorimeters of the MINOS experiment},
   collaboration = {MINOS Collaboration},
   volume={596},
   ISSN={0168-9002},
   url={http://dx.doi.org/10.1016/j.nima.2008.08.003},
   DOI={10.1016/j.nima.2008.08.003},
   number={2},
   journal={Nucl. Instrum. Methods Phys. Res., Sect.~A},
   publisher={Elsevier BV},
   author={Michael, D.G. and others},
   year={2008},
   pages={190}
}

@misc{MINERvAWebsite,
  author       = {{MINER$\nu$A Collaboration}},
  howpublished = {\url{https://minerva.fnal.gov/}},
  year = {2026}
}

@article{Adamson_2016,
   title={The NuMI neutrino beam},
   volume={806},
   ISSN={0168-9002},
   url={http://dx.doi.org/10.1016/j.nima.2015.08.063},
   DOI={10.1016/j.nima.2015.08.063},
   journal={Nucl. Instrum. Methods Phys. Res., Sect.~A},
   publisher={Elsevier BV},
   author={Adamson, P. and others},
   year={2016},
   pages={279–306} 
}

@article{Akbar_2022,
doi = {10.1088/1748-0221/17/08/T08013},
url = {https://doi.org/10.1088/1748-0221/17/08/T08013},
year = {2022},
month = {aug},
publisher = {IOP Publishing},
volume = {17},
number = {08},
pages = {T08013},
author = {Akbar, F. and others},
collaboration = {MINER\ensuremath{\nu}A Collaboration},
title = {Vertex finding in neutrino-nucleus interaction: a model architecture comparison},
journal = {J. Instrum.},

}

@article{DAGOSTINI1995487,
title = {A multidimensional unfolding method based on Bayes' theorem},
journal = {Nucl. Instrum. Methods Phys. Res., Sect. A},
volume = {362},
number = {2},
pages = {487-498},
year = {1995},
issn = {0168-9002},
doi = {https://doi.org/10.1016/0168-9002(95)00274-X},
url = {https://www.sciencedirect.com/science/article/pii/016890029500274X},
author = {G. D'Agostini},
}

@misc{fine2022datapreservationminerva,
      title={{Data Preservation at MINERvA}}, 
      collaboration = {$\mathrm{MINER}\ensuremath{\nu}\mathrm{A}$ Collaboration},
      author={R. Fine and B. Messerly and K. S. McFarland and others},
      year={2022},
      eprint={2009.04548},
      archivePrefix={arXiv},
      primaryClass={hep-ex},
      url={https://arxiv.org/abs/2009.04548}, 
}

@article{Stowell:2016jfr,
  author = "Stowell, P. and others",
  title = "{NUISANCE: a neutrino cross-section generator tuning and comparison framework}",
  eprint = "1612.07393",
  archivePrefix = "arXiv",
  primaryClass = "hep-ex",
  doi = "10.1088/1748-0221/12/01/P01016",
  journal = "J. Instrum.",
  volume = "12",
  number = "01",
  pages = "P01016",
  year = "2017"
}

@article{NEUT2021,
  author  = {Hayato, Y. and Pickering, L. and Wret, C. and others},
  title   = {The NEUT neutrino interaction simulation program library},
  journal = {Eur. Phys. J. Special Topics},
  volume  = {230},
  pages   = {4469--4481},
  year    = {2021},
  doi     = {10.1140/epjs/s11734-021-00287-7}
}

@article{PhysRevD.108.032018,
  title = {High-statistics measurement of antineutrino quasielasticlike scattering at ${E}_{\overline{\ensuremath{\nu}}}$ 6 GeV on a hydrocarbon target},
  author = {Bashyal, A. and others},
  collaboration = {$\mathrm{MINER}\ensuremath{\nu}\mathrm{A}$ Collaboration},
  journal = {Phys. Rev. D},
  volume = {108},
  issue = {3},
  pages = {032018},
  numpages = {14},
  year = {2023},
  month = {Aug},
  publisher = {American Physical Society},
  doi = {10.1103/PhysRevD.108.032018},
  url = {https://link.aps.org/doi/10.1103/PhysRevD.108.032018}
}

@article{NatureMINERvA,
	author = {Cai, T. and others},
    collaboration = {MINER$\nu$A Collaboration},
	date = {2023/02/01},
	doi = {10.1038/s41586-022-05478-3},
	id = {Cai2023},
	isbn = {1476-4687},
	journal = {Nature (London)},
	number = {7946},
	pages = {48--53},
	title = {Measurement of the axial vector form factor from antineutrino--proton scattering},
	url = {https://doi.org/10.1038/s41586-022-05478-3},
	volume = {614},
	year = {2023}}

@article{BETANCOURT20181,
title = {Comparisons and challenges of modern neutrino scattering experiments (TENSIONS2016 report)},
journal = {Phys. Rep.},
volume = {773-774},
pages = {1-28},
year = {2018},
issn = {0370-1573},
doi = {https://doi.org/10.1016/j.physrep.2018.08.003},
url = {https://www.sciencedirect.com/science/article/pii/S0370157318302175},
author = {M. Betancourt and others},
}

@article{PhysRevD.105.092004,
  title = {Comparisons and challenges of modern neutrino-scattering experiments},
  author = {Avanzini, M. Buizza and others},
  journal = {Phys. Rev. D},
  volume = {105},
  issue = {9},
  pages = {092004},
  numpages = {55},
  year = {2022},
  month = {May},
  publisher = {American Physical Society},
  doi = {10.1103/PhysRevD.105.092004},
  url = {https://link.aps.org/doi/10.1103/PhysRevD.105.092004}
}

@article{10.1063/1.4919465,
    author = {Katori, Teppei},
    title = {Meson exchange current (MEC) models in neutrino interaction generators},
    journal = {AIP Conf. Proc.},
    volume = {1663},
    number = {1},
    pages = {030001},
    year = {2015},
    month = {05},
    issn = {0094-243X},
    doi = {10.1063/1.4919465},
    url = {https://doi.org/10.1063/1.4919465},
    eprint = {1304.6014},
    archivePrefix = {arXiv},
    primaryClass  = {nucl-th}
}

@article{PhysRevD.76.113004,
  title = {Lepton mass effects in single pion production by neutrinos},
  author = {Berger, Ch. and Sehgal, L. M.},
  journal = {Phys. Rev. D},
  volume = {76},
  issue = {11},
  pages = {113004},
  numpages = {8},
  year = {2007},
  month = {Dec},
  publisher = {American Physical Society},
  doi = {10.1103/PhysRevD.76.113004},
  url = {https://link.aps.org/doi/10.1103/PhysRevD.76.113004}
}

@article{PhysRevD.79.053003,
  title = {Partially conserved axial vector current and coherent pion production by low energy neutrinos},
  author = {Berger, Ch. and Sehgal, L. M.},
  journal = {Phys. Rev. D},
  volume = {79},
  issue = {5},
  pages = {053003},
  numpages = {6},
  year = {2009},
  month = {Mar},
  publisher = {American Physical Society},
  doi = {10.1103/PhysRevD.79.053003},
  url = {https://link.aps.org/doi/10.1103/PhysRevD.79.053003}
}

@article{BENHAR1994493,
title = {Spectral function of finite nuclei and scattering of GeV electrons},
journal = {Nucl. Phys. A},
volume = {579},
number = {3},
pages = {493-517},
year = {1994},
issn = {0375-9474},
doi = {https://doi.org/10.1016/0375-9474(94)90920-2},
url = {https://www.sciencedirect.com/science/article/pii/0375947494909202},
author = {O. Benhar and A. Fabrocini and S. Fantoni and I. Sick},
}

@ARTICLE{2006NuPhS.159..211J,
       author = {{Juszczak}, Cezary and {Nowak}, Jaros{\l}aw A. and {Sobczyk}, Jan T.},
        title = "{Simulations from a new neutrino event generator}",
      journal = {Nucl. Phys. B, Proc. Suppl.},
         year = 2006,
        month = sep,
       volume = {159},
        pages = {211-216},
          doi = {10.1016/j.nuclphysbps.2006.08.069},
archivePrefix = {arXiv},
       eprint = {hep-ph/0512365},
 primaryClass = {hep-ph},
       adsurl = {https://ui.adsabs.harvard.edu/abs/2006NuPhS.159..211J}
}

@article{PhysRevD.104.072009,
  title = {Neutrino-nucleon cross-section model tuning in GENIE v3},
  author = {Tena-Vidal, J\'ulia and Andreopoulos and others},
  collaboration = {GENIE Collaboration},
  journal = {Phys. Rev. D},
  volume = {104},
  issue = {7},
  pages = {072009},
  numpages = {28},
  year = {2021},
  month = {Oct},
  publisher = {American Physical Society},
  doi = {10.1103/PhysRevD.104.072009},
  url = {https://link.aps.org/doi/10.1103/PhysRevD.104.072009}
}

@article{PhysRevD.77.053001,
  title = {Form factors in the quark resonance model},
  author = {Graczyk, Krzysztof M. and Sobczyk, Jan T.},
  journal = {Phys. Rev. D},
  volume = {77},
  issue = {5},
  pages = {053001},
  numpages = {12},
  year = {2008},
  month = {Mar},
  publisher = {American Physical Society},
  doi = {10.1103/PhysRevD.77.053001},
  url = {https://link.aps.org/doi/10.1103/PhysRevD.77.053001}
}

@article{lozano2025measurementchargedcurrentnumubarnumu,
      title={Measurement of charged-current $\nu_\mu$ and $\bar{\nu}_\mu$ cross sections on hydrocarbon in a shallow inelastic scattering region}, 
      author={A. Lozano and others},
        collaboration = {MINERvA Collaboration},
      journal = {Phys. Rev. D},
      volume = {114},
      issue = {7},
      pages = {072001},
      numpages = {23},
      year = {2026},
      publisher = {American Physical Society},
      doi = {10.1103/vq38-kmxq},
      url = {https://link.aps.org/doi/10.1103/vq38-kmxq}
}

@article{icaruscollaboration2026measurementmuonantineutrinochargedcurrent,
      title={{Measurement of muon (anti-)neutrino charged-current quasielastic-like cross section using off-axis NuMI beam at ICARUS}}, 
      author={F. Abd Alrahman and others},
      collaboration = {ICARUS Collaboration},
      journal = {Phys. Rev. D},
      volume = {114},
      issue = {5},
      pages = {052002},
      numpages = {23},
      year = {2026},
      month = {Sep},
      publisher = {American Physical Society},
      doi = {10.1103/3jfl-dczc},
      url = {https://link.aps.org/doi/10.1103/3jfl-dczc}
}

@article{Liu:2026wlw,
    author = "Liu, Liang and Gardiner, Steven and Dytman, Steven",
    title = "{Benchmarking state-of-the-art theory and empirical models of pionless neutrino-argon scattering in GENIE}",
    eprint = "2605.14196",
    archivePrefix = "arXiv",
    primaryClass = "hep-ph",
    reportNumber = "FERMILAB-PUB-26-0211-CSAID",
    doi = "10.1016/j.physletb.2026.140750",
    journal = "Phys. Lett. B",
    volume = "880",
    pages = "140750",
    year = "2026"
}

@article{meyer2026nucleonaxialformfactor,
      title={The Nucleon Axial Form Factor from Elementary Target Data}, 
      author={A. S. Meyer and others},
      year={2026},
      collaboration = {MINER$\nu$A Collaboration},
      eprint={2512.14097},
      archivePrefix={arXiv},
      primaryClass={hep-ex},
      url={https://arxiv.org/abs/2512.14097}, 
}

@misc{supp,
  note = {See Supplemental Material for details of the analysis steps, the neutrino flux, the systematic uncertainties, and the generator configurations of the interaction model predictions, as well as tabulated cross-section and cross-section-ratio results with their covariance matrices and tabulated $\chi^2$ values for the model comparisons. The Supplemental Material includes Refs.~\cite{BETANCOURT20181,PhysRevD.105.092004,ANDREOPOULOS201087,RFG_SMITH1972605,BodekRitchie1981_PhysRevD.24.1400,LLEWELLYNSMITH1972261,10.1063/1.4919465,PhysRevD.76.113004,PhysRevD.79.053003,BodekYang_2003,BODEK2005113,AGKY_Yang_2009,KNO_KOBA1972317,Pythia_Sjostrand_2001,Dytman:2009zz,PhysRevD.104.072009,RPANieves2004_PhysRevC.70.055503,gran2017model,PhysRevC.83.045501,PhysRevD.88.113007,NEUT2021,REIN198179,BENHAR1994493,2006NuPhS.159..211J}.}
}

@misc{MINERvAOpenData,
  author       = {{MINERvA Collaboration}},
  title        = {{MINERvA Experiment Open Data Release}},
  howpublished = {Fermilab Report No.~FERMILAB-DATA-2025-07},
  year         = {2025},
  doi          = {10.15484/3022562},
  note         = {\href{https://doi.org/10.15484/3022562}{doi:10.15484/3022562}}
}

@misc{ZenodoRelease,
  author        = {Klustov\'{a}, A. and others},
  collaboration = {{MINERvA Collaboration}},
  title         = {{Data release: Measurement of Inclusive Charged-Current
                    $\bar{\nu}_\mu$ Scattering on C, CH, Fe, and Pb at
                    $\langle E_{\bar{\nu}} \rangle \sim 6$~GeV with MINERvA}},
  howpublished  = {Zenodo},
  year          = {2026},
  doi           = {10.5281/zenodo.24306934},
  note          = {\href{https://doi.org/10.5281/zenodo.24306934}{doi:10.5281/zenodo.24306934}}
}

@misc{MINERvASoftware,
  author       = {{MINERvA Collaboration}},
  title        = {{MINERvA software}},
  howpublished = {GitHub repository},
  year         = {2026},
  note         = {\url{https://github.com/MinervaExpt}}
}
 %\input{main.bbl}

% -----------------------------------------------------------------------------
% SUPPLEMENTAL MATERIAL
\newpage

%\documentclass[%
%onecolumn,
%superscriptaddress,
%preprintnumbers,
%nofootinbib,
% amsmath,amssymb,
% aps, prl,
%]{revtex4-2}

%\usepackage{graphicx}% Include figure files
%\usepackage{subcaption} % For subfigures (PRL allows this; avoid the older subfigure pkg)
%\usepackage{dcolumn}% Align table columns on decimal point
%\usepackage{bm}% bold math
%\usepackage{comment}
%\usepackage{xcolor}
%\usepackage{rotating}
%\usepackage{hyperref}% add hypertext capabilities

%\begin{document}

\setcounter{page}{1}
\renewcommand{\thepage}{S\arabic{page}}
\renewcommand\thefigure{Supp. \arabic{figure}}
\setcounter{figure}{0}
\renewcommand\thetable{Supp. \Roman{table}}
\setcounter{table}{0}

\onecolumngrid   % one column
\newpage

\section*{Supplemental Material}

\maketitle

% X/Y/Z distributions for the selected data candidates
\begin{figure}[h]
  \centering
  \begin{subfigure}[b]{0.5\textwidth}
    \centering
    \includegraphics[width=\linewidth]{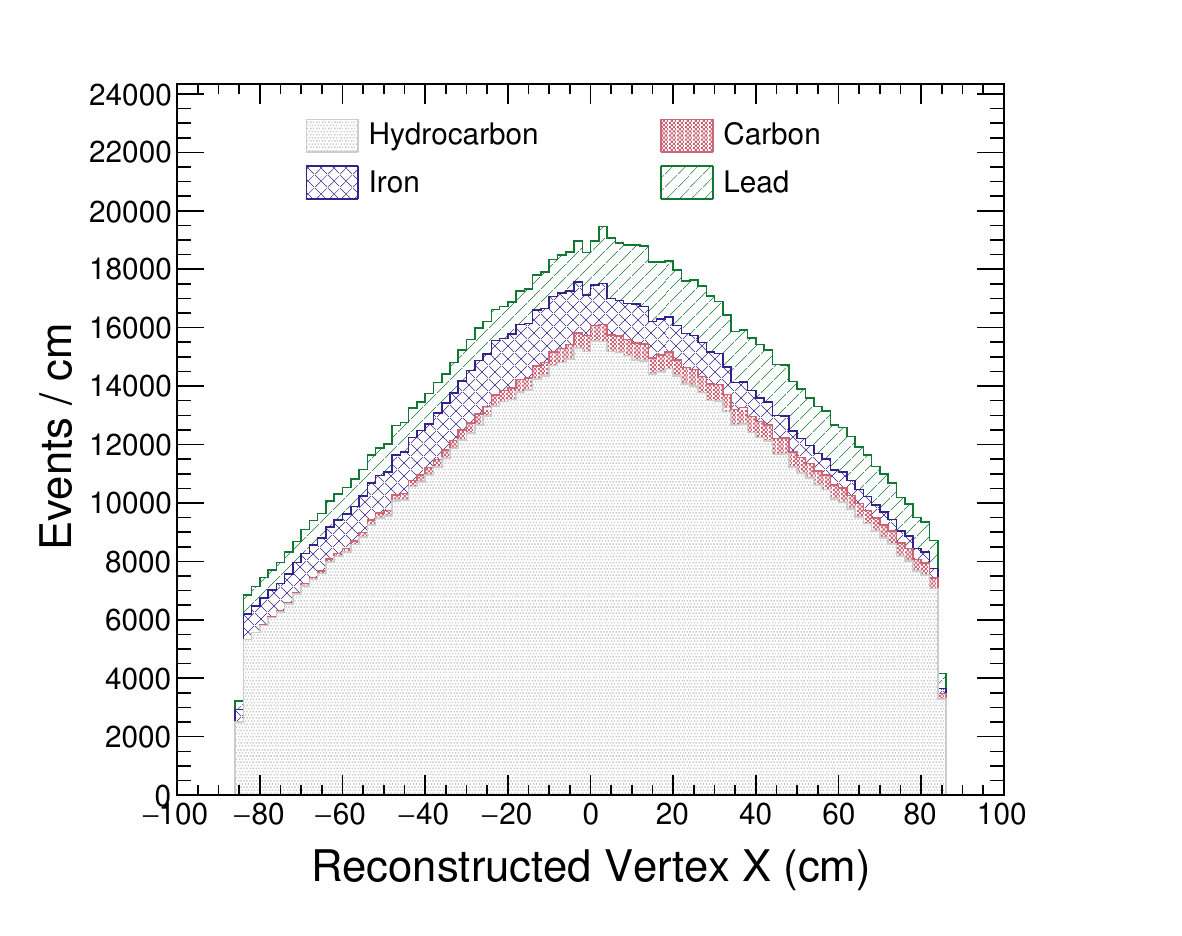}
  \end{subfigure}
  \hspace{-2em}
  \begin{subfigure}[b]{0.5\textwidth}
    \centering
    \includegraphics[width=\linewidth]{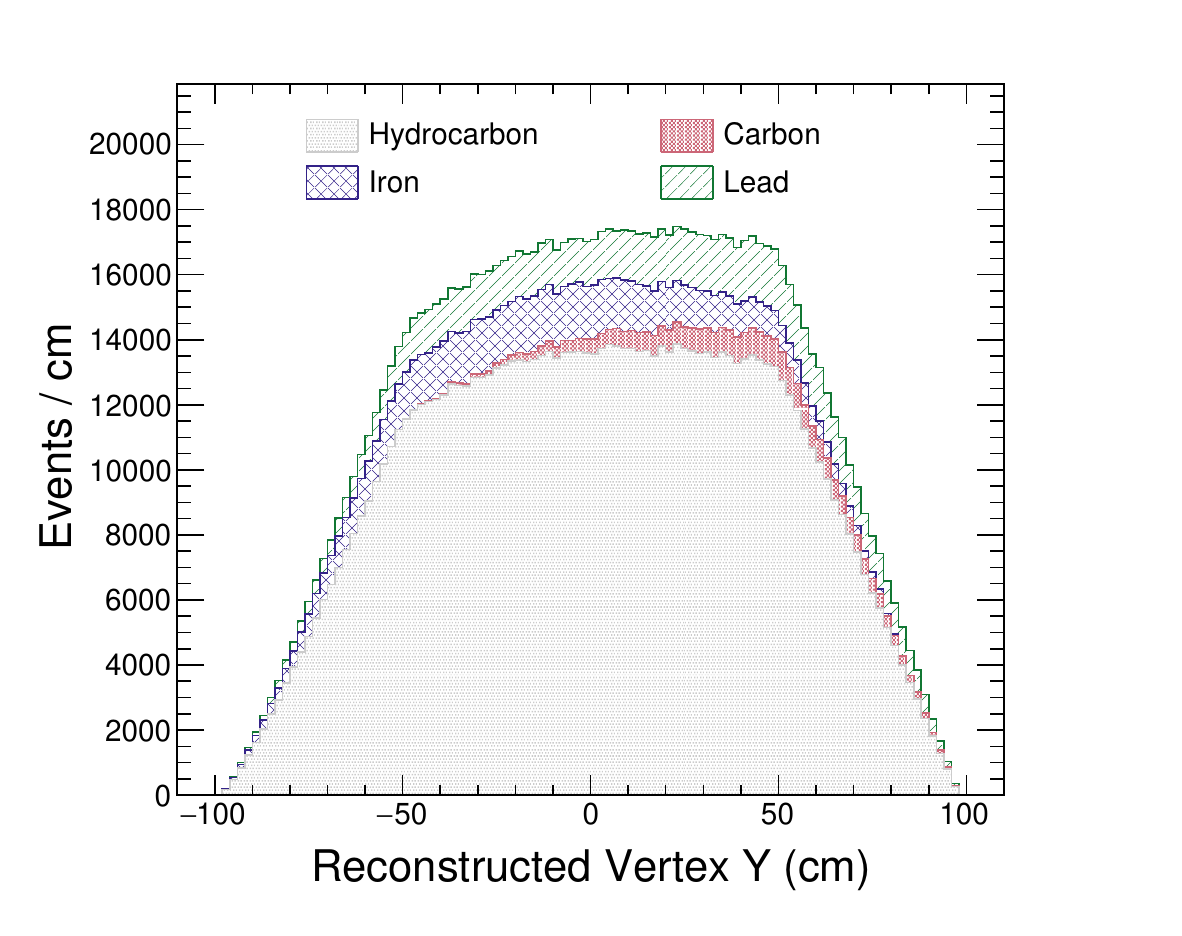}
  \end{subfigure}
  \begin{subfigure}[b]{0.5\textwidth}
    \centering
    \includegraphics[width=\linewidth]{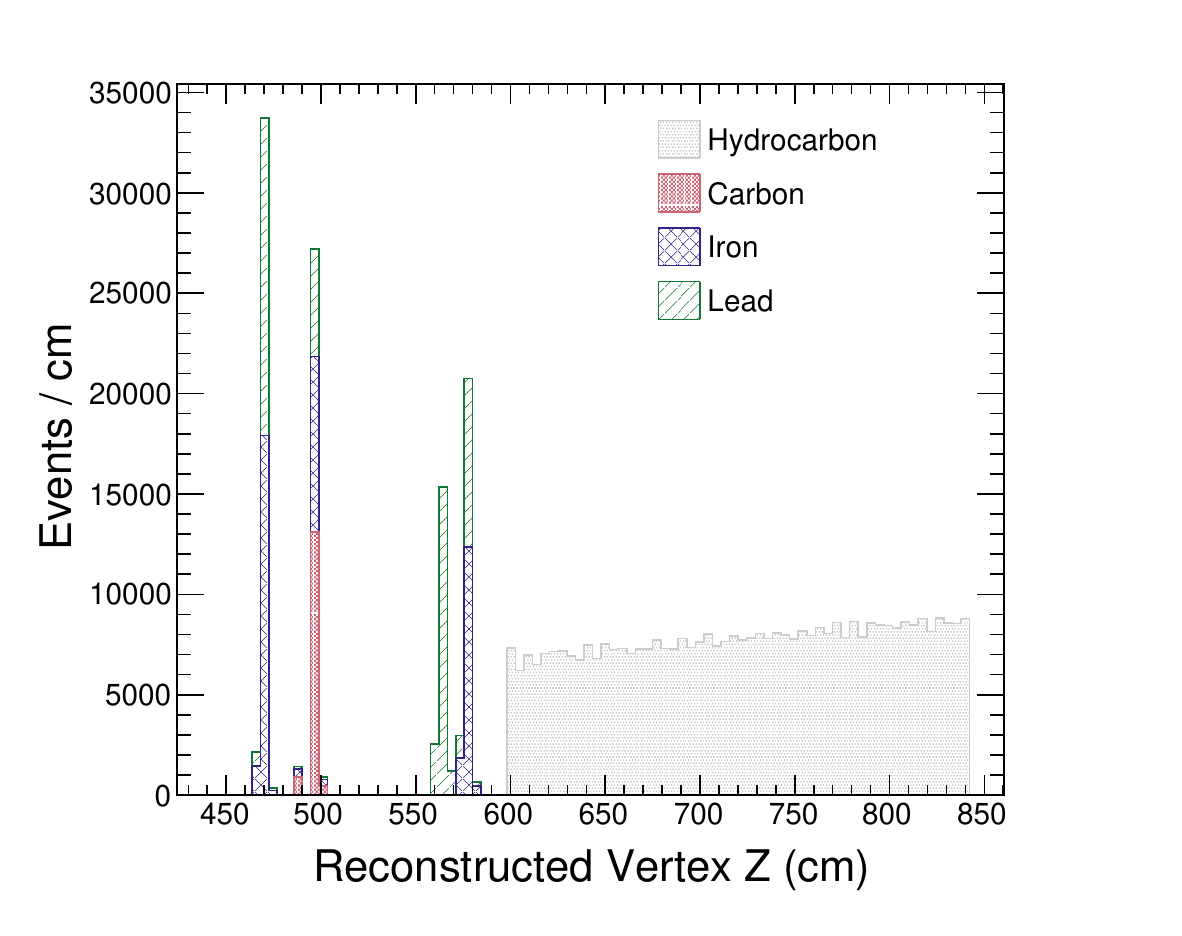}
  \end{subfigure}
  \caption{Reconstructed $X$, $Y$, and $Z$ vertex distributions for the selected candidate data events. The $X$ and $Y$ distributions are shown for events over all $Z$.}
  \label{fig:vertexDistributions}
\end{figure}

% Plane probability
\begin{figure}[h]
  \centering
  \includegraphics[width=0.5\textwidth]{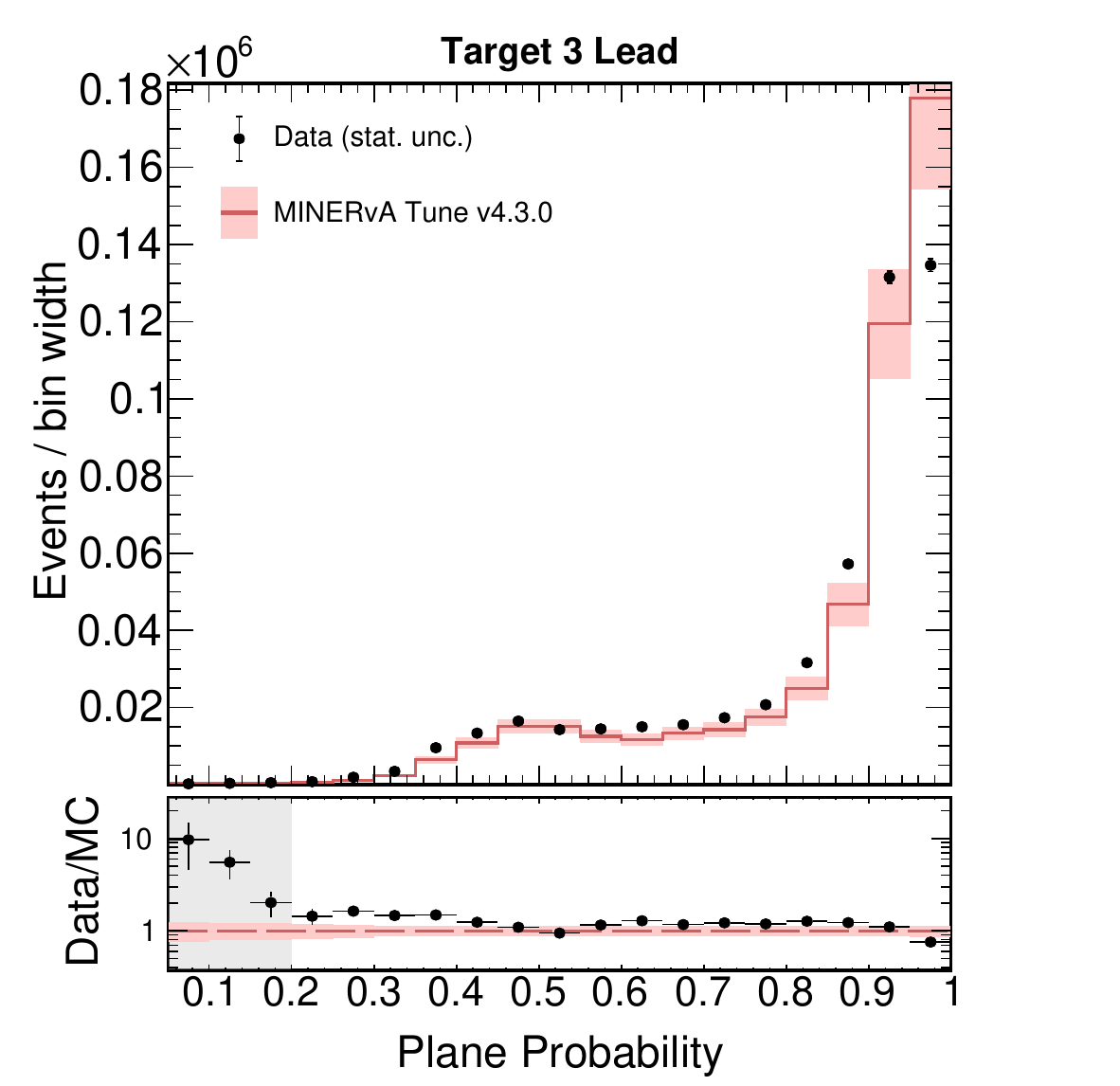}
  \caption{Example CNN plane-probability distribution for target 3 lead, prior to the requirement on plane probability; all other selection criteria are applied. Upper panel: selected data candidates (points, with statistical uncertainties) and the simulation prediction (red line), with the pink band representing the combined statistical and systematic uncertainty on the simulation. Lower panel: data-to-simulation ratio, with error bars showing statistical uncertainties, horizontal bars indicating bin widths, and the band representing simulation systematic uncertainty. The shaded (greyed-out) region indicates events rejected by the plane-probability cut at 0.2.}
  \label{fig:plane_probability}
\end{figure}

% EVENT SELECTION INTERACTION BREAKDOWN: signal and background
\begin{figure}[h]
  \centering
  \begin{subfigure}[b]{0.4\textwidth}
    \centering
    \includegraphics[width=\linewidth]{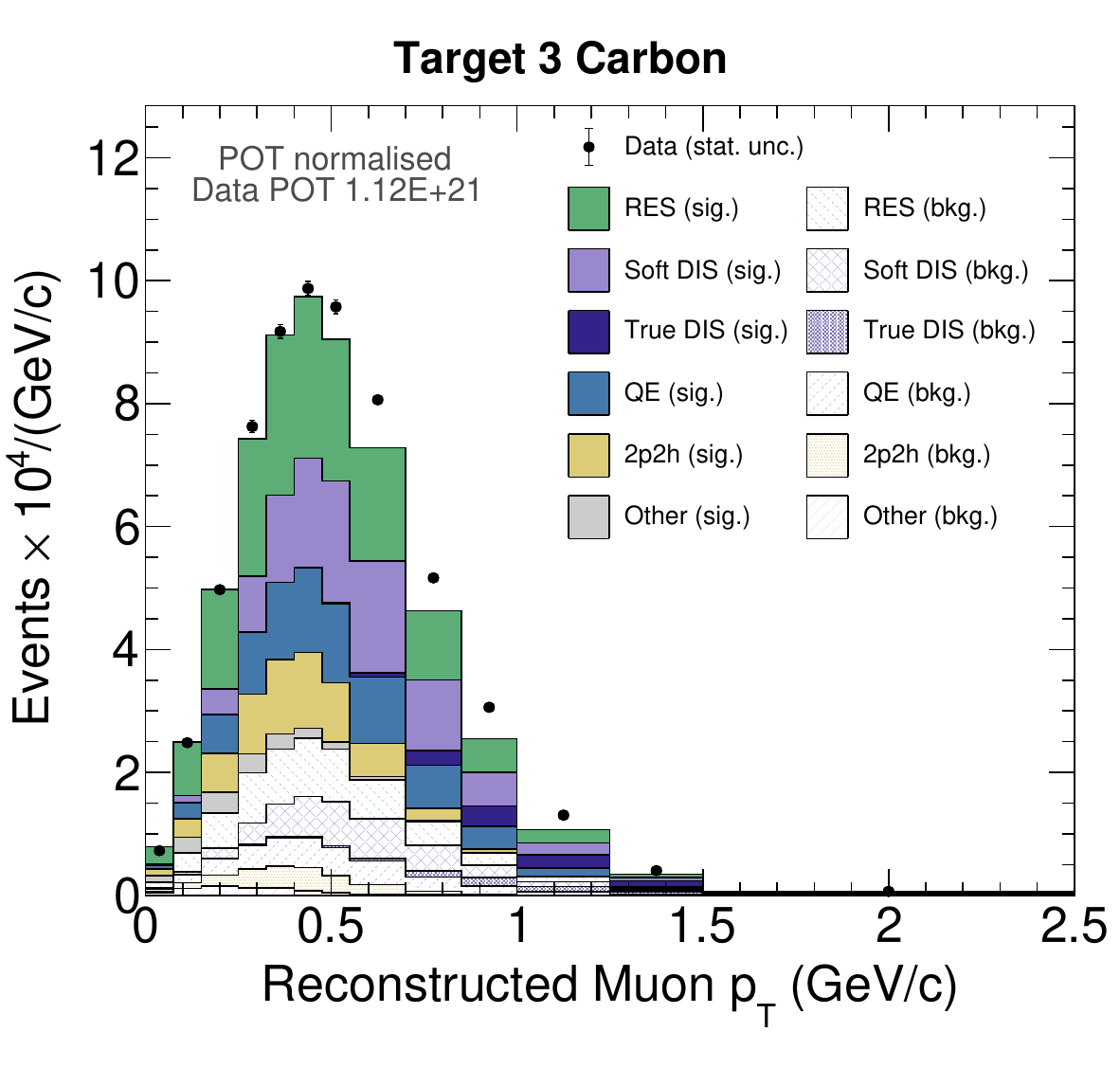}
  \vspace{-2em}
  \end{subfigure}
  \hspace{-0.5em}
  \begin{subfigure}[b]{0.4\textwidth}
    \centering
    \includegraphics[width=\linewidth]{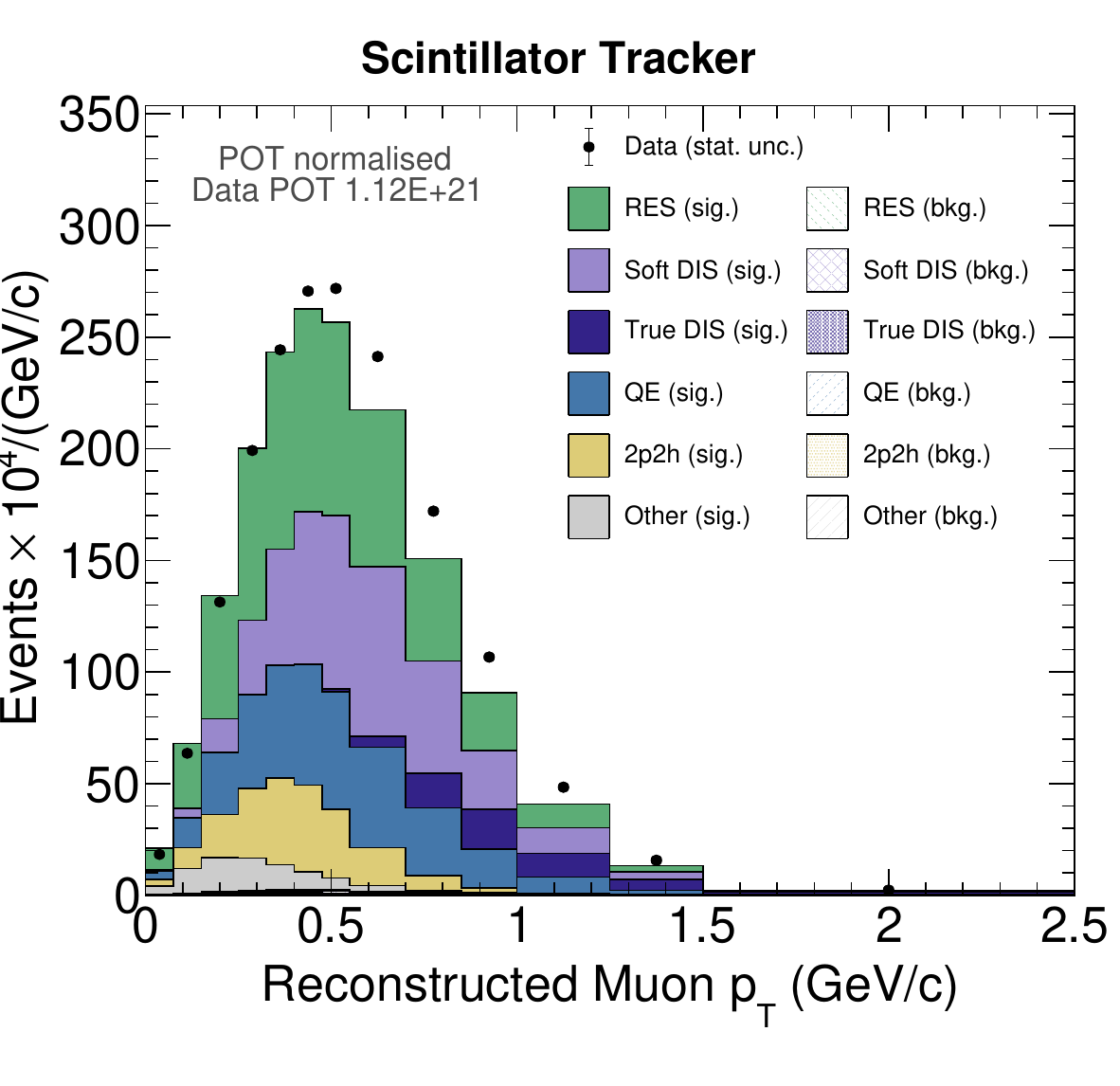}
  \vspace{-2em}
  \end{subfigure}
  \begin{subfigure}[b]{0.4\textwidth}
    \centering
    \includegraphics[width=\linewidth]{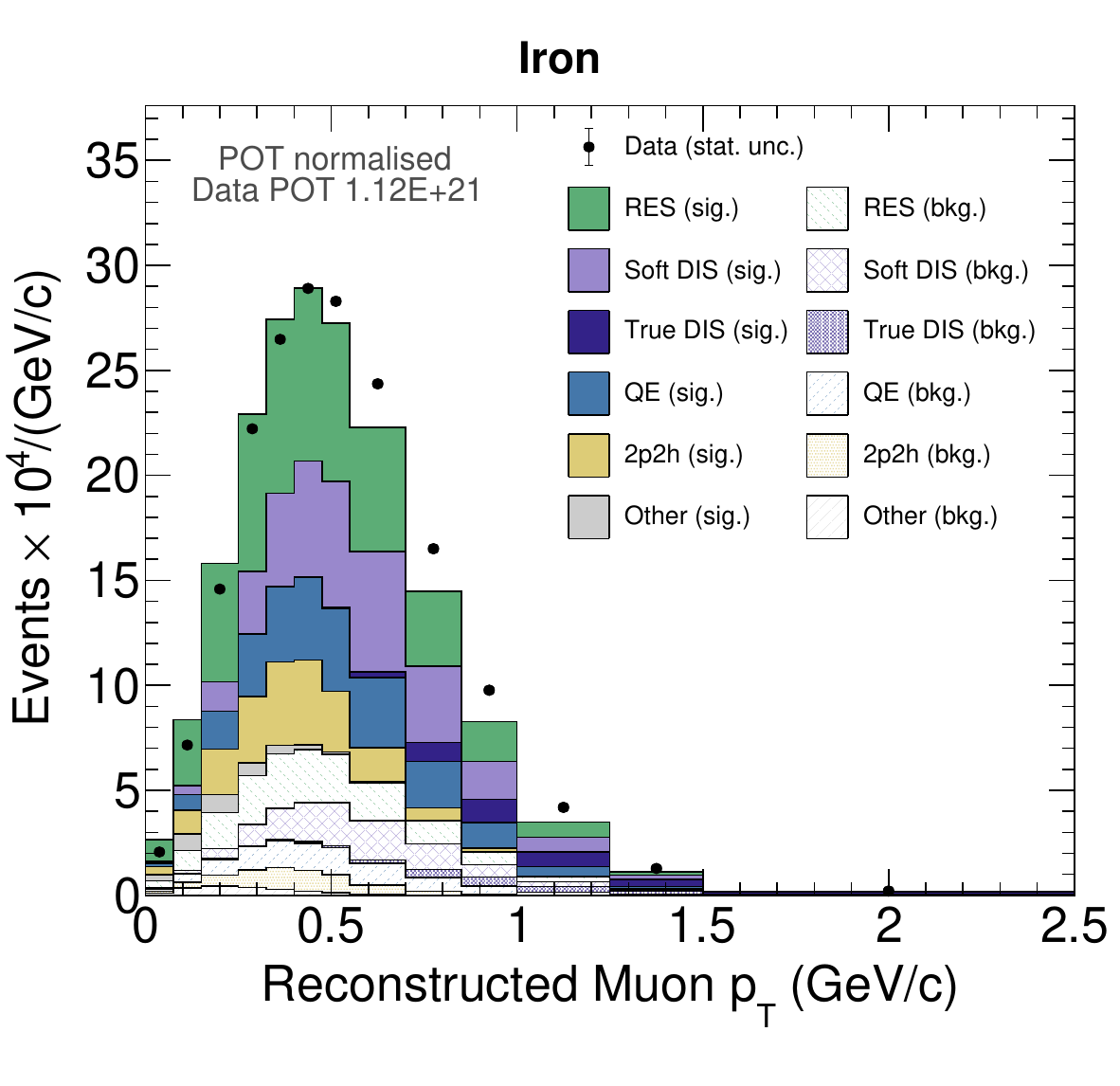}
  \end{subfigure}
  \hspace{-0.5em}
  \begin{subfigure}[b]{0.4\textwidth}
    \centering
    \includegraphics[width=\linewidth]{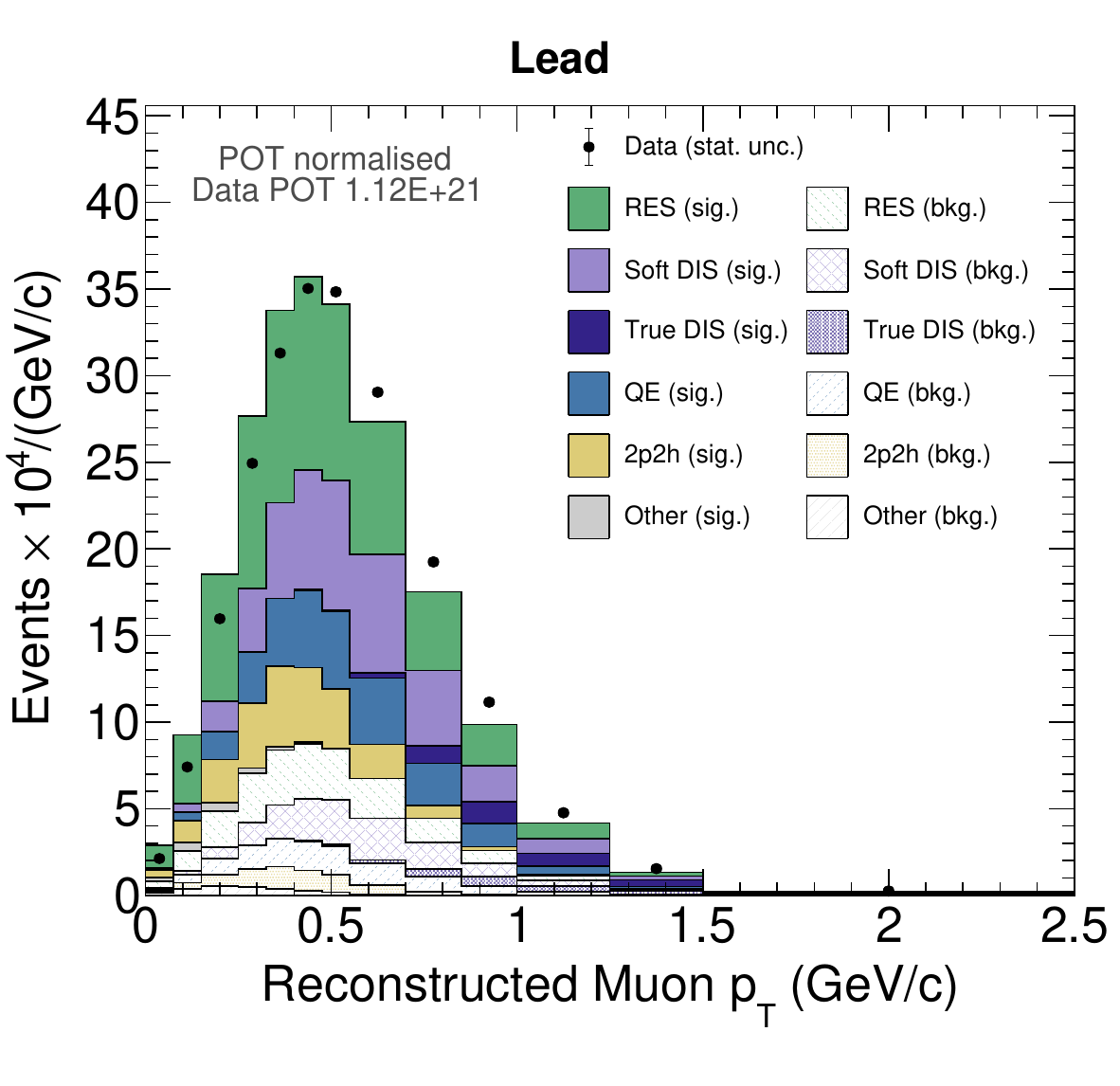}
  \end{subfigure}
  \caption{Selected events in C, CH, Fe, and Pb as a function of $p_{\mathrm{T}}$, with the simulation predictions for both signal and background broken down by contribution from each interaction type.}
  \label{fig:pt_evselCandFe}
\end{figure}

% EVENT SELECTION (C, Fe)
\begin{figure}[h]
  \centering
  \begin{subfigure}[b]{0.5\textwidth}
    \centering
    \includegraphics[width=\linewidth]{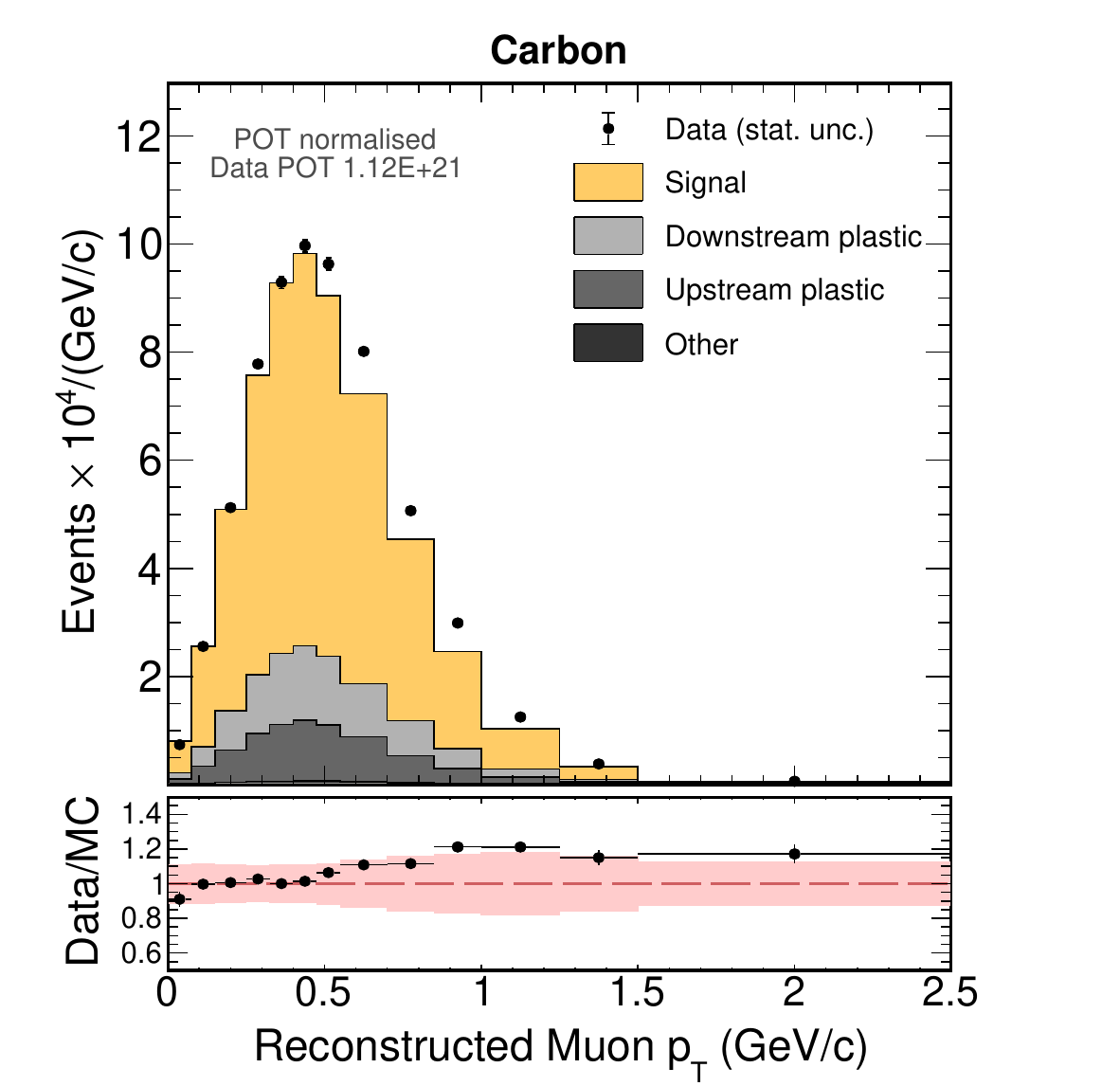}
  \end{subfigure}
  \hspace{-1.7em}
  \begin{subfigure}[b]{0.5\textwidth}
    \centering
    \includegraphics[width=\linewidth]{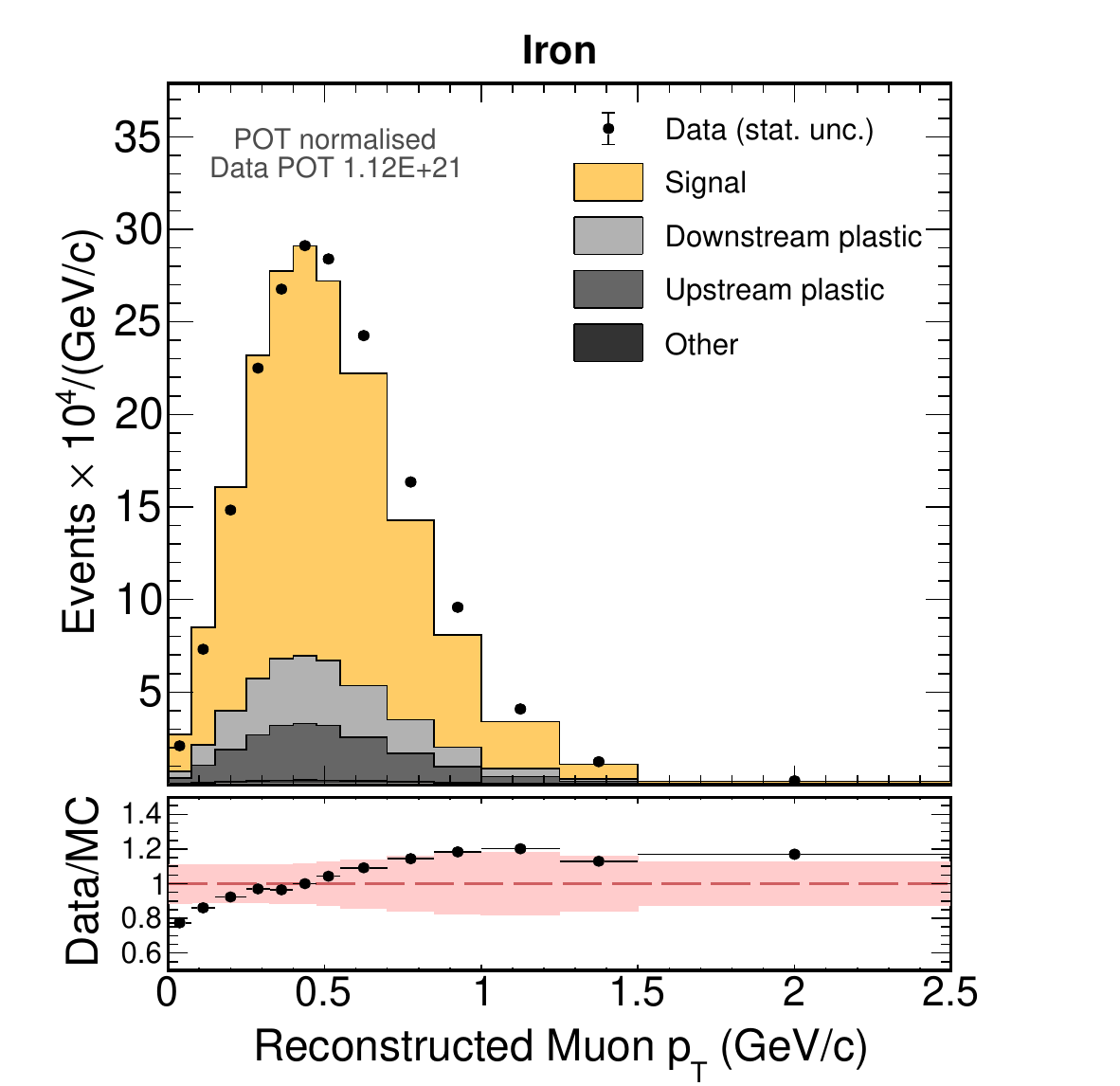}
  \end{subfigure}
  \caption{Selected events in C and Fe as a function of $p_{\mathrm{T}}$. For each target, the upper panel shows data with statistical uncertainties and stacked simulated signal and background; the lower panel shows the data-to-simulation ratio, with a band for simulation systematics, error bars for statistical uncertainties (statistical uncertainty is smaller than the marker size), and horizontal bars indicating bin widths.}
  \label{fig:pt_evselCandFe}
\end{figure}

\begin{table}[h]
  \caption{
    Upstream (US) and downstream (DS) per-material normalisation scale
    factors and their statistical and systematic uncertainties.
  }
  \label{tab:up_down_scales}
  \begin{ruledtabular}
  \begin{tabular}{lcccc}
    Material & Region & Scale factor & Stat. unc. & Syst. unc. \\
\hline
    C  & US & 1.029 & 0.005 & 0.133 \\
       & DS & 1.036 & 0.005 & 0.127 \\
    Fe & US & 1.065 & 0.004 & 0.133 \\
       & DS & 1.061 & 0.003 & 0.125 \\
    Pb & US & 1.069 & 0.003 & 0.137 \\
       & DS & 1.046 & 0.003 & 0.128 \\
  \end{tabular}
  \end{ruledtabular}
\end{table}

% DOWNSTREAM SIDEBAND TUNING FOR LEAD EXAMPLE
\begin{figure}[h]
  \centering
  \begin{subfigure}[b]{0.4\textwidth}
    \centering
    \includegraphics[width=\linewidth]{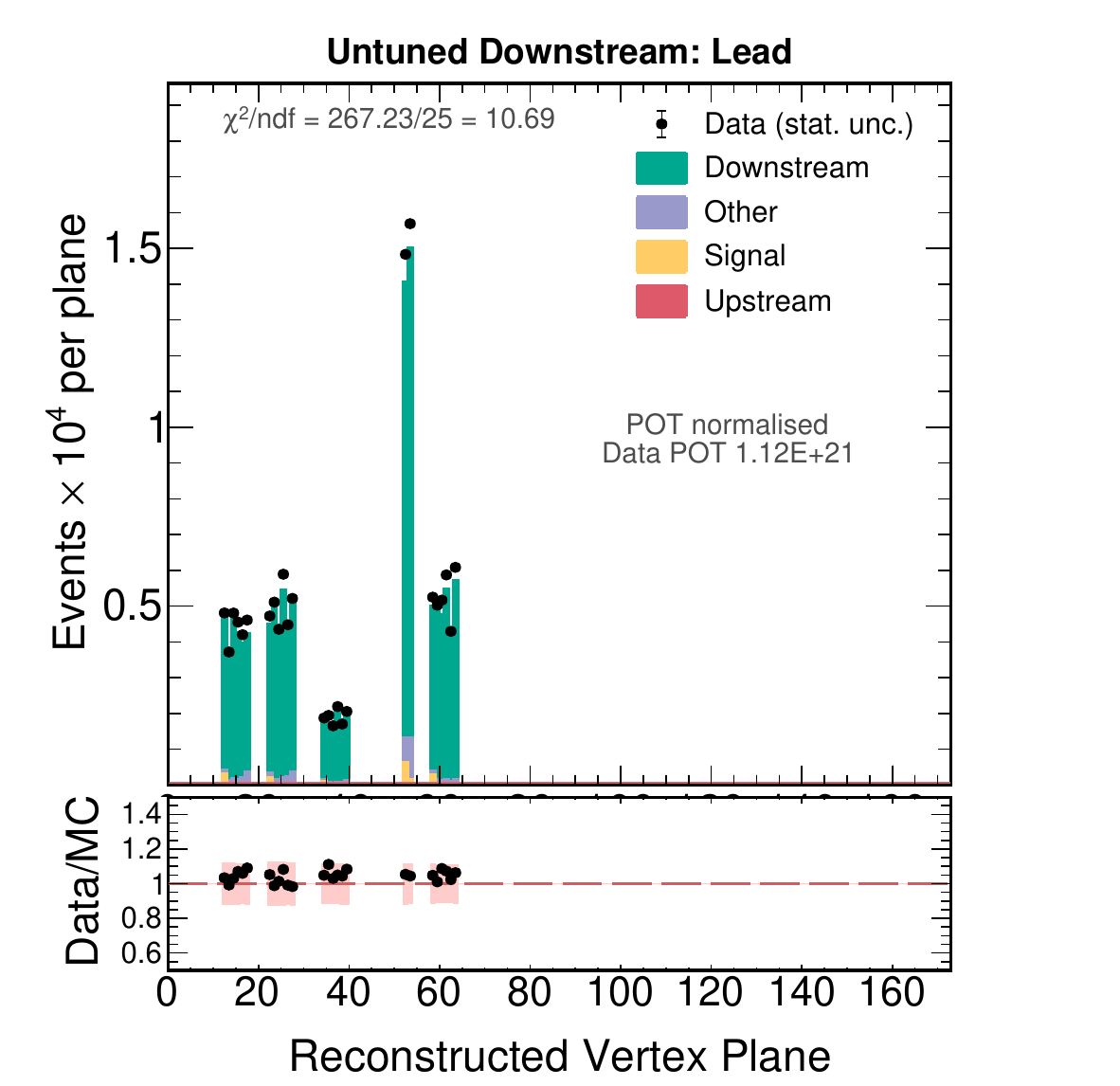}
  \end{subfigure}
  \hspace{-1.7em}
  \begin{subfigure}[b]{0.4\textwidth}
    \centering
    \includegraphics[width=\linewidth]{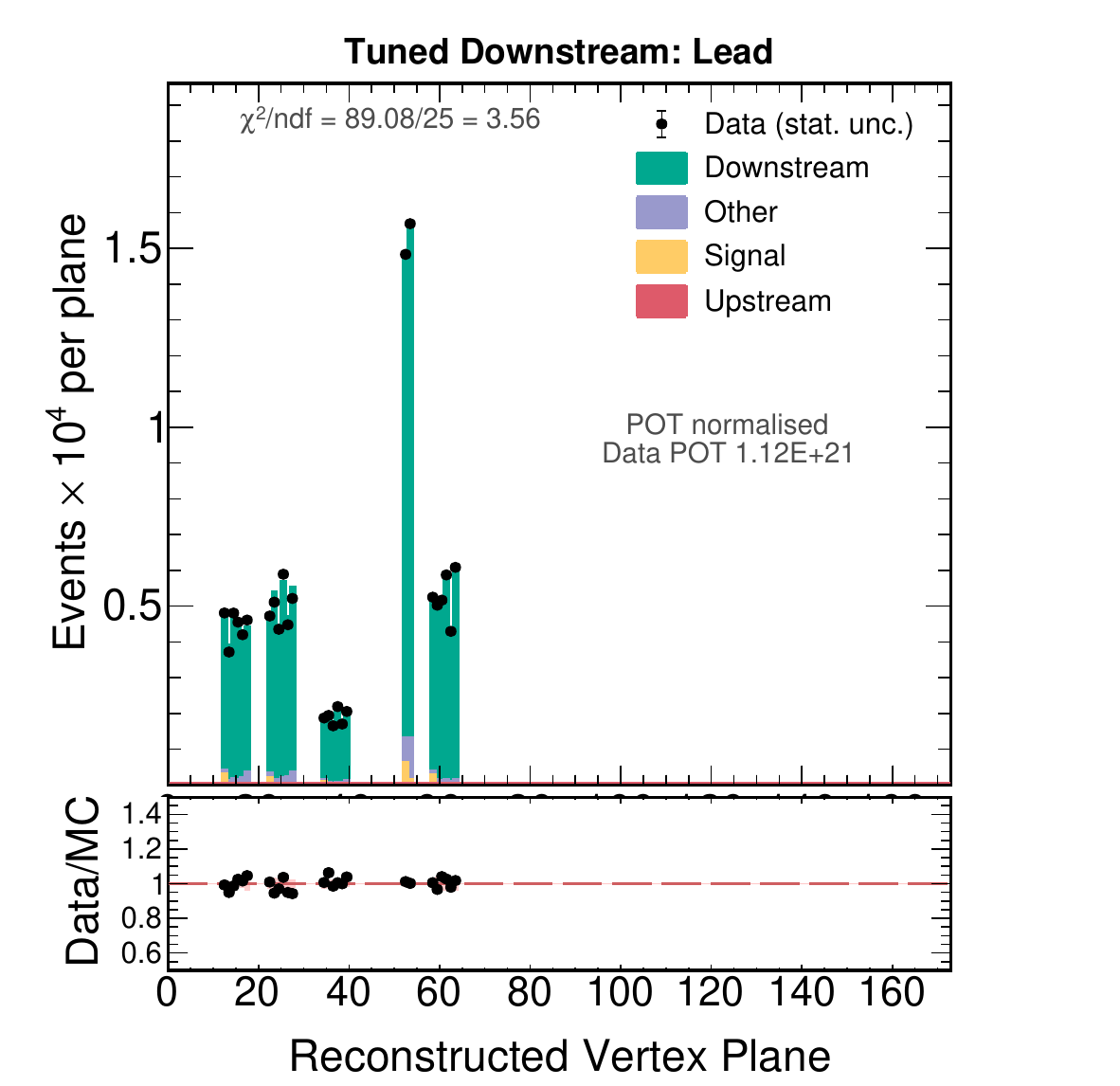}
  \end{subfigure}
  \begin{subfigure}[b]{0.4\textwidth}
    \centering
    \includegraphics[width=\linewidth]{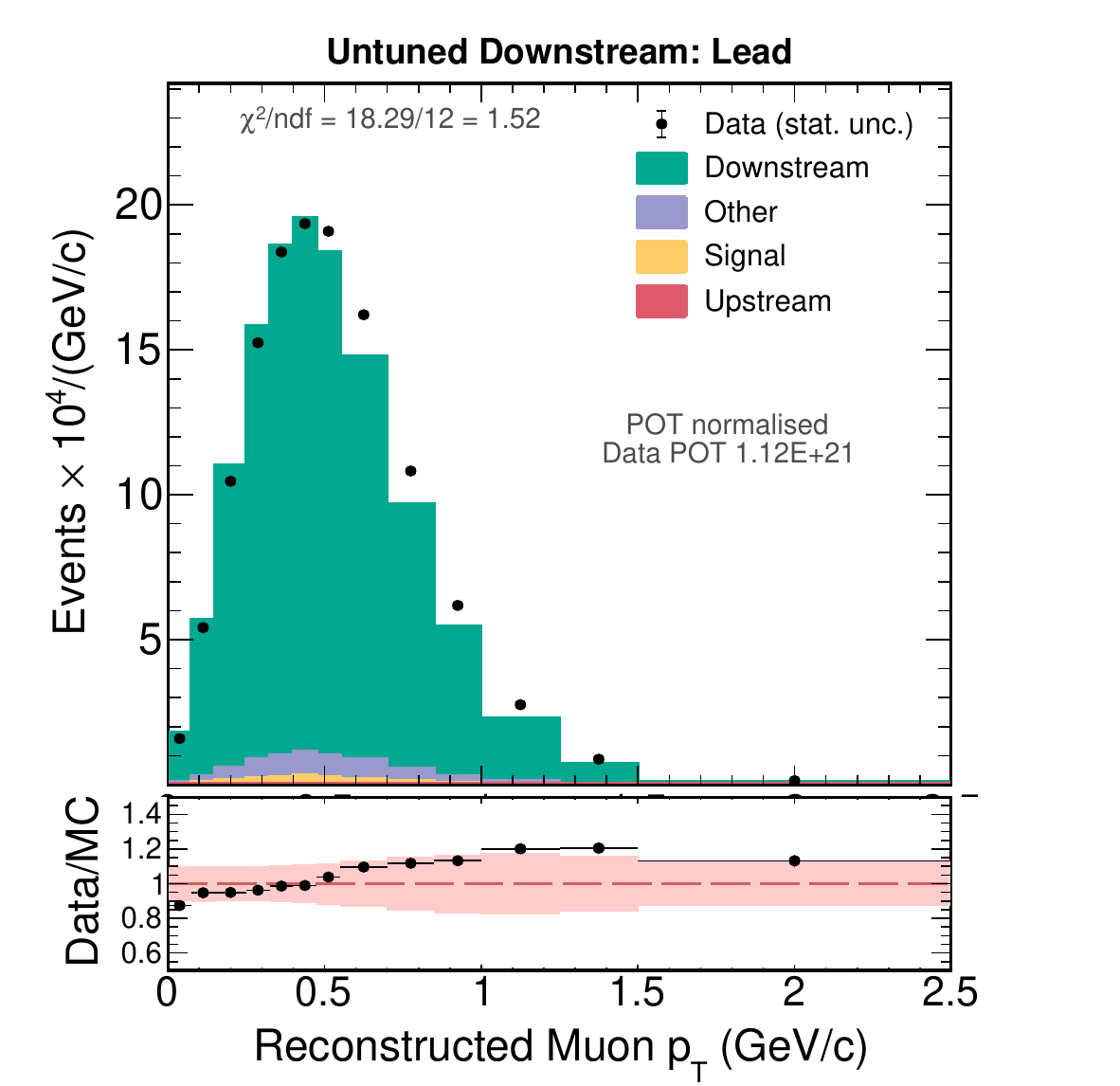}
  \end{subfigure}
  \hspace{-1.7em}
  \begin{subfigure}[b]{0.4\textwidth}
    \centering
    \includegraphics[width=\linewidth]{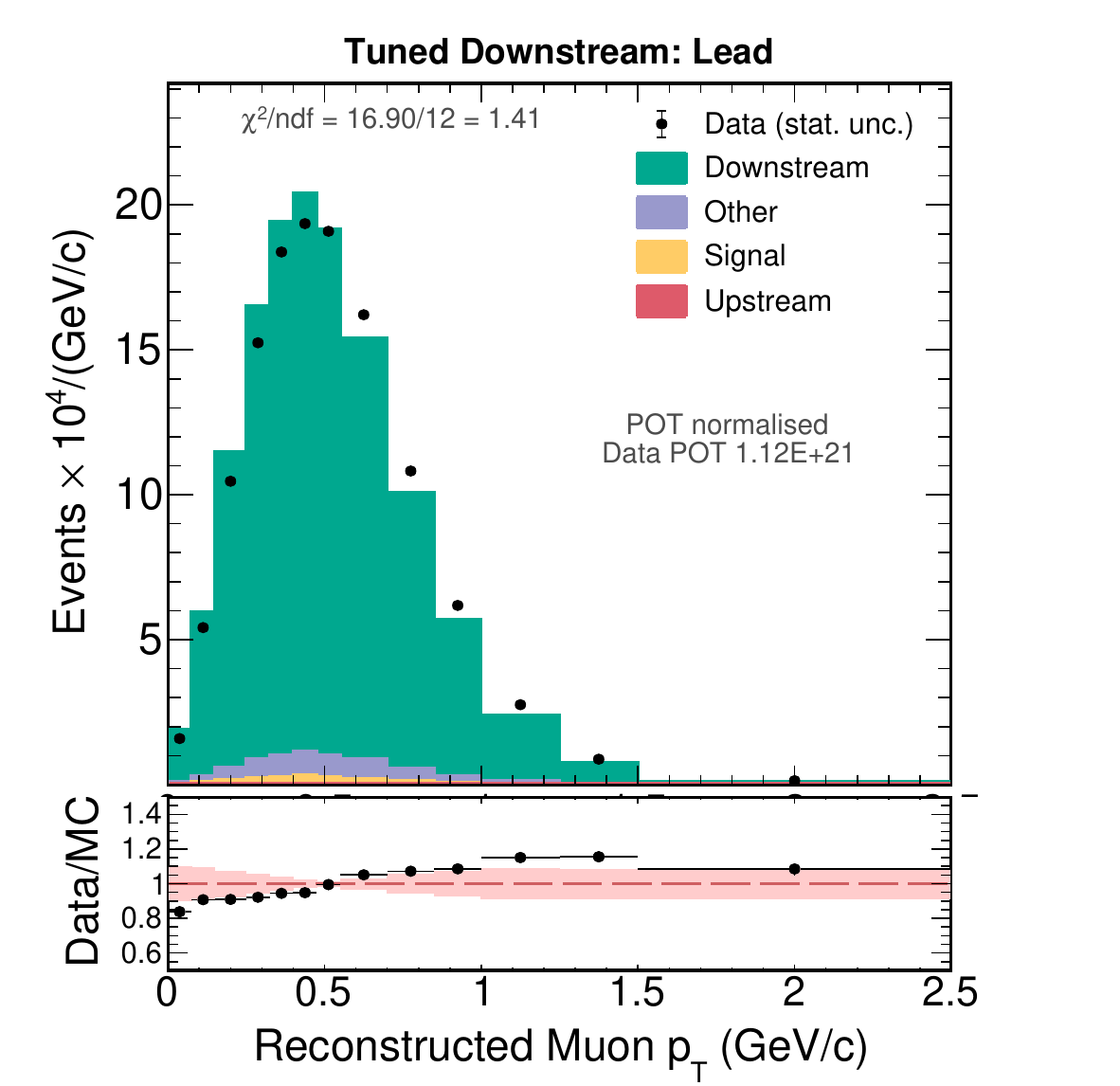}
  \end{subfigure}
    \caption{Downstream sideband fit for Pb. Top: Selected events in the downstream sideband of Pb as a function of plane number before (left) and after (right) applying the normalisation factors extracted from the $\chi^2$ fit. Bottom: Corresponding $p_{\mathrm{T}}$ distributions before (left) and after (right) applying these normalisation factors. In each panel, the lower sub-panel shows the data-to-simulation ratio; the band indicates simulation systematic uncertainties, error bars indicate statistical uncertainties (statistical uncertainty is smaller than the marker size), and horizontal bars indicate bin widths.}
  \label{fig:downstream_sideband_tuning}
\end{figure}

% FLUX
\begin{figure}[h]
  \centering
  \includegraphics[width=0.7\textwidth]{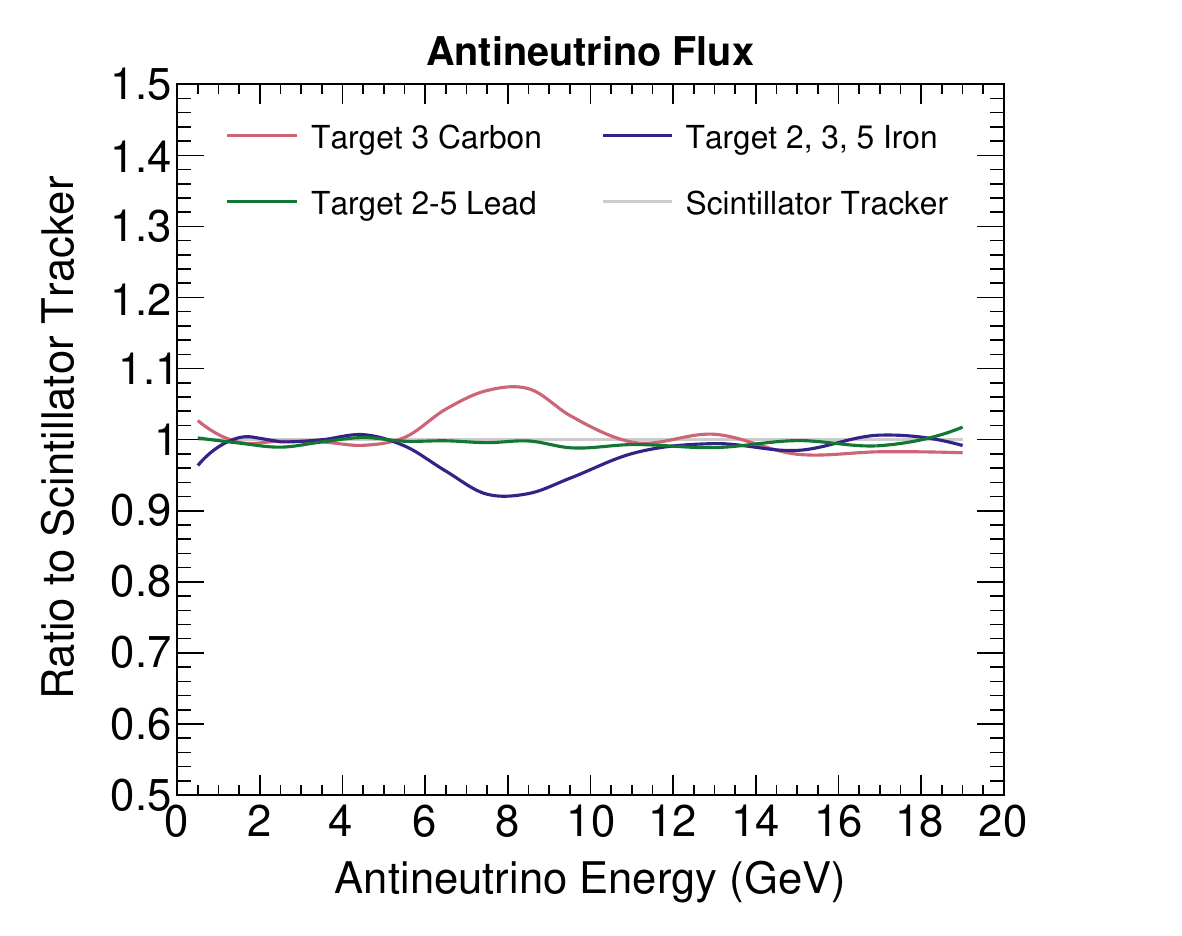}
  \caption{Ratio of the antineutrino flux in each target material relative to the hydrocarbon (scintillator) tracker.}
  \label{fig:flux_targets}
\end{figure}

% Low nu flux parametrisation
\begin{figure}[h]
  \centering
  \includegraphics[width=0.7\textwidth]{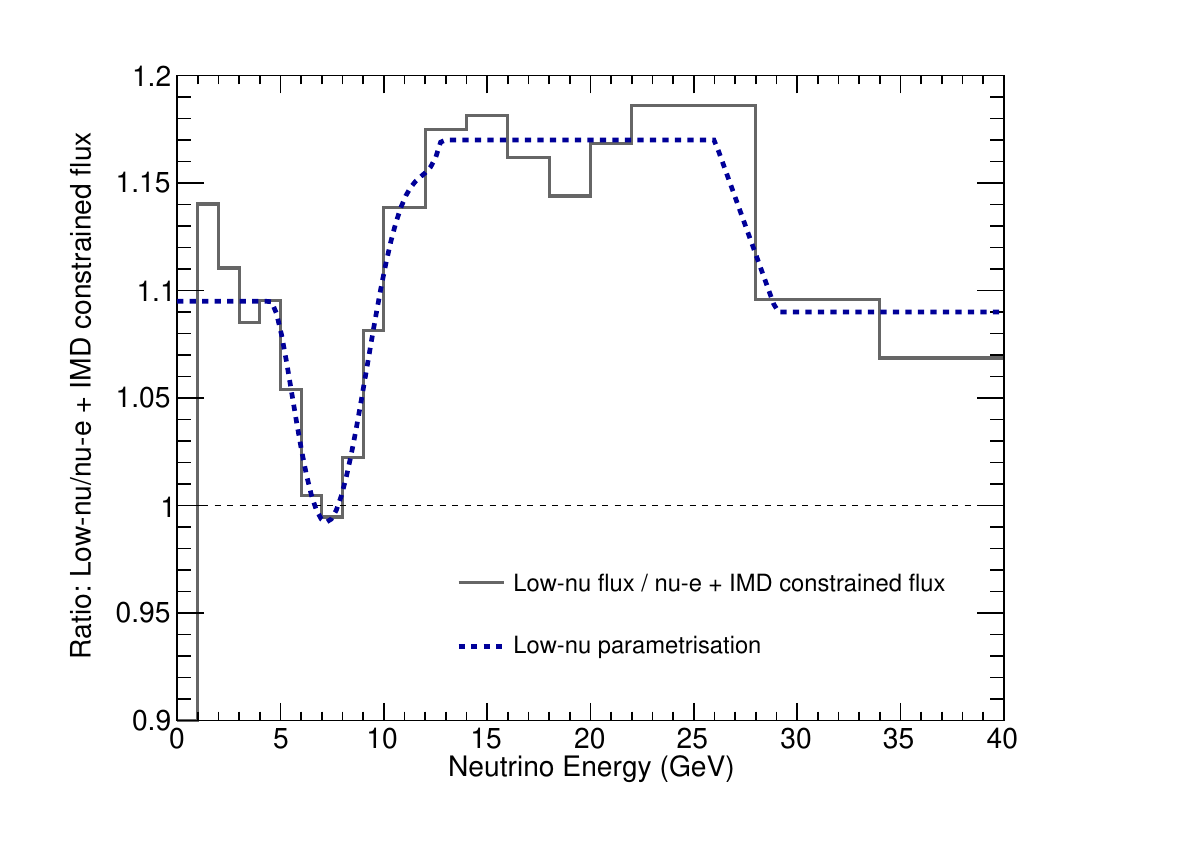}
  \caption{Ratio of the neutrino flux constrained by $\nu/\bar{\nu}$--$e$ scattering and inverse muon decay (IMD)~\cite{PhysRevD.100.092001,PhysRevD.104.092010}, with an additional constraint from a parametrised reweighting to MINERvA low-$\nu$ data above 7.5~GeV~\cite{Fine:2020knh,Srivastava:2023zyx}, to the flux constrained by $\nu/\bar{\nu}$--$e$ and IMD only, both in the neutrino-dominated beam. The low-$\nu$ method uses the approximate energy independence of the cross section at low hadronic energy transfer. The low-$\nu$ constraint increases the flux above 7.5~GeV, reaching up to 17\% in the 13--26~GeV range and anchoring the spectrum to higher-energy data from previous experiments.}
  \label{fig:low-nu_flux_param}
\end{figure}

% EFFICIENCIES
\begin{figure}[h]
  \centering
  \begin{subfigure}[b]{0.4\textwidth}
    \centering
    \includegraphics[width=\linewidth]{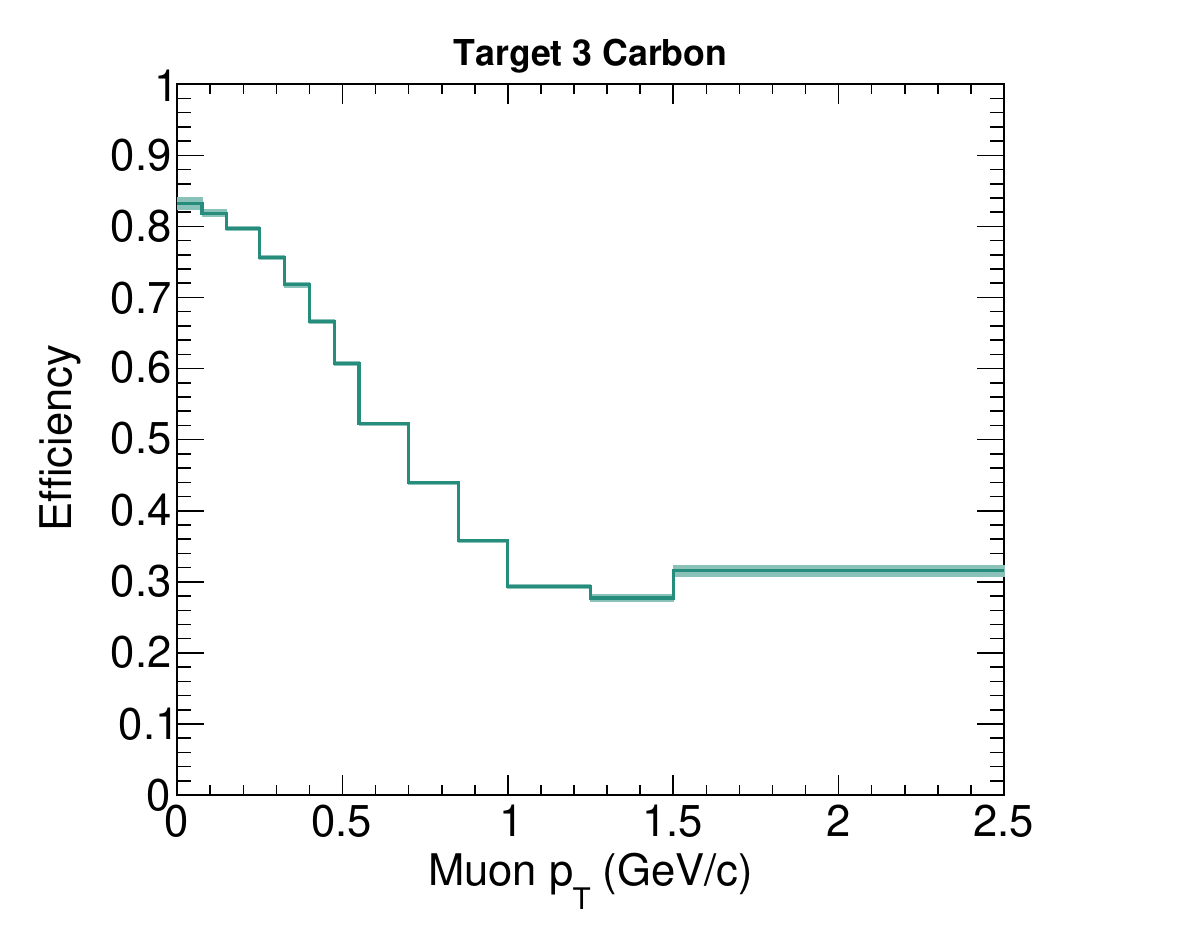}
  \end{subfigure}
  \hspace{-1.7em}
  \begin{subfigure}[b]{0.4\textwidth}
    \centering
    \includegraphics[width=\linewidth]{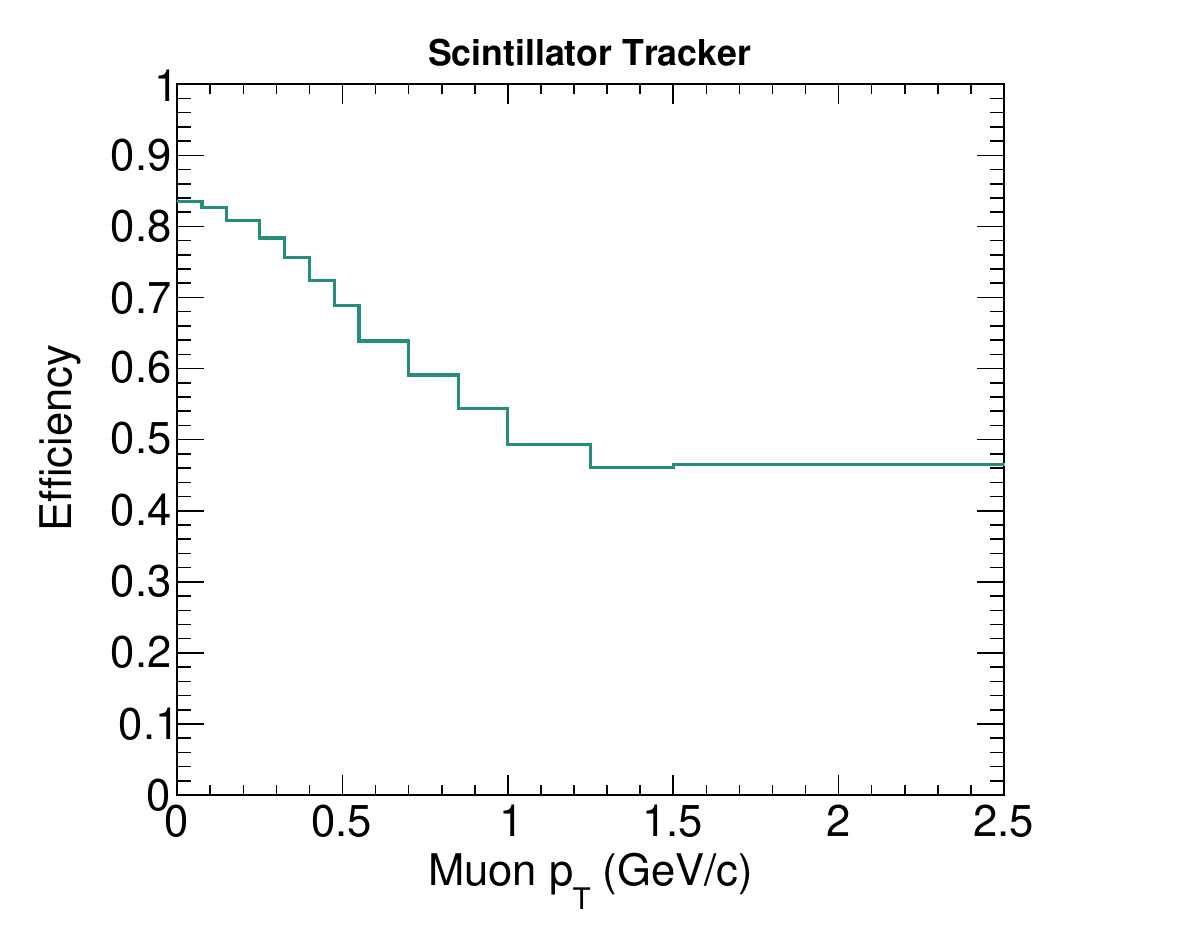}
  \end{subfigure}
  \begin{subfigure}[b]{0.4\textwidth}
    \centering
    \includegraphics[width=\linewidth]{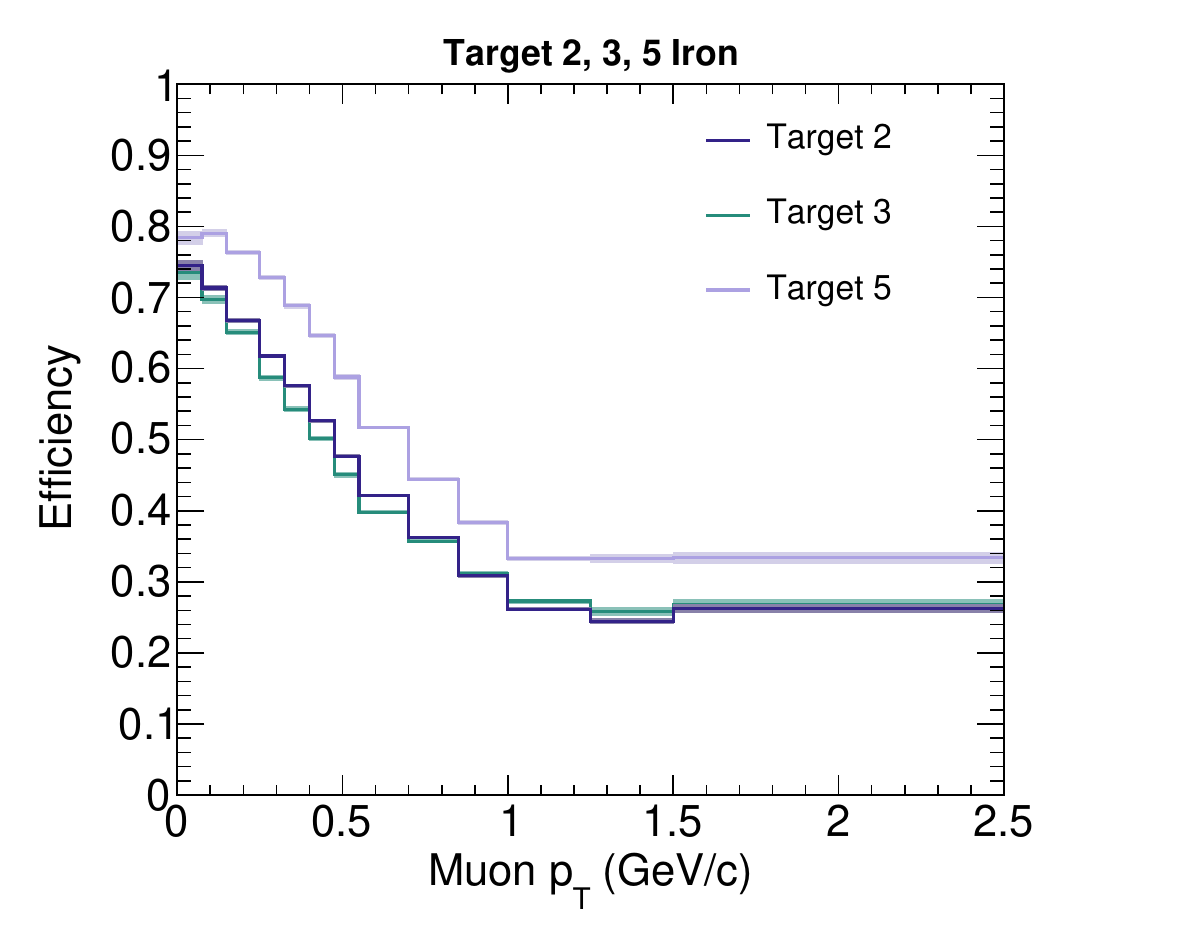}
  \end{subfigure}
  \hspace{-1.7em}
  \begin{subfigure}[b]{0.4\textwidth}
    \centering
    \includegraphics[width=\linewidth]{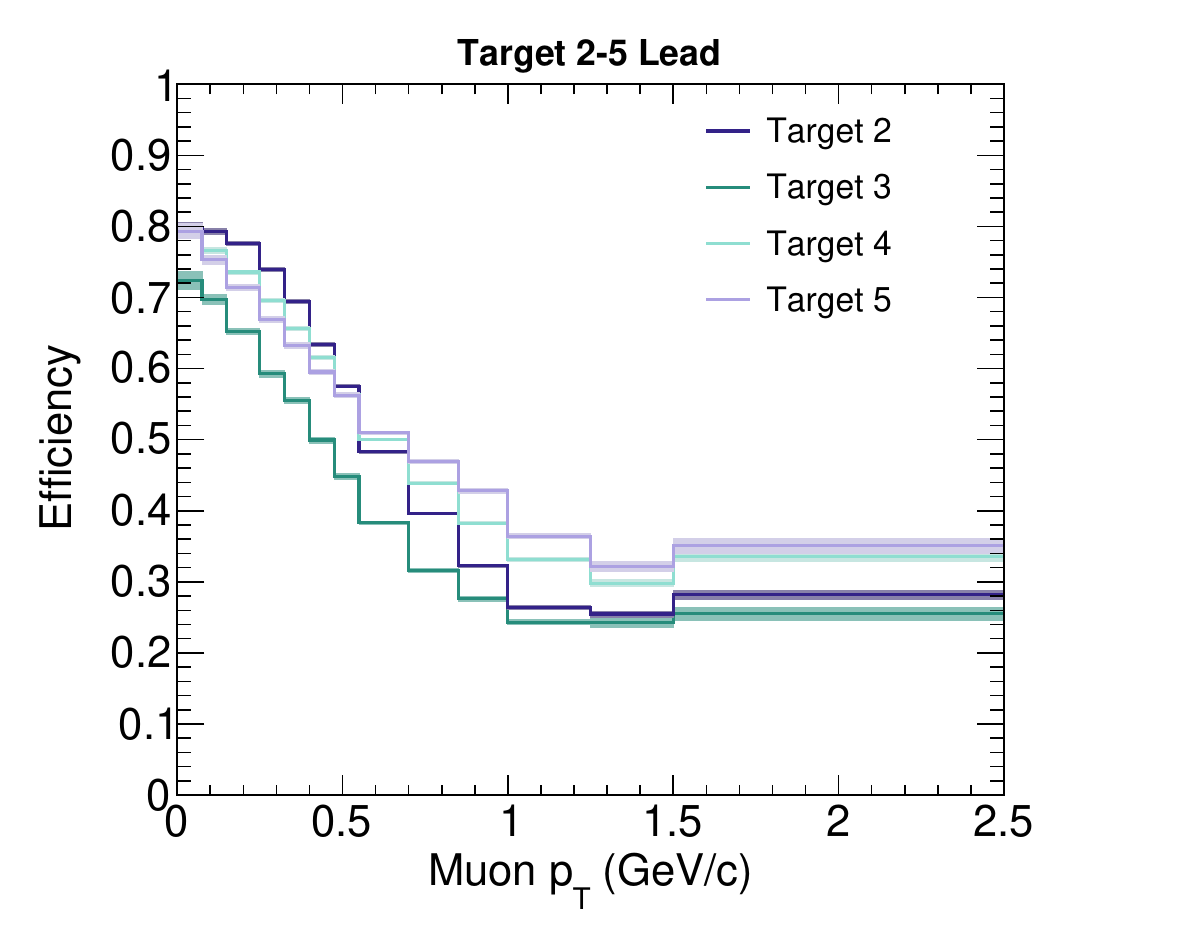}
  \end{subfigure}
  \caption{Per-target-material efficiency correction distributions as a function of $p_{\text{T}}$.}
  \label{fig:pt_eff}
\end{figure}

% CROSS-SECTIONS: Fractional Uncertainties
\begin{figure}[h]
  \centering
  \begin{subfigure}[b]{0.4\textwidth}
    \centering
    \includegraphics[width=\linewidth]{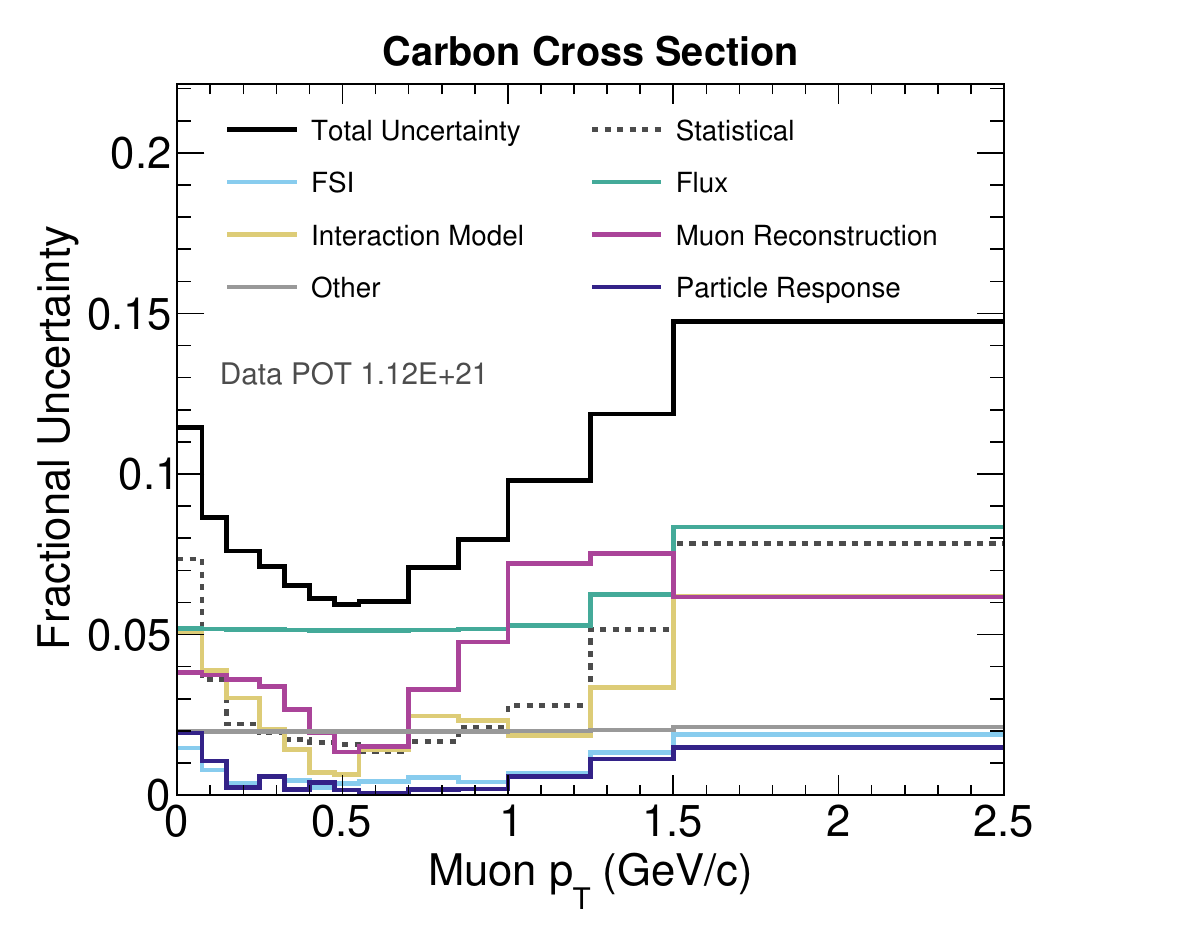}
  \end{subfigure}
  \hspace{-1.7em}
  \begin{subfigure}[b]{0.4\textwidth}
    \centering
    \includegraphics[width=\linewidth]{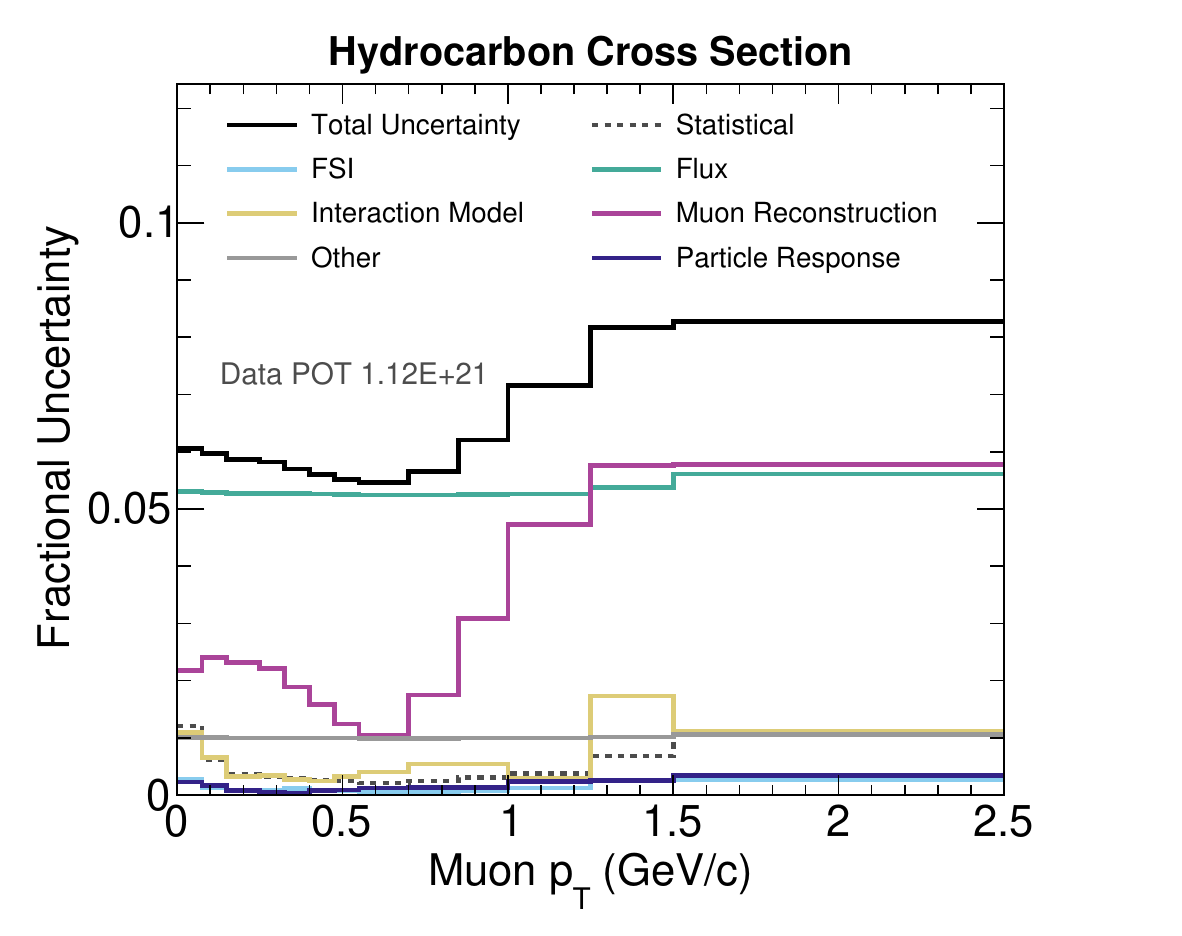}
  \end{subfigure}
  \begin{subfigure}[b]{0.4\textwidth}
    \centering
    \includegraphics[width=\linewidth]{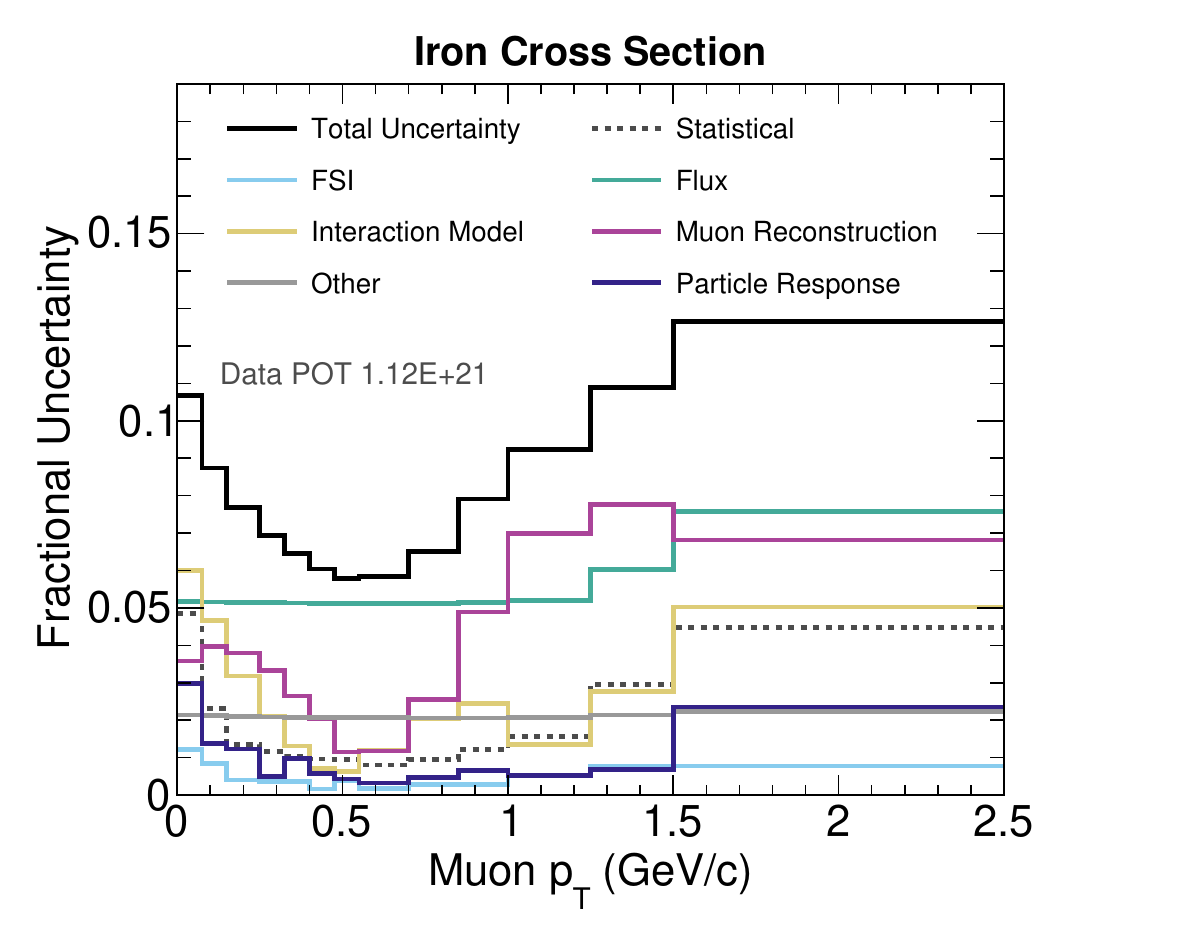}
  \end{subfigure}
  \hspace{-1.7em}
  \begin{subfigure}[b]{0.4\textwidth}
    \centering
    \includegraphics[width=\linewidth]{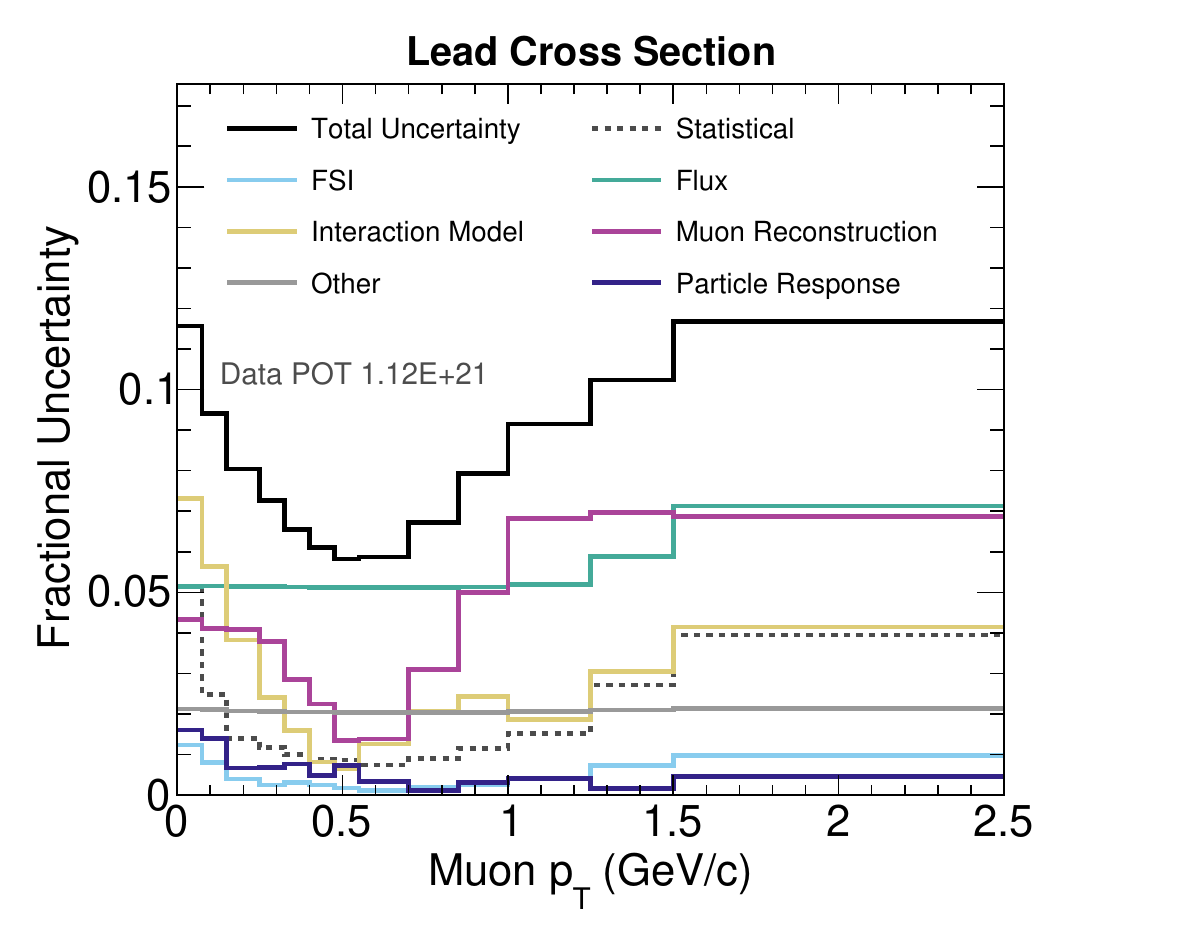}
  \end{subfigure}
  \caption{Breakdown of the fractional uncertainties on the data cross section as a function of $p_\text{T}$ for carbon, hydrocarbon, iron, and lead. The total uncertainty is dominated by uncertainties in flux and antimuon reconstruction. ``Other'' groups target mass uncertainties and a normalisation uncertainty to account for the data/MC difference in the ML vertex reconstruction performance.}
  \label{fig:pt_xsec_fracerr}
\end{figure}

% CROSS-SECTION RATIOS: Fractional Uncertainties
\begin{figure}[h]
  \centering
  \begin{subfigure}[b]{0.4\textwidth}
    \centering
    \includegraphics[width=\linewidth]{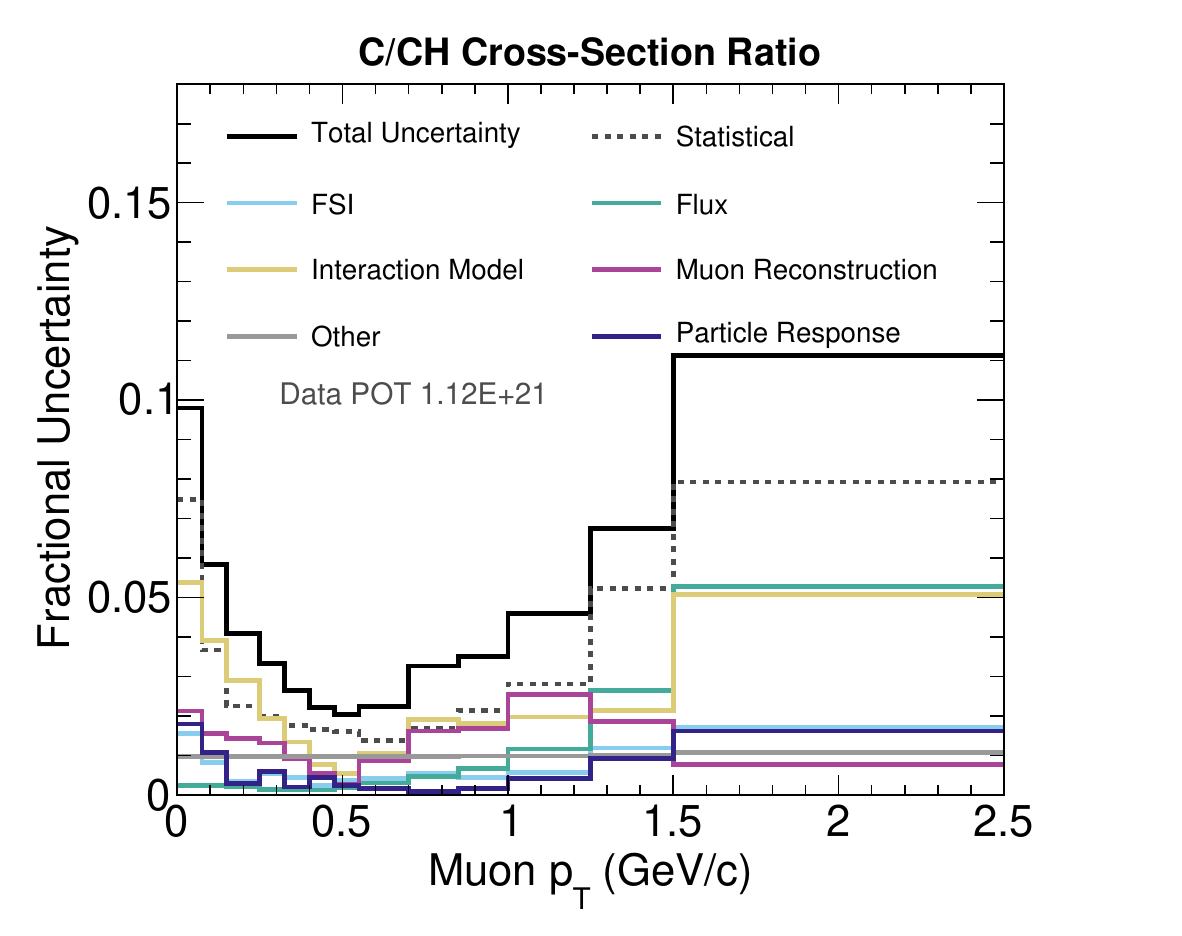}
  \end{subfigure}
  \hspace{-1.7em}
  \begin{subfigure}[b]{0.4\textwidth}
    \centering
    \includegraphics[width=\linewidth]{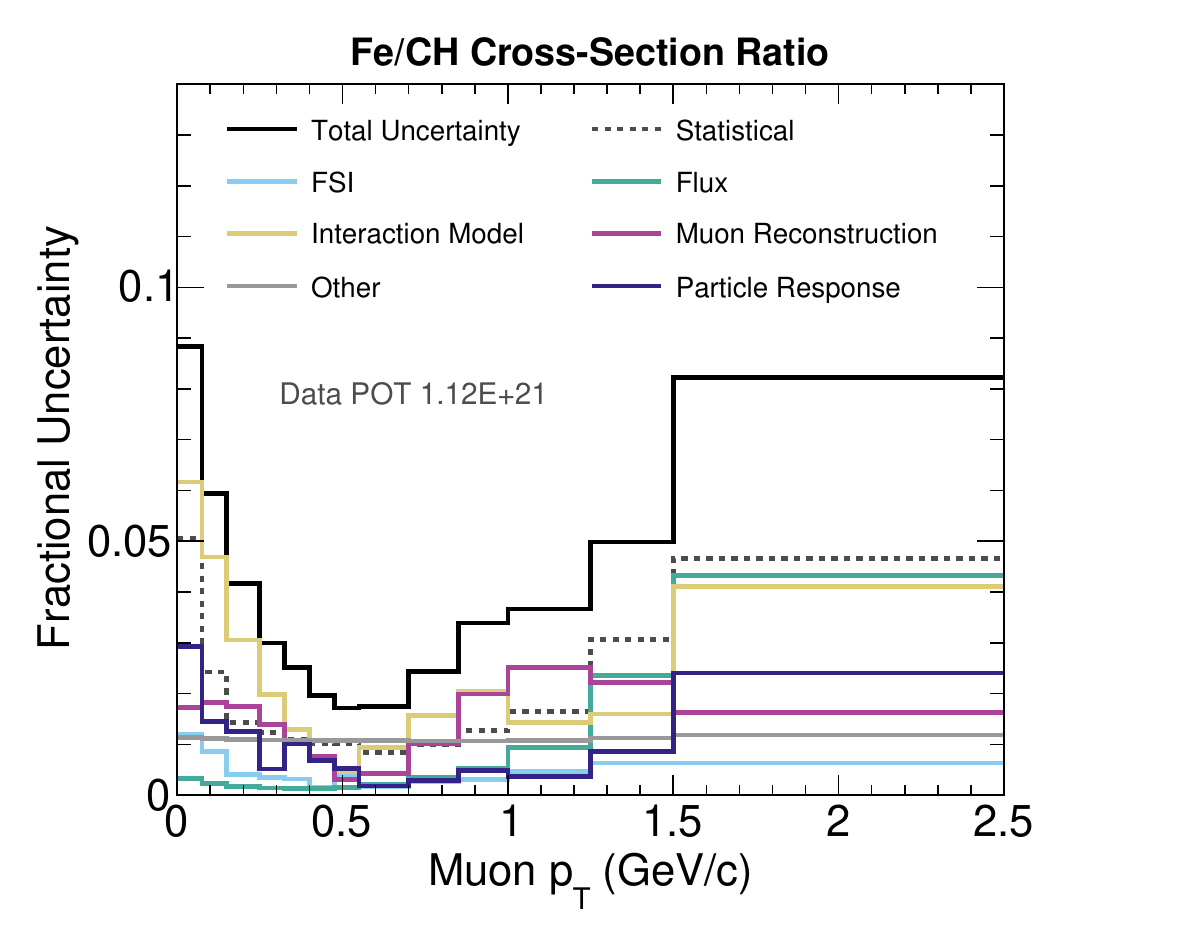}
  \end{subfigure}
  \begin{subfigure}[b]{0.4\textwidth}
    \centering
    \includegraphics[width=\linewidth]{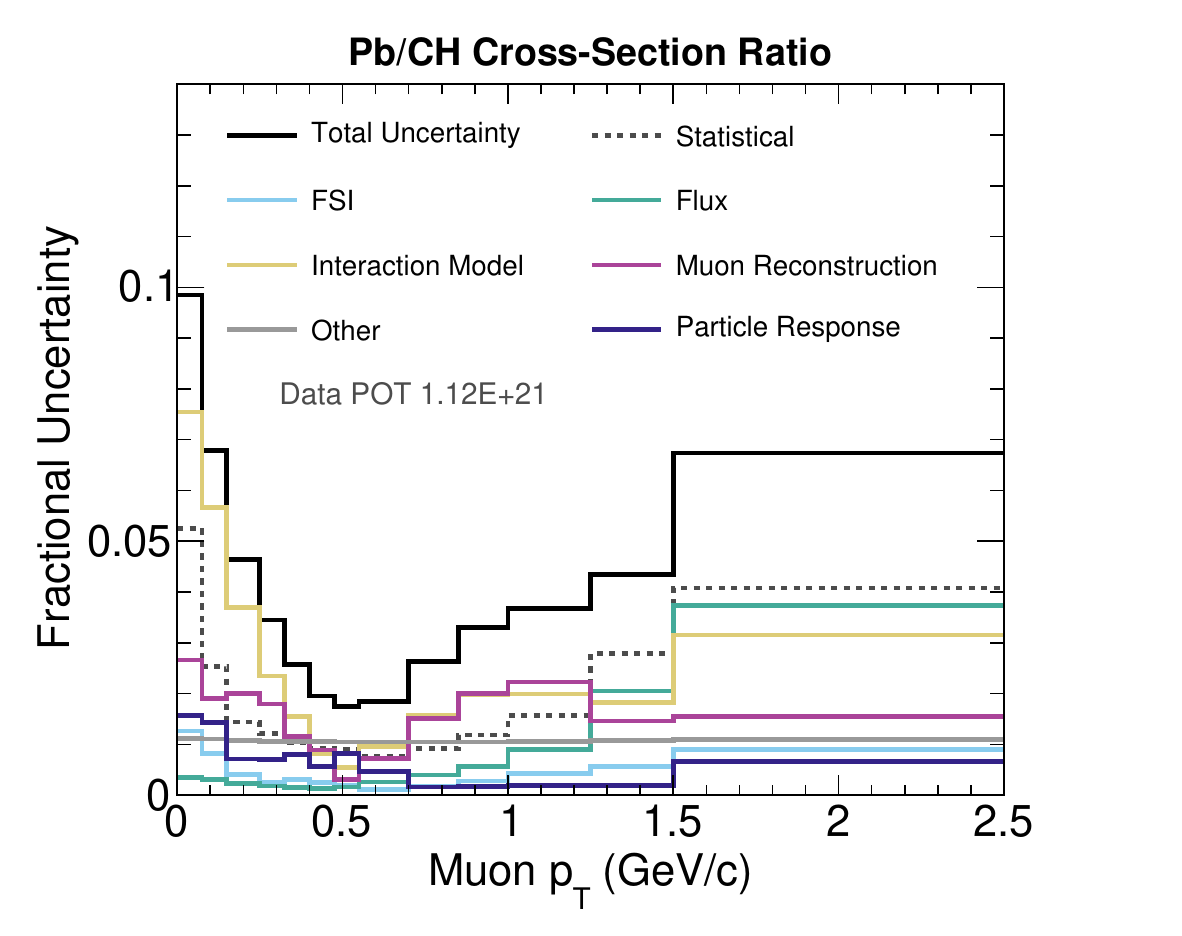}
  \end{subfigure}
  \caption{Breakdown of the fractional uncertainties on the data cross-section ratios to scintillator, C/CH, Fe/CH, and Pb/CH, as a function of $p_\text{T}$. Flux and detector-related uncertainties largely cancel. ``Other'' groups target mass uncertainties and a normalisation uncertainty to account for the data/MC difference in the ML vertex reconstruction performance.}
  \label{fig:pt_xsec_ratio_fracerr}
\end{figure}

\begin{table}[t]
\centering
\caption{Breakdown of interaction mode fractions for the predicted cross sections of all target materials, as predicted by the MINERvA Tune v4.3.0 based on GENIE v2.12.6.}
\begin{ruledtabular}
\begin{tabular}{ccccc}
\textbf{Interaction type} & \textbf{C} & \textbf{CH} & \textbf{Fe} & \textbf{Pb} \\
\hline
QE                                                                          & 17.3\% & 19.4\% & 16.6\% & 14.8\% \\
2p2h                                                                        & 12.0\% & 10.8\% & 12.5\% & 12.0\% \\
RES                                                                         & 35.3\% & 34.1\% & 36.8\% & 40.1\% \\
Soft DIS ($Q^2 < 1~(\mathrm{GeV}/c)^2$ or $W < 2~\mathrm{GeV}/c^2$)       & 27.3\% & 26.8\% & 27.3\% & 27.7\% \\
True DIS ($Q^2 > 1~(\mathrm{GeV}/c)^2$ and $W > 2~\mathrm{GeV}/c^2$)      &  5.0\% &  4.9\% &  4.8\% &  4.6\% \\
Other                                                                       &  3.1\% &  4.0\% &  2.0\% &  0.8\% \\
\end{tabular}
\end{ruledtabular}
\label{tab:interaction_fractions}
\end{table}

% CROSS-SECTION VALUES AND ERRS
\begin{table}[t]
\centering
\caption{Carbon differential cross section as a function of $p_{\text{T}}$ in units of $10^{-39} \text{ cm}^2/(\text{GeV}/c)/\text{nucleon}$.}
\begin{ruledtabular}
\begin{tabular}{cccc}
\textbf{Bin range} [GeV/$c$] & \textbf{Central value} & \textbf{Stat. error} & \textbf{Total error} \\
\hline
$[0.00, 0.075)$  & 1.0819 & 0.0795 & 0.1238 \\
$[0.075, 0.15)$  & 3.9703 & 0.1432 & 0.3430 \\
$[0.15, 0.25)$   & 8.6362 & 0.1916 & 0.6568 \\
$[0.25, 0.325)$  & 14.0135 & 0.2729 & 0.9976 \\
$[0.325, 0.40)$  & 17.8258 & 0.3092 & 1.1653 \\
$[0.40, 0.475)$  & 20.5654 & 0.3359 & 1.2590 \\
$[0.475, 0.55)$  & 23.1440 & 0.3671 & 1.3719 \\
$[0.55, 0.70)$   & 22.8710 & 0.3112 & 1.3813 \\
$[0.70, 0.85)$   & 17.6372 & 0.2955 & 1.2520 \\
$[0.85, 1.00)$   & 12.7239 & 0.2687 & 1.0135 \\
$[1.00, 1.25)$   & 5.9080 & 0.1644 & 0.5788 \\
$[1.25, 1.50)$   & 1.6573 & 0.0856 & 0.1967 \\
$[1.50, 2.50)$   & 0.2184 & 0.0171 & 0.0322 \\
\end{tabular}
\end{ruledtabular}
\label{tab:ptmu_xsec}
\end{table}

\begin{table}[t]
\centering
\caption{Hydrocarbon differential cross section as a function of $p_{\text{T}}$ in units of $10^{-39} \text{ cm}^2/(\text{GeV}/c)/\text{nucleon}$.}
\begin{ruledtabular}
\begin{tabular}{cccc}
\textbf{Bin range} [GeV/$c$] & \textbf{Central value} & \textbf{Stat. error} & \textbf{Total error} \\
\hline
$[0.00, 0.075)$  & 1.1661 & 0.0141 & 0.0702 \\
$[0.075, 0.15)$  & 4.1455 & 0.0255 & 0.2458 \\
$[0.15, 0.25)$   & 8.8613 & 0.0328 & 0.5168 \\
$[0.25, 0.325)$  & 13.9857 & 0.0461 & 0.8091 \\
$[0.325, 0.40)$  & 18.1435 & 0.0517 & 1.0272 \\
$[0.40, 0.475)$  & 21.2999 & 0.0559 & 1.1842 \\
$[0.475, 0.55)$  & 22.8834 & 0.0587 & 1.2524 \\
$[0.55, 0.70)$   & 22.2461 & 0.0469 & 1.2068 \\
$[0.70, 0.85)$   & 17.4553 & 0.0429 & 0.9804 \\
$[0.85, 1.00)$   & 11.5525 & 0.0351 & 0.7130 \\
$[1.00, 1.25)$   & 5.3908 & 0.0204 & 0.3845 \\
$[1.25, 1.50)$   & 1.6185 & 0.0111 & 0.1318 \\
$[1.50, 2.50)$   & 0.2076 & 0.0023 & 0.0171 \\
\end{tabular}
\end{ruledtabular}
\label{tab:ptmu_xsec_ch}
\end{table}

\begin{table}[t]
\centering
\caption{Iron differential cross section as a function of $p_{\text{T}}$ in units of $10^{-39} \text{ cm}^2/(\text{GeV}/c)/\text{nucleon}$.}
\begin{ruledtabular}
\begin{tabular}{cccc}
\textbf{Bin range} [GeV/$c$] & \textbf{Central value} & \textbf{Stat. error} & \textbf{Total error} \\
\hline
$[0.00, 0.075)$  & 0.8104 & 0.0393 & 0.0865 \\
$[0.075, 0.15)$  & 3.0965 & 0.0714 & 0.2707 \\
$[0.15, 0.25)$   & 7.3454 & 0.0998 & 0.5641 \\
$[0.25, 0.325)$  & 12.4850 & 0.1453 & 0.8657 \\
$[0.325, 0.40)$  & 16.3860 & 0.1699 & 1.0570 \\
$[0.40, 0.475)$  & 19.7177 & 0.1896 & 1.1916 \\
$[0.475, 0.55)$  & 21.9530 & 0.2093 & 1.2712 \\
$[0.55, 0.70)$   & 21.6789 & 0.1736 & 1.2667 \\
$[0.70, 0.85)$   & 17.5074 & 0.1665 & 1.1397 \\
$[0.85, 1.00)$   & 11.6881 & 0.1420 & 0.9242 \\
$[1.00, 1.25)$   & 5.4017 & 0.0849 & 0.4986 \\
$[1.25, 1.50)$   & 1.5044 & 0.0443 & 0.1638 \\
$[1.50, 2.50)$   & 0.2073 & 0.0093 & 0.0262 \\
\end{tabular}
\end{ruledtabular}
\label{tab:ptmu_xsec_fe}
\end{table}

\begin{table}[t]
\centering
\caption{Lead differential cross section as a function of $p_{\text{T}}$ in units of $10^{-39} \text{ cm}^2/(\text{GeV}/c)/\text{nucleon}$.}
\begin{ruledtabular}
\begin{tabular}{cccc}
\textbf{Bin range} [GeV/$c$] & \textbf{Central value} & \textbf{Stat. error} & \textbf{Total error} \\
\hline
$[0.00, 0.075)$  & 0.6644 & 0.0341 & 0.0769 \\
$[0.075, 0.15)$  & 2.4643 & 0.0610 & 0.2317 \\
$[0.15, 0.25)$   & 6.1880 & 0.0865 & 0.4976 \\
$[0.25, 0.325)$  & 10.7127 & 0.1252 & 0.7779 \\
$[0.325, 0.40)$  & 14.9061 & 0.1480 & 0.9757 \\
$[0.40, 0.475)$  & 18.7317 & 0.1671 & 1.1433 \\
$[0.475, 0.55)$  & 21.1347 & 0.1835 & 1.2310 \\
$[0.55, 0.70)$   & 20.6342 & 0.1527 & 1.2123 \\
$[0.70, 0.85)$   & 16.2199 & 0.1451 & 1.0898 \\
$[0.85, 1.00)$   & 10.6060 & 0.1221 & 0.8415 \\
$[1.00, 1.25)$   & 4.8483 & 0.0738 & 0.4437 \\
$[1.25, 1.50)$   & 1.5111 & 0.0410 & 0.1547 \\
$[1.50, 2.50)$   & 0.2105 & 0.0083 & 0.0246 \\
\end{tabular}
\end{ruledtabular}
\label{tab:ptmu_xsec_pb}
\end{table}

\begin{table}[t]
\centering
\caption{C/CH cross-section ratio as a function of $p_{\text{T}}$.}
\begin{ruledtabular}
\begin{tabular}{cccc}
\textbf{Bin range} [GeV/$c$] & \textbf{Central value} & \textbf{Stat. error} & \textbf{Total error} \\
\hline
$[0.00, 0.075)$  & 0.9324 & 0.0697 & 0.0914 \\
$[0.075, 0.15)$  & 0.9582 & 0.0352 & 0.0560 \\
$[0.15, 0.25)$   & 0.9763 & 0.0220 & 0.0400 \\
$[0.25, 0.325)$  & 1.0017 & 0.0199 & 0.0333 \\
$[0.325, 0.40)$  & 0.9794 & 0.0173 & 0.0259 \\
$[0.40, 0.475)$  & 0.9637 & 0.0160 & 0.0213 \\
$[0.475, 0.55)$  & 1.0064 & 0.0162 & 0.0205 \\
$[0.55, 0.70)$   & 1.0203 & 0.0141 & 0.0229 \\
$[0.70, 0.85)$   & 1.0014 & 0.0170 & 0.0327 \\
$[0.85, 1.00)$   & 1.0923 & 0.0234 & 0.0383 \\
$[1.00, 1.25)$   & 1.0917 & 0.0307 & 0.0502 \\
$[1.25, 1.50)$   & 1.0257 & 0.0536 & 0.0693 \\
$[1.50, 2.50)$   & 1.0778 & 0.0855 & 0.1200 \\
\end{tabular}
\end{ruledtabular}
\label{tab:pt_ratio_c_ch}
\end{table}

\begin{table}[t]
\centering
\caption{Fe/CH cross-section ratio as a function of $p_{\text{T}}$.}
\begin{ruledtabular}
\begin{tabular}{cccc}
\textbf{Bin range} [GeV/$c$] & \textbf{Central value} & \textbf{Stat. error} & \textbf{Total error} \\
\hline
$[0.00, 0.075)$  & 0.6944 & 0.0352 & 0.0614 \\
$[0.075, 0.15)$  & 0.7492 & 0.0181 & 0.0445 \\
$[0.15, 0.25)$   & 0.8299 & 0.0119 & 0.0346 \\
$[0.25, 0.325)$  & 0.8958 & 0.0110 & 0.0268 \\
$[0.325, 0.40)$  & 0.9091 & 0.0099 & 0.0228 \\
$[0.40, 0.475)$  & 0.9311 & 0.0095 & 0.0182 \\
$[0.475, 0.55)$  & 0.9681 & 0.0097 & 0.0166 \\
$[0.55, 0.70)$   & 0.9862 & 0.0083 & 0.0172 \\
$[0.70, 0.85)$   & 1.0138 & 0.0101 & 0.0246 \\
$[0.85, 1.00)$   & 1.0210 & 0.0130 & 0.0346 \\
$[1.00, 1.25)$   & 1.0072 & 0.0165 & 0.0369 \\
$[1.25, 1.50)$   & 0.9320 & 0.0286 & 0.0464 \\
$[1.50, 2.50)$   & 0.9754 & 0.0455 & 0.0802 \\
\end{tabular}
\end{ruledtabular}
\label{tab:pt_ratio_fe_ch}
\end{table}

\begin{table}[t]
\centering
\caption{Pb/CH cross-section ratio as a function of $p_{\text{T}}$.}
\begin{ruledtabular}
\begin{tabular}{cccc}
\textbf{Bin range} [GeV/$c$] & \textbf{Central value} & \textbf{Stat. error} & \textbf{Total error} \\
\hline
$[0.00, 0.075)$  & 0.5706 & 0.0300 & 0.0562 \\
$[0.075, 0.15)$  & 0.5952 & 0.0151 & 0.0404 \\
$[0.15, 0.25)$   & 0.6990 & 0.0101 & 0.0325 \\
$[0.25, 0.325)$  & 0.7667 & 0.0093 & 0.0264 \\
$[0.325, 0.40)$  & 0.8225 & 0.0085 & 0.0212 \\
$[0.40, 0.475)$  & 0.8806 & 0.0081 & 0.0172 \\
$[0.475, 0.55)$  & 0.9246 & 0.0083 & 0.0161 \\
$[0.55, 0.70)$   & 0.9284 & 0.0071 & 0.0171 \\
$[0.70, 0.85)$   & 0.9287 & 0.0086 & 0.0244 \\
$[0.85, 1.00)$   & 0.9172 & 0.0109 & 0.0302 \\
$[1.00, 1.25)$   & 0.8991 & 0.0140 & 0.0331 \\
$[1.25, 1.50)$   & 0.9350 & 0.0261 & 0.0407 \\
$[1.50, 2.50)$   & 1.0143 & 0.0414 & 0.0683 \\
\end{tabular}
\end{ruledtabular}
\label{tab:pt_ratio_pb_ch}
\end{table}

% CROSS-SECTION COVARIANCE MATRICES
\begin{sidewaystable*}[t]
\centering
\caption{Carbon cross-section ($p_{\text{T}}$) covariance and correlation matrices. The underflow bin (bin 0) is omitted because it is zero.}
\scriptsize
\begin{ruledtabular}
\begin{tabular}{ccccccccccccccc}
\multicolumn{15}{c}{\textbf{Covariance matrix}} \\[0.3em]
Bin & $1$ & $2$ & $3$ & $4$ & $5$ & $6$ & $7$ & $8$ & $9$ & $10$ & $11$ & $12$ & $13$ & $14$ \\
\hline
$1$ & 1.533E-80 & 3.052E-80 & 5.242E-80 & 7.986E-80 & 8.690E-80 & 9.015E-80 & 8.346E-80 & 5.344E-80 & 1.944E-80 & 7.342E-81 & -1.469E-81 & 1.750E-81 & 6.439E-82 & 5.602E-83 \\
$2$ & 3.052E-80 & 1.176E-79 & 1.933E-79 & 2.776E-79 & 3.124E-79 & 3.244E-79 & 3.069E-79 & 2.054E-79 & 8.396E-80 & 3.569E-80 & -4.510E-82 & 6.216E-81 & 2.144E-81 & 1.701E-82 \\
$3$ & 5.242E-80 & 1.933E-79 & 4.314E-79 & 5.916E-79 & 6.487E-79 & 6.808E-79 & 6.680E-79 & 4.668E-79 & 2.143E-79 & 1.008E-79 & 9.622E-81 & 1.333E-80 & 4.304E-81 & 3.385E-82 \\
$4$ & 7.986E-80 & 2.776E-79 & 5.916E-79 & 9.951E-79 & 1.071E-78 & 1.072E-78 & 1.090E-78 & 8.138E-79 & 4.169E-79 & 2.182E-79 & 4.002E-80 & 2.225E-80 & 5.838E-81 & 4.200E-82 \\
$5$ & 8.690E-80 & 3.124E-79 & 6.487E-79 & 1.071E-78 & 1.358E-78 & 1.352E-78 & 1.365E-78 & 1.098E-78 & 6.587E-79 & 3.921E-79 & 1.230E-79 & 4.100E-80 & 7.963E-81 & 5.798E-82 \\
$6$ & 9.015E-80 & 3.244E-79 & 6.808E-79 & 1.072E-78 & 1.352E-78 & 1.585E-78 & 1.606E-78 & 1.339E-78 & 8.979E-79 & 5.817E-79 & 2.208E-79 & 6.376E-80 & 1.043E-80 & 6.104E-82 \\
$7$ & 8.346E-80 & 3.069E-79 & 6.680E-79 & 1.090E-78 & 1.365E-78 & 1.606E-78 & 1.882E-78 & 1.644E-78 & 1.163E-78 & 8.080E-79 & 3.388E-79 & 8.724E-80 & 1.197E-80 & 6.032E-82 \\
$8$ & 5.344E-80 & 2.054E-79 & 4.668E-79 & 8.138E-79 & 1.098E-78 & 1.339E-78 & 1.644E-78 & 1.908E-78 & 1.545E-78 & 1.134E-78 & 5.471E-79 & 1.315E-79 & 1.446E-80 & 4.335E-82 \\
$9$ & 1.944E-80 & 8.396E-80 & 2.143E-79 & 4.169E-79 & 6.587E-79 & 8.979E-79 & 1.163E-78 & 1.545E-78 & 1.568E-78 & 1.165E-78 & 5.840E-79 & 1.361E-79 & 1.347E-80 & 3.610E-82 \\
$10$ & 7.342E-81 & 3.569E-80 & 1.008E-79 & 2.182E-79 & 3.921E-79 & 5.817E-79 & 8.080E-79 & 1.134E-78 & 1.165E-78 & 1.027E-78 & 5.214E-79 & 1.263E-79 & 1.342E-80 & 3.943E-82 \\
$11$ & -1.469E-81 & -4.510E-82 & 9.622E-81 & 4.002E-80 & 1.230E-79 & 2.208E-79 & 3.388E-79 & 5.471E-79 & 5.840E-79 & 5.214E-79 & 3.350E-79 & 8.863E-80 & 9.551E-81 & 3.084E-82 \\
$12$ & 1.750E-81 & 6.216E-81 & 1.333E-80 & 2.225E-80 & 4.100E-80 & 6.376E-80 & 8.724E-80 & 1.315E-79 & 1.361E-79 & 1.263E-79 & 8.863E-80 & 3.867E-80 & 4.605E-81 & 1.756E-82 \\
$13$ & 6.439E-82 & 2.144E-81 & 4.304E-81 & 5.838E-81 & 7.963E-81 & 1.043E-80 & 1.197E-80 & 1.446E-80 & 1.347E-80 & 1.342E-80 & 9.551E-81 & 4.605E-81 & 1.039E-81 & 4.841E-83 \\
$14$ & 5.602E-83 & 1.701E-82 & 3.385E-82 & 4.200E-82 & 5.798E-82 & 6.104E-82 & 6.032E-82 & 4.335E-82 & 3.610E-82 & 3.943E-82 & 3.084E-82 & 1.756E-82 & 4.841E-83 & 1.913E-83 \\
\end{tabular}
\end{ruledtabular}
\vspace{1em}
\begin{ruledtabular}
\begin{tabular}{ccccccccccccccc}
\multicolumn{15}{c}{\textbf{Correlation matrix}} \\[0.3em]
Bin & $1$ & $2$ & $3$ & $4$ & $5$ & $6$ & $7$ & $8$ & $9$ & $10$ & $11$ & $12$ & $13$ & $14$ \\
\hline
$1$ & 1.000E+00 & 7.187E-01 & 6.446E-01 & 6.466E-01 & 6.023E-01 & 5.783E-01 & 4.913E-01 & 3.125E-01 & 1.254E-01 & 5.851E-02 & -2.050E-02 & 7.188E-02 & 1.613E-01 & 1.034E-01 \\
$2$ & 7.187E-01 & 1.000E+00 & 8.581E-01 & 8.112E-01 & 7.815E-01 & 7.512E-01 & 6.522E-01 & 4.335E-01 & 1.955E-01 & 1.027E-01 & -2.272E-03 & 9.215E-02 & 1.939E-01 & 1.134E-01 \\
$3$ & 6.446E-01 & 8.581E-01 & 1.000E+00 & 9.029E-01 & 8.475E-01 & 8.232E-01 & 7.413E-01 & 5.144E-01 & 2.606E-01 & 1.514E-01 & 2.531E-02 & 1.032E-01 & 2.033E-01 & 1.178E-01 \\
$4$ & 6.466E-01 & 8.112E-01 & 9.029E-01 & 1.000E+00 & 9.214E-01 & 8.537E-01 & 7.964E-01 & 5.906E-01 & 3.338E-01 & 2.159E-01 & 6.931E-02 & 1.134E-01 & 1.816E-01 & 9.626E-02 \\
$5$ & 6.023E-01 & 7.815E-01 & 8.475E-01 & 9.214E-01 & 1.000E+00 & 9.215E-01 & 8.536E-01 & 6.822E-01 & 4.515E-01 & 3.320E-01 & 1.824E-01 & 1.789E-01 & 2.120E-01 & 1.137E-01 \\
$6$ & 5.783E-01 & 7.512E-01 & 8.232E-01 & 8.537E-01 & 9.215E-01 & 1.000E+00 & 9.297E-01 & 7.702E-01 & 5.696E-01 & 4.559E-01 & 3.030E-01 & 2.575E-01 & 2.570E-01 & 1.108E-01 \\
$7$ & 4.913E-01 & 6.522E-01 & 7.413E-01 & 7.964E-01 & 8.536E-01 & 9.297E-01 & 1.000E+00 & 8.676E-01 & 6.773E-01 & 5.811E-01 & 4.267E-01 & 3.234E-01 & 2.706E-01 & 1.005E-01 \\
$8$ & 3.125E-01 & 4.335E-01 & 5.144E-01 & 5.906E-01 & 6.822E-01 & 7.702E-01 & 8.676E-01 & 1.000E+00 & 8.933E-01 & 8.099E-01 & 6.844E-01 & 4.842E-01 & 3.247E-01 & 7.174E-02 \\
$9$ & 1.254E-01 & 1.955E-01 & 2.606E-01 & 3.338E-01 & 4.515E-01 & 5.696E-01 & 6.773E-01 & 8.933E-01 & 1.000E+00 & 9.180E-01 & 8.059E-01 & 5.530E-01 & 3.338E-01 & 6.591E-02 \\
$10$ & 5.851E-02 & 1.027E-01 & 1.514E-01 & 2.159E-01 & 3.320E-01 & 4.559E-01 & 5.811E-01 & 8.099E-01 & 9.180E-01 & 1.000E+00 & 8.890E-01 & 6.338E-01 & 4.109E-01 & 8.894E-02 \\
$11$ & -2.050E-02 & -2.272E-03 & 2.531E-02 & 6.931E-02 & 1.824E-01 & 3.030E-01 & 4.267E-01 & 6.844E-01 & 8.059E-01 & 8.890E-01 & 1.000E+00 & 7.787E-01 & 5.119E-01 & 1.218E-01 \\
$12$ & 7.188E-02 & 9.215E-02 & 1.032E-01 & 1.134E-01 & 1.789E-01 & 2.575E-01 & 3.234E-01 & 4.842E-01 & 5.530E-01 & 6.338E-01 & 7.787E-01 & 1.000E+00 & 7.265E-01 & 2.041E-01 \\
$13$ & 1.613E-01 & 1.939E-01 & 2.033E-01 & 1.816E-01 & 2.120E-01 & 2.570E-01 & 2.706E-01 & 3.247E-01 & 3.338E-01 & 4.109E-01 & 5.119E-01 & 7.265E-01 & 1.000E+00 & 3.433E-01 \\
$14$ & 1.034E-01 & 1.134E-01 & 1.178E-01 & 9.626E-02 & 1.137E-01 & 1.108E-01 & 1.005E-01 & 7.174E-02 & 6.591E-02 & 8.894E-02 & 1.218E-01 & 2.041E-01 & 3.433E-01 & 1.000E+00 \\
\end{tabular}
\end{ruledtabular}
\normalsize
\label{tab:ptmu_cov_corr_carbon}
\end{sidewaystable*}

\begin{sidewaystable*}[t]
\centering
\caption{Hydrocarbon cross-section ($p_{\text{T}}$) covariance and correlation matrices. The underflow bin (bin 0) is omitted because it is zero.}
\scriptsize
\begin{ruledtabular}
\begin{tabular}{ccccccccccccccc}
\multicolumn{15}{c}{\textbf{Covariance matrix}} \\[0.3em]
Bin & $1$ & $2$ & $3$ & $4$ & $5$ & $6$ & $7$ & $8$ & $9$ & $10$ & $11$ & $12$ & $13$ & $14$ \\
\hline
$1$ & 4.932E-81 & 1.669E-80 & 3.479E-80 & 5.431E-80 & 6.898E-80 & 7.901E-80 & 8.226E-80 & 7.519E-80 & 5.434E-80 & 3.216E-80 & 1.309E-80 & 3.635E-81 & 4.997E-82 & 2.172E-83 \\
$2$ & 1.669E-80 & 6.042E-80 & 1.255E-79 & 1.955E-79 & 2.475E-79 & 2.828E-79 & 2.931E-79 & 2.658E-79 & 1.898E-79 & 1.100E-79 & 4.334E-80 & 1.168E-80 & 1.627E-81 & 7.145E-83 \\
$3$ & 3.479E-80 & 1.255E-79 & 2.671E-79 & 4.155E-79 & 5.258E-79 & 6.018E-79 & 6.248E-79 & 5.684E-79 & 4.076E-79 & 2.380E-79 & 9.457E-80 & 2.549E-80 & 3.561E-81 & 1.573E-82 \\
$4$ & 5.431E-80 & 1.955E-79 & 4.155E-79 & 6.547E-79 & 8.271E-79 & 9.457E-79 & 9.844E-79 & 8.983E-79 & 6.467E-79 & 3.813E-79 & 1.533E-79 & 4.154E-80 & 5.790E-81 & 2.544E-82 \\
$5$ & 6.898E-80 & 2.475E-79 & 5.258E-79 & 8.271E-79 & 1.055E-78 & 1.210E-78 & 1.266E-78 & 1.169E-78 & 8.565E-79 & 5.174E-79 & 2.151E-79 & 5.948E-80 & 8.203E-81 & 3.552E-82 \\
$6$ & 7.901E-80 & 2.828E-79 & 6.018E-79 & 9.457E-79 & 1.210E-78 & 1.402E-78 & 1.474E-78 & 1.377E-78 & 1.027E-78 & 6.354E-79 & 2.725E-79 & 7.687E-80 & 1.047E-80 & 4.445E-82 \\
$7$ & 8.226E-80 & 2.931E-79 & 6.248E-79 & 9.844E-79 & 1.266E-78 & 1.474E-78 & 1.568E-78 & 1.487E-78 & 1.133E-78 & 7.219E-79 & 3.208E-79 & 9.230E-80 & 1.245E-80 & 5.194E-82 \\
$8$ & 7.519E-80 & 2.658E-79 & 5.684E-79 & 8.983E-79 & 1.169E-78 & 1.377E-78 & 1.487E-78 & 1.456E-78 & 1.153E-78 & 7.712E-79 & 3.617E-79 & 1.073E-79 & 1.427E-80 & 5.786E-82 \\
$9$ & 5.434E-80 & 1.898E-79 & 4.076E-79 & 6.467E-79 & 8.565E-79 & 1.027E-78 & 1.133E-78 & 1.153E-78 & 9.611E-79 & 6.774E-79 & 3.348E-79 & 1.020E-79 & 1.341E-80 & 5.291E-82 \\
$10$ & 3.216E-80 & 1.100E-79 & 2.380E-79 & 3.813E-79 & 5.174E-79 & 6.354E-79 & 7.219E-79 & 7.712E-79 & 6.774E-79 & 5.084E-79 & 2.652E-79 & 8.345E-80 & 1.087E-80 & 4.195E-82 \\
$11$ & 1.309E-80 & 4.334E-80 & 9.457E-80 & 1.533E-79 & 2.151E-79 & 2.725E-79 & 3.208E-79 & 3.617E-79 & 3.348E-79 & 2.652E-79 & 1.479E-79 & 4.889E-80 & 6.288E-81 & 2.399E-82 \\
$12$ & 3.635E-81 & 1.168E-80 & 2.549E-80 & 4.154E-80 & 5.948E-80 & 7.687E-80 & 9.230E-80 & 1.073E-79 & 1.020E-79 & 8.345E-80 & 4.889E-80 & 1.737E-80 & 2.205E-81 & 8.233E-83 \\
$13$ & 4.997E-82 & 1.627E-81 & 3.561E-81 & 5.790E-81 & 8.203E-81 & 1.047E-80 & 1.245E-80 & 1.427E-80 & 1.341E-80 & 1.087E-80 & 6.288E-81 & 2.205E-81 & 2.935E-82 & 1.116E-83 \\
$14$ & 2.172E-83 & 7.145E-83 & 1.573E-82 & 2.544E-82 & 3.552E-82 & 4.445E-82 & 5.194E-82 & 5.786E-82 & 5.291E-82 & 4.195E-82 & 2.399E-82 & 8.233E-83 & 1.116E-83 & 6.926E-85 \\
\end{tabular}
\end{ruledtabular}
\vspace{1em}
\begin{ruledtabular}
\begin{tabular}{ccccccccccccccc}
\multicolumn{15}{c}{\textbf{Correlation matrix}} \\[0.3em]
Bin & $1$ & $2$ & $3$ & $4$ & $5$ & $6$ & $7$ & $8$ & $9$ & $10$ & $11$ & $12$ & $13$ & $14$ \\
\hline
$1$ & 1.000E+00 & 9.671E-01 & 9.585E-01 & 9.558E-01 & 9.562E-01 & 9.500E-01 & 9.353E-01 & 8.872E-01 & 7.892E-01 & 6.423E-01 & 4.846E-01 & 3.927E-01 & 4.153E-01 & 3.717E-01 \\
$2$ & 9.671E-01 & 1.000E+00 & 9.881E-01 & 9.828E-01 & 9.803E-01 & 9.716E-01 & 9.522E-01 & 8.961E-01 & 7.875E-01 & 6.279E-01 & 4.586E-01 & 3.606E-01 & 3.865E-01 & 3.493E-01 \\
$3$ & 9.585E-01 & 9.881E-01 & 1.000E+00 & 9.936E-01 & 9.904E-01 & 9.833E-01 & 9.653E-01 & 9.112E-01 & 8.045E-01 & 6.459E-01 & 4.758E-01 & 3.742E-01 & 4.021E-01 & 3.656E-01 \\
$4$ & 9.558E-01 & 9.828E-01 & 9.936E-01 & 1.000E+00 & 9.951E-01 & 9.870E-01 & 9.714E-01 & 9.199E-01 & 8.153E-01 & 6.609E-01 & 4.926E-01 & 3.895E-01 & 4.177E-01 & 3.777E-01 \\
$5$ & 9.562E-01 & 9.803E-01 & 9.904E-01 & 9.951E-01 & 1.000E+00 & 9.950E-01 & 9.840E-01 & 9.430E-01 & 8.505E-01 & 7.065E-01 & 5.444E-01 & 4.393E-01 & 4.661E-01 & 4.154E-01 \\
$6$ & 9.500E-01 & 9.716E-01 & 9.833E-01 & 9.870E-01 & 9.950E-01 & 1.000E+00 & 9.940E-01 & 9.634E-01 & 8.846E-01 & 7.525E-01 & 5.984E-01 & 4.925E-01 & 5.160E-01 & 4.510E-01 \\
$7$ & 9.353E-01 & 9.522E-01 & 9.653E-01 & 9.714E-01 & 9.840E-01 & 9.940E-01 & 1.000E+00 & 9.835E-01 & 9.224E-01 & 8.085E-01 & 6.660E-01 & 5.592E-01 & 5.804E-01 & 4.983E-01 \\
$8$ & 8.872E-01 & 8.961E-01 & 9.112E-01 & 9.199E-01 & 9.430E-01 & 9.634E-01 & 9.835E-01 & 1.000E+00 & 9.744E-01 & 8.962E-01 & 7.794E-01 & 6.743E-01 & 6.902E-01 & 5.761E-01 \\
$9$ & 7.892E-01 & 7.875E-01 & 8.045E-01 & 8.153E-01 & 8.505E-01 & 8.846E-01 & 9.224E-01 & 9.744E-01 & 1.000E+00 & 9.691E-01 & 8.882E-01 & 7.893E-01 & 7.983E-01 & 6.485E-01 \\
$10$ & 6.423E-01 & 6.279E-01 & 6.459E-01 & 6.609E-01 & 7.065E-01 & 7.525E-01 & 8.085E-01 & 8.962E-01 & 9.691E-01 & 1.000E+00 & 9.673E-01 & 8.880E-01 & 8.898E-01 & 7.070E-01 \\
$11$ & 4.846E-01 & 4.586E-01 & 4.758E-01 & 4.926E-01 & 5.444E-01 & 5.984E-01 & 6.660E-01 & 7.794E-01 & 8.882E-01 & 9.673E-01 & 1.000E+00 & 9.646E-01 & 9.543E-01 & 7.497E-01 \\
$12$ & 3.927E-01 & 3.606E-01 & 3.742E-01 & 3.895E-01 & 4.393E-01 & 4.925E-01 & 5.592E-01 & 6.743E-01 & 7.893E-01 & 8.880E-01 & 9.646E-01 & 1.000E+00 & 9.765E-01 & 7.505E-01 \\
$13$ & 4.153E-01 & 3.865E-01 & 4.021E-01 & 4.177E-01 & 4.661E-01 & 5.160E-01 & 5.804E-01 & 6.902E-01 & 7.983E-01 & 8.898E-01 & 9.543E-01 & 9.765E-01 & 1.000E+00 & 7.825E-01 \\
$14$ & 3.717E-01 & 3.493E-01 & 3.656E-01 & 3.777E-01 & 4.154E-01 & 4.510E-01 & 4.983E-01 & 5.761E-01 & 6.485E-01 & 7.070E-01 & 7.497E-01 & 7.505E-01 & 7.825E-01 & 1.000E+00 \\
\end{tabular}
\end{ruledtabular}
\normalsize
\label{tab:ptmu_cov_corr_hydrocarbon}
\end{sidewaystable*}

\begin{sidewaystable*}[t]
\centering
\caption{Iron cross-section ($p_{\text{T}}$) covariance and correlation matrices. The underflow bin (bin 0) is omitted because it is zero.}
\scriptsize
\begin{ruledtabular}
\begin{tabular}{ccccccccccccccc}
\multicolumn{15}{c}{\textbf{Covariance matrix}} \\[0.3em]
Bin & $1$ & $2$ & $3$ & $4$ & $5$ & $6$ & $7$ & $8$ & $9$ & $10$ & $11$ & $12$ & $13$ & $14$ \\
\hline
$1$ & 7.489E-81 & 1.986E-80 & 3.869E-80 & 5.531E-80 & 6.581E-80 & 6.735E-80 & 6.055E-80 & 4.193E-80 & 1.833E-80 & 1.473E-81 & 2.202E-83 & 1.797E-81 & 5.986E-82 & 2.340E-83 \\
$2$ & 1.986E-80 & 7.329E-80 & 1.417E-79 & 2.042E-79 & 2.405E-79 & 2.506E-79 & 2.251E-79 & 1.675E-79 & 7.883E-80 & 1.088E-80 & -7.070E-82 & 4.182E-81 & 1.675E-81 & 7.527E-83 \\
$3$ & 3.869E-80 & 1.417E-79 & 3.182E-79 & 4.635E-79 & 5.443E-79 & 5.799E-79 & 5.395E-79 & 4.165E-79 & 2.209E-79 & 5.923E-80 & 8.218E-81 & 8.743E-81 & 3.281E-81 & 1.475E-82 \\
$4$ & 5.531E-80 & 2.042E-79 & 4.635E-79 & 7.495E-79 & 8.804E-79 & 9.470E-79 & 9.086E-79 & 7.483E-79 & 4.453E-79 & 1.732E-79 & 4.580E-80 & 1.743E-80 & 4.881E-81 & 2.149E-82 \\
$5$ & 6.581E-80 & 2.405E-79 & 5.443E-79 & 8.804E-79 & 1.117E-78 & 1.216E-78 & 1.192E-78 & 1.019E-78 & 6.667E-79 & 3.167E-79 & 1.102E-79 & 3.522E-80 & 7.545E-81 & 3.023E-82 \\
$6$ & 6.735E-80 & 2.506E-79 & 5.799E-79 & 9.470E-79 & 1.216E-78 & 1.420E-78 & 1.439E-78 & 1.277E-78 & 9.071E-79 & 4.990E-79 & 1.948E-79 & 5.487E-80 & 9.630E-81 & 3.668E-82 \\
$7$ & 6.055E-80 & 2.251E-79 & 5.395E-79 & 9.086E-79 & 1.192E-78 & 1.439E-78 & 1.616E-78 & 1.508E-78 & 1.173E-78 & 7.523E-79 & 3.260E-79 & 8.673E-80 & 1.258E-80 & 4.456E-82 \\
$8$ & 4.193E-80 & 1.675E-79 & 4.165E-79 & 7.483E-79 & 1.019E-78 & 1.277E-78 & 1.508E-78 & 1.605E-78 & 1.347E-78 & 9.404E-79 & 4.299E-79 & 1.067E-79 & 1.320E-80 & 4.330E-82 \\
$9$ & 1.833E-80 & 7.883E-80 & 2.209E-79 & 4.453E-79 & 6.667E-79 & 9.071E-79 & 1.173E-78 & 1.347E-78 & 1.299E-78 & 9.849E-79 & 4.721E-79 & 1.172E-79 & 1.332E-80 & 4.221E-82 \\
$10$ & 1.473E-81 & 1.088E-80 & 5.923E-80 & 1.732E-79 & 3.167E-79 & 4.990E-79 & 7.523E-79 & 9.404E-79 & 9.849E-79 & 8.541E-79 & 4.303E-79 & 1.099E-79 & 1.231E-80 & 4.059E-82 \\
$11$ & 2.202E-83 & -7.070E-82 & 8.218E-81 & 4.580E-80 & 1.102E-79 & 1.948E-79 & 3.260E-79 & 4.299E-79 & 4.721E-79 & 4.303E-79 & 2.486E-79 & 7.059E-80 & 8.541E-81 & 3.001E-82 \\
$12$ & 1.797E-81 & 4.182E-81 & 8.743E-81 & 1.743E-80 & 3.522E-80 & 5.487E-80 & 8.673E-80 & 1.067E-79 & 1.172E-79 & 1.099E-79 & 7.059E-80 & 2.683E-80 & 3.674E-81 & 1.340E-82 \\
$13$ & 5.986E-82 & 1.675E-81 & 3.281E-81 & 4.881E-81 & 7.545E-81 & 9.630E-81 & 1.258E-80 & 1.320E-80 & 1.332E-80 & 1.231E-80 & 8.541E-81 & 3.674E-81 & 6.890E-82 & 2.602E-83 \\
$14$ & 2.340E-83 & 7.527E-83 & 1.475E-82 & 2.149E-82 & 3.023E-82 & 3.668E-82 & 4.456E-82 & 4.330E-82 & 4.221E-82 & 4.059E-82 & 3.001E-82 & 1.340E-82 & 2.602E-83 & 4.031E-84 \\
\end{tabular}
\end{ruledtabular}
\vspace{1em}
\begin{ruledtabular}
\begin{tabular}{ccccccccccccccc}
\multicolumn{15}{c}{\textbf{Correlation matrix}} \\[0.3em]
Bin & $1$ & $2$ & $3$ & $4$ & $5$ & $6$ & $7$ & $8$ & $9$ & $10$ & $11$ & $12$ & $13$ & $14$ \\
\hline
$1$ & 1.000E+00 & 8.478E-01 & 7.925E-01 & 7.383E-01 & 7.195E-01 & 6.531E-01 & 5.504E-01 & 3.825E-01 & 1.858E-01 & 1.841E-02 & 5.104E-04 & 1.267E-01 & 2.635E-01 & 1.347E-01 \\
$2$ & 8.478E-01 & 1.000E+00 & 9.278E-01 & 8.711E-01 & 8.406E-01 & 7.768E-01 & 6.543E-01 & 4.886E-01 & 2.555E-01 & 4.347E-02 & -5.238E-03 & 9.432E-02 & 2.357E-01 & 1.385E-01 \\
$3$ & 7.925E-01 & 9.278E-01 & 1.000E+00 & 9.490E-01 & 9.129E-01 & 8.627E-01 & 7.524E-01 & 5.829E-01 & 3.437E-01 & 1.136E-01 & 2.922E-02 & 9.463E-02 & 2.216E-01 & 1.302E-01 \\
$4$ & 7.383E-01 & 8.711E-01 & 9.490E-01 & 1.000E+00 & 9.620E-01 & 9.179E-01 & 8.256E-01 & 6.824E-01 & 4.513E-01 & 2.164E-01 & 1.061E-01 & 1.229E-01 & 2.148E-01 & 1.236E-01 \\
$5$ & 7.195E-01 & 8.406E-01 & 9.129E-01 & 9.620E-01 & 1.000E+00 & 9.652E-01 & 8.874E-01 & 7.607E-01 & 5.534E-01 & 3.242E-01 & 2.091E-01 & 2.034E-01 & 2.719E-01 & 1.425E-01 \\
$6$ & 6.531E-01 & 7.768E-01 & 8.627E-01 & 9.179E-01 & 9.652E-01 & 1.000E+00 & 9.502E-01 & 8.457E-01 & 6.679E-01 & 4.531E-01 & 3.278E-01 & 2.811E-01 & 3.079E-01 & 1.533E-01 \\
$7$ & 5.504E-01 & 6.543E-01 & 7.524E-01 & 8.256E-01 & 8.874E-01 & 9.502E-01 & 1.000E+00 & 9.367E-01 & 8.098E-01 & 6.404E-01 & 5.144E-01 & 4.165E-01 & 3.772E-01 & 1.746E-01 \\
$8$ & 3.825E-01 & 4.886E-01 & 5.829E-01 & 6.824E-01 & 7.607E-01 & 8.457E-01 & 9.367E-01 & 1.000E+00 & 9.334E-01 & 8.033E-01 & 6.806E-01 & 5.141E-01 & 3.969E-01 & 1.703E-01 \\
$9$ & 1.858E-01 & 2.555E-01 & 3.437E-01 & 4.513E-01 & 5.534E-01 & 6.679E-01 & 8.098E-01 & 9.334E-01 & 1.000E+00 & 9.351E-01 & 8.308E-01 & 6.278E-01 & 4.454E-01 & 1.845E-01 \\
$10$ & 1.841E-02 & 4.347E-02 & 1.136E-01 & 2.164E-01 & 3.242E-01 & 4.531E-01 & 6.404E-01 & 8.033E-01 & 9.351E-01 & 1.000E+00 & 9.338E-01 & 7.257E-01 & 5.073E-01 & 2.187E-01 \\
$11$ & 5.104E-04 & -5.238E-03 & 2.922E-02 & 1.061E-01 & 2.091E-01 & 3.278E-01 & 5.144E-01 & 6.806E-01 & 8.308E-01 & 9.338E-01 & 1.000E+00 & 8.643E-01 & 6.526E-01 & 2.997E-01 \\
$12$ & 1.267E-01 & 9.432E-02 & 9.463E-02 & 1.229E-01 & 2.034E-01 & 2.811E-01 & 4.165E-01 & 5.141E-01 & 6.278E-01 & 7.257E-01 & 8.643E-01 & 1.000E+00 & 8.546E-01 & 4.074E-01 \\
$13$ & 2.635E-01 & 2.357E-01 & 2.216E-01 & 2.148E-01 & 2.719E-01 & 3.079E-01 & 3.772E-01 & 3.969E-01 & 4.454E-01 & 5.073E-01 & 6.526E-01 & 8.546E-01 & 1.000E+00 & 4.938E-01 \\
$14$ & 1.347E-01 & 1.385E-01 & 1.302E-01 & 1.236E-01 & 1.425E-01 & 1.533E-01 & 1.746E-01 & 1.703E-01 & 1.845E-01 & 2.187E-01 & 2.997E-01 & 4.074E-01 & 4.938E-01 & 1.000E+00 \\
\end{tabular}
\end{ruledtabular}
\normalsize
\label{tab:ptmu_cov_corr_iron}
\end{sidewaystable*}

\begin{sidewaystable*}[t]
\centering
\caption{Lead cross-section ($p_{\text{T}}$) covariance and correlation matrices. The underflow bin (bin 0) is omitted because it is zero.}
\scriptsize
\begin{ruledtabular}
\begin{tabular}{ccccccccccccccc}
\multicolumn{15}{c}{\textbf{Covariance matrix}} \\[0.3em]
Bin & $1$ & $2$ & $3$ & $4$ & $5$ & $6$ & $7$ & $8$ & $9$ & $10$ & $11$ & $12$ & $13$ & $14$ \\
\hline
$1$ & 5.907E-81 & 1.552E-80 & 3.046E-80 & 4.498E-80 & 5.234E-80 & 5.480E-80 & 4.900E-80 & 2.955E-80 & 8.003E-81 & -2.326E-81 & -2.504E-81 & 1.308E-81 & 3.178E-82 & 2.620E-83 \\
$2$ & 1.552E-80 & 5.368E-80 & 1.065E-79 & 1.570E-79 & 1.851E-79 & 1.981E-79 & 1.815E-79 & 1.162E-79 & 4.044E-80 & -3.284E-83 & -5.671E-81 & 4.292E-81 & 1.060E-81 & 8.740E-83 \\
$3$ & 3.046E-80 & 1.065E-79 & 2.476E-79 & 3.694E-79 & 4.357E-79 & 4.806E-79 & 4.518E-79 & 3.092E-79 & 1.327E-79 & 2.582E-80 & -4.515E-81 & 8.158E-81 & 2.005E-81 & 1.622E-82 \\
$4$ & 4.498E-80 & 1.570E-79 & 3.694E-79 & 6.052E-79 & 7.285E-79 & 8.100E-79 & 7.821E-79 & 5.693E-79 & 2.832E-79 & 9.190E-80 & 1.335E-80 & 1.458E-80 & 3.246E-81 & 2.689E-82 \\
$5$ & 5.234E-80 & 1.851E-79 & 4.357E-79 & 7.285E-79 & 9.521E-79 & 1.078E-78 & 1.074E-78 & 8.466E-79 & 5.010E-79 & 2.345E-79 & 7.554E-80 & 3.155E-80 & 5.533E-81 & 4.204E-82 \\
$6$ & 5.480E-80 & 1.981E-79 & 4.806E-79 & 8.100E-79 & 1.078E-78 & 1.307E-78 & 1.346E-78 & 1.124E-78 & 7.413E-79 & 3.981E-79 & 1.479E-79 & 4.801E-80 & 7.565E-81 & 4.859E-82 \\
$7$ & 4.900E-80 & 1.815E-79 & 4.518E-79 & 7.821E-79 & 1.074E-78 & 1.346E-78 & 1.515E-78 & 1.369E-78 & 1.004E-78 & 6.135E-79 & 2.577E-79 & 7.431E-80 & 1.068E-80 & 6.594E-82 \\
$8$ & 2.955E-80 & 1.162E-79 & 3.092E-79 & 5.693E-79 & 8.466E-79 & 1.124E-78 & 1.369E-78 & 1.470E-78 & 1.227E-78 & 8.418E-79 & 3.893E-79 & 1.057E-79 & 1.402E-80 & 6.813E-82 \\
$9$ & 8.003E-81 & 4.044E-80 & 1.327E-79 & 2.832E-79 & 5.010E-79 & 7.413E-79 & 1.004E-78 & 1.227E-78 & 1.188E-78 & 8.757E-79 & 4.254E-79 & 1.131E-79 & 1.450E-80 & 6.127E-82 \\
$10$ & -2.326E-81 & -3.284E-83 & 2.582E-80 & 9.190E-80 & 2.345E-79 & 3.981E-79 & 6.135E-79 & 8.418E-79 & 8.757E-79 & 7.081E-79 & 3.553E-79 & 9.534E-80 & 1.222E-80 & 4.654E-82 \\
$11$ & -2.504E-81 & -5.671E-81 & -4.515E-81 & 1.335E-80 & 7.554E-80 & 1.479E-79 & 2.577E-79 & 3.893E-79 & 4.254E-79 & 3.553E-79 & 1.969E-79 & 5.767E-80 & 7.612E-81 & 2.864E-82 \\
$12$ & 1.308E-81 & 4.292E-81 & 8.158E-81 & 1.458E-80 & 3.155E-80 & 4.801E-80 & 7.431E-80 & 1.057E-79 & 1.131E-79 & 9.534E-80 & 5.767E-80 & 2.393E-80 & 3.377E-81 & 1.309E-82 \\
$13$ & 3.178E-82 & 1.060E-81 & 2.005E-81 & 3.246E-81 & 5.533E-81 & 7.565E-81 & 1.068E-80 & 1.402E-80 & 1.450E-80 & 1.222E-80 & 7.612E-81 & 3.377E-81 & 6.050E-82 & 2.478E-83 \\
$14$ & 2.620E-83 & 8.740E-83 & 1.622E-82 & 2.689E-82 & 4.204E-82 & 4.859E-82 & 6.594E-82 & 6.813E-82 & 6.127E-82 & 4.654E-82 & 2.864E-82 & 1.309E-82 & 2.478E-83 & 5.731E-84 \\
\end{tabular}
\end{ruledtabular}
\vspace{1em}
\begin{ruledtabular}
\begin{tabular}{ccccccccccccccc}
\multicolumn{15}{c}{\textbf{Correlation matrix}} \\[0.3em]
Bin & $1$ & $2$ & $3$ & $4$ & $5$ & $6$ & $7$ & $8$ & $9$ & $10$ & $11$ & $12$ & $13$ & $14$ \\
\hline
$1$ & 1.000E+00 & 8.717E-01 & 7.965E-01 & 7.523E-01 & 6.979E-01 & 6.236E-01 & 5.179E-01 & 3.172E-01 & 9.556E-02 & -3.596E-02 & -7.344E-02 & 1.100E-01 & 1.681E-01 & 1.424E-01 \\
$2$ & 8.717E-01 & 1.000E+00 & 9.239E-01 & 8.709E-01 & 8.188E-01 & 7.479E-01 & 6.363E-01 & 4.138E-01 & 1.602E-01 & -1.684E-04 & -5.516E-02 & 1.197E-01 & 1.861E-01 & 1.576E-01 \\
$3$ & 7.965E-01 & 9.239E-01 & 1.000E+00 & 9.544E-01 & 8.975E-01 & 8.449E-01 & 7.376E-01 & 5.126E-01 & 2.447E-01 & 6.167E-02 & -2.045E-02 & 1.060E-01 & 1.638E-01 & 1.362E-01 \\
$4$ & 7.523E-01 & 8.709E-01 & 9.544E-01 & 1.000E+00 & 9.598E-01 & 9.107E-01 & 8.167E-01 & 6.036E-01 & 3.340E-01 & 1.404E-01 & 3.868E-02 & 1.211E-01 & 1.696E-01 & 1.444E-01 \\
$5$ & 6.979E-01 & 8.188E-01 & 8.975E-01 & 9.598E-01 & 1.000E+00 & 9.666E-01 & 8.940E-01 & 7.157E-01 & 4.712E-01 & 2.856E-01 & 1.745E-01 & 2.090E-01 & 2.305E-01 & 1.800E-01 \\
$6$ & 6.236E-01 & 7.479E-01 & 8.449E-01 & 9.107E-01 & 9.666E-01 & 1.000E+00 & 9.566E-01 & 8.113E-01 & 5.949E-01 & 4.138E-01 & 2.915E-01 & 2.714E-01 & 2.690E-01 & 1.775E-01 \\
$7$ & 5.179E-01 & 6.363E-01 & 7.376E-01 & 8.167E-01 & 8.940E-01 & 9.566E-01 & 1.000E+00 & 9.173E-01 & 7.481E-01 & 5.923E-01 & 4.718E-01 & 3.902E-01 & 3.528E-01 & 2.237E-01 \\
$8$ & 3.172E-01 & 4.138E-01 & 5.126E-01 & 6.036E-01 & 7.157E-01 & 8.113E-01 & 9.173E-01 & 1.000E+00 & 9.288E-01 & 8.253E-01 & 7.238E-01 & 5.634E-01 & 4.702E-01 & 2.348E-01 \\
$9$ & 9.556E-02 & 1.602E-01 & 2.447E-01 & 3.340E-01 & 4.712E-01 & 5.949E-01 & 7.481E-01 & 9.288E-01 & 1.000E+00 & 9.549E-01 & 8.798E-01 & 6.707E-01 & 5.409E-01 & 2.348E-01 \\
$10$ & -3.596E-02 & -1.684E-04 & 6.167E-02 & 1.404E-01 & 2.856E-01 & 4.138E-01 & 5.923E-01 & 8.253E-01 & 9.549E-01 & 1.000E+00 & 9.515E-01 & 7.324E-01 & 5.906E-01 & 2.310E-01 \\
$11$ & -7.344E-02 & -5.516E-02 & -2.045E-02 & 3.868E-02 & 1.745E-01 & 2.915E-01 & 4.718E-01 & 7.238E-01 & 8.798E-01 & 9.515E-01 & 1.000E+00 & 8.403E-01 & 6.975E-01 & 2.697E-01 \\
$12$ & 1.100E-01 & 1.197E-01 & 1.060E-01 & 1.211E-01 & 2.090E-01 & 2.714E-01 & 3.902E-01 & 5.634E-01 & 6.707E-01 & 7.324E-01 & 8.403E-01 & 1.000E+00 & 8.875E-01 & 3.535E-01 \\
$13$ & 1.681E-01 & 1.861E-01 & 1.638E-01 & 1.696E-01 & 2.305E-01 & 2.690E-01 & 3.528E-01 & 4.702E-01 & 5.409E-01 & 5.906E-01 & 6.975E-01 & 8.875E-01 & 1.000E+00 & 4.208E-01 \\
$14$ & 1.424E-01 & 1.576E-01 & 1.362E-01 & 1.444E-01 & 1.800E-01 & 1.775E-01 & 2.237E-01 & 2.348E-01 & 2.348E-01 & 2.310E-01 & 2.697E-01 & 3.535E-01 & 4.208E-01 & 1.000E+00 \\
\end{tabular}
\end{ruledtabular}
\normalsize
\label{tab:ptmu_cov_corr_lead}
\end{sidewaystable*}

% Generator Configurations
\begin{table*}[t]
\caption{Details of generator configurations used for comparison with the data. Additional discussion of interaction physics content
can be found in Refs.~\cite{BETANCOURT20181,PhysRevD.105.092004}.}
\label{tab:generators}
\footnotesize
\newcommand{\desc}[1]{\parbox[t]{13cm}{\raggedright #1}}
\begin{flushleft}
\begin{tabular}{ll}
\hline\hline
Name/Label & Description \\
\hline
(MINERvA) Tune v4.3.0 &
\desc{Central-value Monte Carlo used in this analysis based on GENIE v2.12.6~\cite{ANDREOPOULOS201087} with modifications as described in the main text.} \\[8pt]
GENIE v3 G18\_02a & \desc{GENIE v3.0.6 G18\_02a\_02\_11a~\cite{ANDREOPOULOS201087}. Relativistic Fermi gas nuclear model with a Bodek--Ritchie tail~\cite{RFG_SMITH1972605, BodekRitchie1981_PhysRevD.24.1400}, Llewellyn-Smith quasi-elastic model~\cite{LLEWELLYNSMITH1972261}, empirical 2p2h correlations~\cite{10.1063/1.4919465}, tuned Berger-Sehgal model for single pion~\cite{PhysRevD.76.113004} and coherent pion production~\cite{PhysRevD.79.053003}, Bodek--Yang model~\cite{BodekYang_2003, BODEK2005113} for deep inelastic scattering and AGKY/PYTHIA hadronisation model~\cite{AGKY_Yang_2009, KNO_KOBA1972317, Pythia_Sjostrand_2001}, and hA2018 final state interactions~\cite{Dytman:2009zz}. Note that this configuration includes a tune of resonances, and 1 and 2$\pi$ non-resonant pion production processes (preliminary version of the tune in Ref.~\cite{PhysRevD.104.072009}); resonances are included up to $W$ of 1.93 GeV/$c^2$.} \\[8pt]
GENIE v3 G18\_02b & \desc{GENIE v3.0.6 G18\_02b\_02\_11a. Identical to G18\_02a except that the final state interaction option is hN. Differences with the hA prediction are negligible; lines visually overlap in plots.} \\[8pt]
GENIE v3 G18\_10a & \desc{GENIE v3.0.6 G18\_10a\_02\_11a. Val\`{e}ncia local Fermi gas nuclear model with random phase approximation for quasi-elastic scattering~\cite{RPANieves2004_PhysRevC.70.055503, gran2017model} and the Val\`{e}ncia 2p2h~\cite{PhysRevC.83.045501, PhysRevD.88.113007}. Other components are the same as in GENIE v3 G18\_02a.} \\[8pt]
GENIE v3 G18\_10b & \desc{GENIE v3.0.6 G18\_10b\_02\_11a. Identical to G18\_10a except that the final state interaction option is hN. Differences with the hA prediction are negligible; lines visually overlap in plots.} \\[8pt]
GENIE v3 AR23 (DUNE tune) & \desc{GENIE v3.04.0 AR23\_20i\_00\_000 tune, built on top of G18\_10a. Introduces an extended treatment of the nuclear ground state to provide broader coverage of nucleon removal energies and momenta, replaces the Val\`{e}ncia 2p2h model with SuSAv2, adopts the $z$-expansion parametrisation for nucleon form factors in charged-current quasi-elastic interactions, and includes nuclear de-excitation photon emission for carbon and argon targets.} \\[8pt]
NEUT LFG & \desc{NEUT version 5.4.1~\cite{NEUT2021}. Val\`{e}ncia local Fermi gas nuclear model with random phase approximation for quasi-elastic scattering~\cite{RPANieves2004_PhysRevC.70.055503, gran2017model}, Val\`{e}ncia 2p2h~\cite{PhysRevC.83.045501, PhysRevD.88.113007}, the updated Rein--Sehgal model for single-pion production~\cite{REIN198179}, and the Berger--Sehgal model for coherent pion production~\cite{PhysRevD.79.053003}. In this NEUT configuration, the coherent pion production scales as $A$. A custom multi-pion model is used for $W < 2\ \mathrm{GeV}/c^2$, based on the Bodek--Yang model~\cite{BodekYang_2003, BODEK2005113} with KNO scaling~\cite{KNO_KOBA1972317}. For $W > 2\ \mathrm{GeV}/c^2$, the simulation is purely DIS, with the kinematics of the produced particles determined by PYTHIA~\cite{Pythia_Sjostrand_2001}.} \\[8pt]
NEUT SF & \desc{NEUT version 5.4.1~\cite{NEUT2021}. Same as NEUT LFG except the quasi-elastic scattering model uses the spectral function by Benhar et al.~\cite{BENHAR1994493} with the implementation based on the one in NuWro~\cite{2006NuPhS.159..211J}. Note that there is no spectral function prediction for lead.} \\[8pt]
\hline\hline
\end{tabular}
\end{flushleft}
\normalsize
\end{table*}

\begin{table}[t]
\centering
\caption{$\chi^2$ comparison of models and data for $p_\text{T}$ cross section (ndf = 13).}
\begin{ruledtabular}
\begin{tabular}{lcccc}
\textbf{Model} & \textbf{C} & \textbf{CH} & \textbf{Fe} & \textbf{Pb} \\
\hline
MINERvA Tune v4.3.0       & 24.08 & 239.47 & 61.04 & 92.86  \\
GENIE v3 G18\_02a         & 20.96 & 72.57  & 16.09 & 34.23  \\
GENIE v3 G18\_02b         & 20.77 & 72.28  & 16.32 & 32.98  \\
GENIE v3 G18\_10a         & 17.29 & 118.36 & 36.14 & 65.79  \\
GENIE v3 G18\_10b         & 17.63 & 110.23 & 35.97 & 67.44  \\
GENIE v3 AR23 (DUNE Tune) & 14.81 & 57.18  & 34.74 & 64.51  \\
NEUT v5.4.1 LFG           & 25.31 & 410.83 & 76.30 & 185.39 \\
NEUT v5.4.1 SF            & 26.07 & 601.36 & 93.84 & --     \\
\end{tabular}
\end{ruledtabular}
\label{tab:pt_chi2_abs}
\end{table}

\begin{table}[t]
\centering
\caption{$\chi^2$ comparison of models and data for $p_\text{T}$ cross-section ratios (ndf = 13).}
\begin{ruledtabular}
\begin{tabular}{lccc}
\textbf{Model} & \textbf{C/CH} & \textbf{Fe/CH} & \textbf{Pb/CH} \\
\hline
MINERvA Tune v4.3.0       & 20.05 & 21.37 & 77.61  \\
GENIE v3 G18\_02a         & 20.53 & 27.00 & 46.76  \\
GENIE v3 G18\_02b         & 20.66 & 27.08 & 43.90  \\
GENIE v3 G18\_10a         & 19.48 & 20.13 & 45.60  \\
GENIE v3 G18\_10b         & 19.33 & 19.71 & 47.06  \\
GENIE v3 AR23 (DUNE Tune) & 18.91 & 27.10 & 47.05  \\
NEUT v5.4.1 LFG           & 19.45 & 25.01 & 155.13 \\
NEUT v5.4.1 SF            & 18.70 & 34.44 & --     \\
\end{tabular}
\end{ruledtabular}
\label{tab:pt_chi2_ratios}
\end{table}

\clearpage
%\bibliographystyle{apsrev4-1}
%\bibliography{biblio}% Produces the bibliography via BibTeX.
%\end{document}

\end{document}